\documentclass[aps,prb,twocolumn,superscriptaddress,eqsecnum,nofootinbib,showpacs]{revtex4}
\usepackage{amssymb}
\usepackage{graphicx}
\usepackage{epsf,epsfig}
\usepackage{bm}
\usepackage{bbm}
\usepackage{amsfonts}

\newcommand{\be}{\begin{eqnarray}}
\newcommand{\ee}{\end{eqnarray}}
\newcommand{\ba}{\begin{array}}
\newcommand{\ea}{\end{array}}
\newcommand{\no}{\nonumber}

\newcommand{\eps}{\varepsilon}

\newcommand{\bfr}{{\bf r}}

\newcommand{\bfn}{{\bf n}}

\newcommand{\bfDelta}{{\bm \Delta}}

\begin{document}

\title{Temperature dependence of the spin susceptibility of a clean Fermi
gas with repulsion.}
\author{G. Schwiete}
\email{georg.schwiete@weizmann.ac.il} \affiliation{Department of
Condensed Matter Physics, The Weizmann Institute of Science, 76100
Rehovot, Israel}
\author{K. B. Efetov}
\affiliation{Theoretische Physik III, Ruhr-Universit\"at Bochum, 44780 Bochum, Germany\\
and L. D. Landau Institute for Theoretical Physics, 117940 Moscow, Russia}
\date{\today}

\begin{abstract}
Spin susceptibility of a clean Fermi gas with repulsion in any dimension is
considered using a supersymmetric low energy theory of interacting spin
excitations and renormalization scheme recently proposed by Aleiner and
Efetov [1]. We generalize this method to include the coupling to the
magnetic field. As a result, we obtain for the correction $\delta \chi $ to
the Pauli susceptibility a non-analytic temperature dependence of the form $%
T^{d-1}\gamma _{b}^{2}(T)$ in dimensions $d=2,3,$ where $\gamma _{b}(T)$ is
an effective $d$-dependent logarithmically renormalized backscattering
amplitude. In one dimension, $\delta \chi $ is proportional to $\gamma
_{b}(T)$, and we reproduce a well known result obtained long ago by a direct
calculation.
\end{abstract}

 \maketitle


\section{\label{sec:Introduction}Introduction}

According to Landau's Fermi liquid theory\cite{landau} low energy properties
of interacting fermion systems are similar to those of an ideal Fermi gas.
For many purposes interaction effects can be incorporated into renormalized
parameters like the effective mass. Using the Landau theory one might expect
that thermodynamic quantities like $C(T)/T$ and $\chi (T)$, where $C(T)$ is
the specific heat and $\chi (T)$ is the spin susceptibility, have an
expansion in powers of $T^{2}$, as known from the standard Sommerfeld
expansion for the ideal Fermi gas.

This would mean extending the analogy too far, however. In fact,
it is known that interaction between fermions can induce so-called
non-analytic corrections, that are absent for the ideal Fermi gas.
The form of these corrections strongly depends on the
dimensionality of the system. As a result of theoretical studies
the leading temperature dependence of $C(T)/T$ was found to be
$T^{2}\ln T$ in $d=3$ and $T$ in $d=2$ at low temperatures, see
Refs.
\onlinecite{eliash,doniach,brink,amit,chubukov4,houghton1,neto,
pethick1} and Refs. \onlinecite{coffey,chubukov1,sarma,Catelani,
chubukov3} respectively. For the spin susceptibility a linear in
$T$ dependence in $d=2$ was obtained in Refs. \onlinecite{baranov,
belitz, hirashima, misawa, chitov, chubukov1, chubukov3, sarma,
betouras}, while for $d=3$ a non-analytic $Q^{2}\ln Q$ dependence
of the wave-vector dependent spin susceptibility\cite
{belitz,baranov,chitov,chubukov2,chubukov3} was found not to be
paralleled by a similar temperature dependence.

The model of a weakly interacting Fermi gas allows for a controlled
perturbative expansion in the strength of the interaction potential. In such
an approach the corrections cited above appear in the second order of
perturbation theory. However, calculations in higher orders are difficult
and a full analysis of high order corrections has not been performed
previously.

Recently, a new low energy supersymmetric field theory for weakly
interacting electrons was introduced in Ref. \cite{aleiner}. This is some
kind of a bosonization that includes not only charge but also spin degrees
of freedom. The scattering processes responsible for the non-analytic
corrections are quasi one-dimensional in character, which leads to
logarithmic contributions originating from interactions between spin
excitations. As concerns the charge excitations in $d>1$ studied previously
within other bosonization schemes \cite%
{luther,haldane,houghton1,houghton,neto,kopietz,kopietzb,
khveshchenko,khveshchenko1,castellani,metzner}, their contribution is less
singular and does not lead to the effects discovered in Ref. \cite{aleiner}.
The low energy theory of Ref. \cite{aleiner} resembles the supersymmetry
approach\cite{review,book}, well known in the theory of disordered systems
and random matrix theory.

In the model of interacting spin excitations, logarithmic corrections to
interaction amplitudes were found in any dimension, motivating a
renormalization group study. The origin of the logarithms can be easily
understood. Consider the interaction of two spin modes with propagators $%
(v_{F}\mathbf{n}\mathbf{p}-i\omega )^{-1}$, but opposite directions $\mathbf{%
n}$, $-\mathbf{n}$ on the Fermi surface. Then one finds for perturbative
corrections to the interaction amplitudes integrals of the type (compare
Fig. \ref{fig:intamplitudes} below)
\be
T\sum_{\omega }\int d\mathbf{p}\frac{1}{(v_{F}\mathbf{n}\mathbf{p}%
)^{2}+\omega ^{2}}\propto \ln (\varepsilon _{\infty }/T) \ee
$\varepsilon _{\infty }$ is the largest energy in the model.

The integral over momenta transverse to $\mathbf{n}$ has to be regularized.
Of course, in dimensions $d>1$, the region of phase space for which the
excitations move in almost antiparallel directions $\mathbf{n}_{1}\sim -%
\mathbf{n}_{2}$ is rather restricted and in general the logarithm is cut by $%
\max [T/\varepsilon _{\infty },|\mathbf{n}_{1}+\mathbf{n}_{2}|]$. The
question arises how the logarithms found at this level affect the
temperature dependence of observable quantities, since eventually one needs
to perform a (weighted) average over all directions. To this end the
specific heat was studied in Ref.~\onlinecite{aleiner}. It was found that $%
C(T)/T$ generally depends on an effective amplitude of backward scattering,
that displays a complicated dependence on $\ln T$, in such a way that the
results reduced to well known $\delta C_{d=2}\propto T^{2}$ and $\delta
C_{d=3}\propto T^{3}\ln T$ when replacing the effective backscattering
constants by the bare ones.

The evident question arises whether similar corrections exist for the spin
susceptibility. We will show in this paper that this is really so. To this
end we generalize the formalism introduced in Ref.\onlinecite{aleiner} to
include the external magnetic field. A convenient diagrammatic
representation makes apparent the relation to conventional diagrammatic
approaches. It turns out that, unlike for the thermodynamic potential, the
contributions to the spin susceptibility in $d>1$ are determined by all the
renormalized interaction amplitudes of the model in various combinations. In
$d=2$ we find non-analytic contributions $\delta \chi (T)\propto T\gamma
_{b}^{2}(T)$ and $\gamma _{b}(T)$ is an effective backscattering constant
that includes logarithmic corrections. In $d=3$ we confirm the absence of $%
T^{2}\ln T$ terms in the second order in the interaction. This does however
not mean the absence of non-analytic corrections in three dimensions. Here
we sum leading logarithmic corrections to the $T^{2}$ behavior and come to
the result $\delta \chi (T)\propto T^{2}\gamma _{b}^{2}(T)$. Within the same
formalism the one-dimensional case can be also considered and we reproduce
the temperature dependence first obtained by Dzyaloshinskii and Larkin\cite%
{dl}. Although the one-dimensional case is not the main focus of the
approach, we consider the result obtained as an important check of the
overall consistency.

The paper is organized as follows. In Sec. \ref{sec:Model} we introduce the
model that serves as a starting point for our subsequent analysis. In Sec. %
\ref{sec:Derivation of the bosonized action} we decouple the interaction
part via a Hubbard-Stratonovich transformation in both the charge and spin
channel and reformulate the partition function in terms of charge and spin
excitations. Charge and the spin excitations decouple from each other and
calculating the spin susceptibility we can concentrate on the spin sector. A
representation for the partition function in this sector is derived in Sec. %
\ref{sec:susyrepresentation} using the supersymmetry technique. A low energy
effective action is obtained and rules of the perturbation theory are
introduced as well as a convenient diagrammatic representation. In Sec. \ref%
{sec:Renormalization and Perturbation Theory} we analyze corrections to the
magnetic field vertices in a renormalization group scheme. In Sec. \ref%
{sec:sus1d} we study in our formalism temperature dependence of
the spin susceptibility in one spatial dimension. Then we turn to
the calculation of non-analytic corrections to the temperature
dependent spin susceptibility in two and three dimensions in Sec.
\ref{sec:Non-analytic corrections to spin susceptibility in
$d=2,3$} before concluding with a discussion of our results in
Sec.~ \ref{sec:discussion}.

\section{\label{sec:Model}Model}

We introduce our model by specifying the partition function in the imaginary
time formalism\cite{agd,negele}
\be
\mathcal{Z}=\int \mathcal{D}(\chi ^{\ast },\chi )\;\exp (-\mathcal{S}).
\ee
The fermionic fields $\chi ,\chi ^{\ast }$ depend on coordinates $\mathbf{r}$
and imaginary time $\tau $ and carry a spin label $\sigma $. They obey the
antiperiodic boundary conditions $\chi _{\sigma }(\mathbf{r},\tau )=-\chi
_{\sigma }^{\ast }(\mathbf{r},\tau +\beta )$, where $\beta =1/T$ and $T$ is
the temperature.

We write the action $\mathcal{S}$ as the sum of three parts
\be
\mathcal{S}=\mathcal{S}_{0}+\mathcal{S}_{b}+\mathcal{S}_{int}.
\ee
$\mathcal{S}_{0}$ describes free motion, $\mathcal{S}_{b}$ stands for the
coupling of the spin to an external field $\mathbf{b}$ and $\mathcal{S}%
_{int} $ is the interaction of fermions,
\begin{eqnarray}
\mathcal{S}_{0} &=&\int dx\;\chi _{\sigma }^{\ast }(x)\Big[-\partial _{\tau
}-\hat{H}_{0}\Big]\chi _{\sigma }(x),  \label{eq:2.3} \\
\mathcal{S}_{b} &=&\int dx\;\chi _{\sigma }^{\ast }(x)\mathbf{b}{\bm\sigma }%
_{\sigma \sigma ^{\prime }}\chi _{\sigma ^{\prime }}(x),  \label{eq:2.4} \\
\mathcal{S}_{int} &=&\frac{1}{2}\int dxdx'\;\chi_{\sigma }^{\ast
}(x)\chi _{\sigma ^{\prime }}^{\ast }(x^{\prime })v(x-x^{\prime
})\chi _{\sigma
^{\prime }}(x^{\prime })\chi _{\sigma }(x).  \nonumber \\
&&
\end{eqnarray}%
Here and in the following summation over repeated spin indices is implied
and $v(x-x^{\prime })=V(\mathbf{r}-\mathbf{r^{\prime }})\delta (\tau -\tau
^{\prime })$, where $V(\mathbf{r}-\mathbf{r^{\prime }})$ is the interaction
potential. We use the notation
\be
x=(\mathbf{r},\tau )\quad \int dx=\int_{0}^{\beta }d\tau \int
d\mathbf{r}\label{eq:2.6}
\ee
and $\hat{H}_{0}=\frac{\hat{\mathbf{p}}^{2}}{2m}-\mu $, where $\hat{\mathbf{p%
}}$ is the momentum operator and $\mu $ is the chemical potential. The
standard relation between the partition function $\mathcal{Z}$ and the
thermodynamic potential $\Omega $ is
\be
\Omega =-T\ln {\mathcal{Z}}.
\ee
The spin susceptibility is then obtained as the second derivative with
respect to $b$
\be
\chi(x-x^{\prime })=-\left. \frac{\delta ^{2}\Omega \lbrack
b]}{\delta b(x)\delta b(x^{\prime })}\right\vert _{b=0}\;.
\ee
We will mostly be interested in the static spin susceptibility $\chi _{s}$
for a spatially homogeneous external field, in which case after Fourier
transform the limit of vanishing frequencies should be taken before taking
the external wave-vector to zero.

Let us also introduce the following convention for the integration over
momenta
\be
\int d\mathbf{q}=\int \frac{d^{d}q}{(2\pi )^{d}}
\ee%
in dimension $d$. For the sake of notational convenience we sometimes write $%
\int_{\mathbf{r}}$, $\int_{\mathbf{q}}$, $\int_{\tau }$, symbolizing $\int d%
\mathbf{q}$, $\int d\mathbf{r}$ and $\int_{0}^{\beta }d\tau $ respectively. $%
\int d\hat{\mathbf{n}}$ stands for the integration over the solid angle
normalized to unity.

\section{\label{sec:Derivation of the bosonized action}Derivation of the
bosonized action}

In this section we present the derivation of the model that will be used for
the further analysis of the spin susceptibility. It describes low lying
charge and spin excitations in the system. A derivation in the absence of
external sources has been presented in Ref. \onlinecite{aleiner}. Here we
include the coupling to the magnetic field, so that we will mainly focus on
the changes introduced by the magnetic field.

\subsection{\label{subsec:Decoupling}Decoupling into slow pairs}

For the interaction part we perform decoupling in two different channels by
singling out slow pairs in the following way
\begin{eqnarray}
\mathcal{S}_{int} &\rightarrow &\tilde{\mathcal{S}}_{int}=\tilde{S}_{int,1}+%
\tilde{S}_{int,2}\;,  \label{eq:3.1} \\
\tilde{S}_{int,1} &=&\int dp_{1}dp_{2}dq\;V(\mathbf{q})  \label{eq:3.2} \\
&&\times \chi _{\sigma }^{\ast }(p_{1})\chi _{\sigma }(p_{1}-q)\chi _{\sigma
^{\prime }}^{\ast }(p_{2})\chi _{\sigma ^{\prime }}(p_{2}+q)\;,  \nonumber \\
\tilde{S}_{int,2} &=&-\int dp_{1}dp_{2}dq\;V(\mathbf{p}_{1}-\mathbf{p}_{2}-%
\mathbf{q})  \label{eq:3.3} \\
&&\times \chi _{\sigma }^{\ast }(p_{1})\chi _{\sigma ^{\prime
}}(p_{1}-q)\chi _{\sigma ^{\prime }}^{\ast }(p_{2})\chi _{\sigma
}(p_{2}+q)\;.  \nonumber
\end{eqnarray}%
Here we denoted
\be
\int dp_{i}=T\sum_{\varepsilon _{n_{i}}}\int d\mathbf{p}_{i}\;,
\ee%
where the sum goes over fermionic frequencies $\varepsilon _{n_{i}}=\pi
T(2n_{i}+1)$ and
\be
\int dq=T\sum_{\Omega _{n}}\int
d\mathbf{q}f(\mathbf{q})\;,\label{eq:3.5}
\ee%
where the sum goes over bosonic frequencies $\Omega _{n}=2\pi nT$, further $%
p_{i}=(\mathbf{p}_{i},\varepsilon _{n_{i}})$ and $q=(\mathbf{q},\Omega _{n})$%
.

The cutoff function $f$, introduced in Eq.~(\ref{eq:3.5}), is defined as
follows
\be
f(\mathbf{p})=f_{0}(pr_{0}),\qquad p=|\mathbf{p}|\label{eq:3.6}
\ee
where $f_{0}(t)$ has the properties $f_{0}(0)=1$ and $f(t)\rightarrow 0$
smoothly for $t\rightarrow \infty $. The function $f$ has been introduced to
avoid double-counting when singling out regions of small momentum transfer
in Eqs.~(\ref{eq:3.1})--(\ref{eq:3.3}), since without the cutoff both $%
\tilde{\mathcal{S}}_{int,1}$ and $\tilde{\mathcal{S}}_{int,2}$ would
identically reproduce the original $\mathcal{S}_{int}$. Accordingly, $%
k_{c}=r_{0}^{-1}$ is a momentum cutoff that is much smaller than
the Fermi momentum $k_{c}\ll p_{F}$, but much larger than typical
momenta for the excitations of the low-energy theory that we wish
to construct. We denote also the cutoff energy $\varepsilon
_{\infty }=r_{0}^{-1}v_{F}$. Additional decoupling in the Cooper
channel is not included, since this would amount to overcounting
of relevant scattering processes (compare the related discussion
in Ref. \onlinecite{aleiner}, Sec.~II A).

For a short range potential we can further simplify our
considerations by setting $V_2=V(\mathbf{q}\ll p_F)$. Since
important momenta are close to the
Fermi surface we can write $V_1(\theta_{12})=V(\mathbf{p}_1-\mathbf{p}%
_2)=V(2p_0\sin(\frac{\theta_{12}}{2}))$, where $\theta_{12}$ is the angle
between momenta $\mathbf{p}_1$ and $\mathbf{p}_2$, $\theta_{12}=\widehat{%
\mathbf{p}_1\mathbf{p}_2}$.

For the further development of the theory it will be crucial to separate
explicitly interactions in the triplet and singlet channel.
\be
V_{s}(\theta _{12})=V_{2}-\frac{1}{2}V_{1}(\theta _{12}),\quad V_{t}(\theta
_{12})=\frac{1}{2}V_{1}(\theta _{12})\qquad
\ee%
The action separates into a charge and a spin sector,
\begin{eqnarray}
\tilde{\mathcal{S}}_{int} &=&\mathcal{S}_{int,s}+\mathcal{S}_{int,t}\;, \\
\mathcal{S}_{int,s} &=&\frac{1}{2}\int dp_{1}dp_{2}dq\;\rho
(p_{1},-q)V_{s}(\theta _{12})\rho (p_{2},q)\;,  \nonumber \\
\mathcal{S}_{int,t} &=&-\frac{1}{2}\int dp_{1}dp_{2}dq\;\mathbf{S}%
(p_{1},-q)V_{t}(\theta _{12})\mathbf{S}(p_{2},q)\;,\quad  \nonumber
\end{eqnarray}%
where the charge $\rho (p,q)$ and spin densities $\mathbf{S}(p,q)$ are
\begin{eqnarray}
\rho (p,q) &=&\chi ^{\dagger }\left( p-\frac{q}{2}\right) \chi \left( p+%
\frac{q}{2}\right) \;, \\
\mathbf{S}(p,q) &=&\chi ^{\dagger }\left( p-\frac{q}{2}\right) {\bm\sigma }%
\chi \left( p+\frac{q}{2}\right) \;,
\end{eqnarray}%
and we turned to a spinor notation $\chi =(\chi _{\uparrow },\chi
_{\downarrow })$.

Finally, one may decouple the interaction term $\tilde{\mathcal{S}}_{int}$
using a Hubbard-Stratonovich transformation with a field $\phi _{\mathbf{n}%
}(x)\equiv i\varphi _{\mathbf{n}}(x)+{{\bm\sigma }\mathbf{h}}_{\mathbf{n}%
}(x) $. Here $\varphi _{\mathbf{n}}(x)$ and $\mathbf{h}_{\mathbf{n}}(x)$ are
real bosonic fields, so that $\phi _{\mathbf{n}}(\mathbf{r},\tau )=\phi _{%
\mathbf{n}}(\mathbf{r},\tau +\beta )$ and $\mathbf{n}$ is the direction of
momentum $\mathbf{p}$ on the Fermi surface, $\mathbf{n}=\mathbf{p}/|\mathbf{p%
}|$. The result is the following representation of the partition function
\be
\mathcal{Z}=\mathcal{N}\int \mathcal{D}\phi \;W_{s}[\varphi ]W_{t}[\mathbf{h}%
]\mathcal{Z}[\mathbf{b},\mathbf{h},\varphi ]\;.
\ee%
The weight functions $W_{s}$, $W_{t}$ are shown below in Eqs.~(\ref{eq:Ws}),
(\ref{eq:Wt}) and $\mathcal{N}$ is a simple normalization constant that will
not be displayed from now on. The partition function $\mathcal{Z}[\mathbf{b},%
\mathbf{h},\varphi ]$ describes the fermion motion for fixed configuration
of fields $\mathbf{b},\mathbf{h},\varphi $,
\be
\mathcal{Z}[\mathbf{b},\mathbf{h},\varphi ]=\int D(\chi ^{\ast },\chi
)\;\exp (-S_{eff}[\mathbf{b},\mathbf{h},\varphi ]).
\ee%
where the effective action $\mathcal{S}_{eff}$ has the form
\begin{eqnarray}
&&S_{eff}[\mathbf{b},\mathbf{h},\varphi ]=\mathcal{S}_{0}+\mathcal{S}_{b}[%
\mathbf{b}]+\int d\mathbf{p}d\mathbf{r}_{1}d\mathbf{r}_{2}  \nonumber \\
&&\qquad \chi ^{\dagger }(\mathbf{r}_{1},\tau )\phi _{\mathbf{n}}\left(
\frac{\mathbf{r}_{1}+\mathbf{r}_{2}}{2}\right) \chi (\mathbf{r}_{2},\tau )%
\mbox{e}^{i\mathbf{p}(\mathbf{r}_{1}-\mathbf{r}_{2})}\;.\qquad
\label{eq:3.13}
\end{eqnarray}%
By comparing Eq.~(\ref{eq:2.4}) with Eq.~(\ref{eq:3.13}) we observe the
following simple relation%
\be
\mathcal{S}_{eff}[\mathbf{b},\mathbf{h},\varphi ]=\mathcal{S}_{eff}[0,%
\mathbf{h}+\mathbf{b},\varphi ],
\ee%
which enables us to remove the field $\mathbf{b}$ from $\mathcal{S}_{eff}$
by a shift in $\mathbf{h}$ at the expense of changing the weight $W_{t}$
accordingly.

Now we can write down a representation of the partition function in the
presence of the magnetic field as a weighted integral over field
configurations
\be
\mathcal{Z}=\int \mathcal{D}\phi \;W_{s}[\varphi ]W_{t}[\mathbf{h}-\mathbf{b}%
]\mathcal{Z}[\phi ]\;,
\ee%
where
\begin{eqnarray}
\mathcal{Z}[\phi ] &=&\int D(\chi ^{\ast },\chi )\;\exp (-\mathcal{S}%
_{eff}[\phi ])  \label{eq:3.16} \\
\mathcal{S}_{eff}[\phi ] &=&\mathcal{S}_{0}+\int d\mathbf{p}d\mathbf{r}_{1}d%
\mathbf{r}_{2}\;\mbox{e}^{i\mathbf{p}(\mathbf{r}_{1}-\mathbf{r}_{2})}
\nonumber \\
&&\chi ^{\dagger }(\mathbf{r}_{1},\tau )\phi _{\mathbf{n}}\left( \frac{%
\mathbf{r}_{1}+\mathbf{r}_{2}}{2}\right) \chi (\mathbf{r}_{2},\tau )\;.
\label{eq:3.17}
\end{eqnarray}%
The weights $W_{s}[\varphi ]$ and $W_{t}[\mathbf{h}]$ are
\begin{eqnarray}
W_{s}[\varphi ] &=&\exp \Big[-\frac{1}{2}\int d\hat{\mathbf{n}}_{1}d\hat{%
\mathbf{n}}_{2}d\mathbf{q}d\tau  \nonumber \\
&&\quad \varphi _{\mathbf{n}_{1}}^{\ast }(\mathbf{q},\tau
)\;V_{s}^{-1}(\theta _{12},\mathbf{q})\;\varphi _{\mathbf{n}_{2}}(\mathbf{q}%
,\tau )\Big]\;,\;  \label{eq:Ws} \\
W_{t}[\mathbf{h}] &=&\exp \Big[-\frac{1}{2}\int d\hat{\mathbf{n}}_{1}d\hat{%
\mathbf{n}}_{2}d\mathbf{q}d\tau  \nonumber \\
&&\quad \mathbf{h}_{\mathbf{n}_{1}}^{\dagger }(\mathbf{q},\tau
)\;V_{t}^{-1}(\theta _{12},\mathbf{q})\;\mathbf{h}_{\mathbf{n}_{2}}(\mathbf{q%
},\tau )\Big]\;,\;  \label{eq:Wt}
\end{eqnarray}%
and $V_{s,t}(\theta _{12},\mathbf{q})=V_{s,t}(\theta _{12})f(\mathbf{q})$.

Eqs. (\ref{eq:3.16}-\ref{eq:Wt}) represent the final result of
this subsection.

The model is still quite general and in order to make progress further
approximations have to be introduced, where the focus will be on calculation
of $\mathcal{Z}[\phi ]$. Since it was possible to remove the field $\mathbf{b%
}$ from this term, the further derivation of the theory parallels that of
Ref. \onlinecite{aleiner} up to the point, where the weighted integral over
the field configurations of $\phi $ is performed. In the following we will
outline the main steps here in order to introduce our notations and prepare
the subsequent discussion of the model.

\subsection{\label{subsec:Bosonized action}Bosonized action}

In Sec. \ref{subsubsec:charge} the derivation of the representation for $%
\mathcal{Z}[\phi ]$ in the so-called quasi-classical approximation is
outlined following Ref. \onlinecite{aleiner}. Sec. \ref{subsubsec:weight}
deals with some additional contribution from the high energy sector and the
resulting change of weights.

\subsubsection{Charge and spin modes}

\label{subsubsec:charge} Integration over the fields $\chi ,\chi ^{\ast }$
in the expression, Eq.~(\ref{eq:3.16}), for $\mathcal{Z}[\phi ]$ results in
\be
\mathcal{Z}[\phi ]=\exp \left( \mbox{Tr}\ln \left( -\partial _{\tau }-%
\mathcal{H}_{0}+\hat{\Phi}\right) \right) .
\ee%
where the symbol $\mbox{Tr}$ includes integration over coordinates as well
as trace $\mbox{tr}_{\sigma }$ in spin space. The operator $\hat{\Phi}$ acts
in accordance with Eq.~(\ref{eq:3.17}). In the next step, we use a standard
trick introducing an auxiliary integration over parameter $u$, which enables
one to formally avoid expanding the logarithm while keeping track of
appropriate combinatorial factors. Instead, the Green's function for fixed
field configuration $G\left( x,x^{\prime }|u\phi \right) $ comes into play.
\begin{eqnarray}
\frac{\mathcal{Z}[\phi ]}{\mathcal{Z}[0]} &=&\exp \left[ \mbox{Tr}%
\int_{0}^{1}du\partial _{u}\ln \left( -\partial _{\tau }-\hat{H}_{0}+u\hat{%
\Phi}\right) \right] \qquad \\
&=&\exp \left( -i\mbox{Tr}\int_{0}^{1}du\left( \hat{\Phi}G\right) \left(
x,x|u\phi \right) \right) \;.
\end{eqnarray}%
The Green's function,
\be
G\left( x,x^{\prime }|\phi \right) =\frac{i}{\mathcal{Z}[\phi ]}\int D(\chi
,\chi ^{\ast })\;\chi (x)\chi ^{\dagger }(x^{\prime })\;\mbox{e}^{-\mathcal{S%
}_{eff}[\phi ]}\quad
\ee%
enters at coinciding points and this is why it is advantageous to Fourier
transform with respect to the difference of coordinates.
\be
G\left( x,x^{\prime }|\phi \right) =\int (d\mathbf{p})\;\mbox{e}^{i\mathbf{p}%
(\mathbf{r}-\mathbf{r^{\prime }})}\;\overline{G}_{\mathbf{p}}\left( \frac{%
\mathbf{r}+\mathbf{r}^{\prime }}{2},\tau ,\tau ^{\prime }|\phi \right)
.\quad
\ee%
Using the fact that $\overline{G}_{\mathbf{p}}$ is sharply peaked at the
Fermi surface one splits $\int d\mathbf{p}\sim \nu \int d\hat{\mathbf{n}}%
\int d\xi _{\mathbf{p}}$, where $\xi _{\mathbf{p}}=\mathbf{p}^{2}/2m-\mu $
and $\nu $ is the single particle density of states at the Fermi surface per
spin direction. After integration over $\xi _{\mathbf{p}}$ the
quasiclassical Green's function
\be
g_{\mathbf{n}}\left( \mathbf{r},\tau ,\tau ^{\prime }|\phi \right)
\equiv  \frac{1}{\pi }\int d\xi
_{\mathbf{p}}\;\overline{G}_{\mathbf{p}}\left( \mathbf{r},\tau
,\tau ^{\prime }|\phi \right)
\ee%
enters the expression
\begin{eqnarray}
&&{\mathcal{Z}[\phi ]}/{\mathcal{Z}[0]}  \label{eq:3.25} \\
&=&\exp \left( -i\pi \nu \int_{\mathbf{r}\tau }\int_{0}^{1}du\;\mbox{tr}%
\left[ \phi _{\mathbf{n}}(x)g_{\mathbf{n}}\left( \mathbf{r},\tau ,\tau
|u\phi \right) \right] \right) .  \nonumber
\end{eqnarray}%
One of the main results of Ref.\onlinecite{aleiner}, which we only cite
here, is a set of \emph{decoupled} differential equations for the charge $%
\rho $ and spin $\mathbf{S}$ components of $g_{\mathbf{n}}(x,\tau ,\tau )$
in the decomposition
\be
i\pi g_{\mathbf{n}}(x,\tau ,\tau )=i\rho _{\mathbf{n}}(x)+\mathbf{S}_{%
\mathbf{n}}(x){\bm\sigma },\label{eq:3.26}
\ee%
namely
\begin{eqnarray}
\hat{L}_{\mathbf{n},u}\mathcal{S}_{\mathbf{n}}(x,u) &=&-u\partial _{\tau }%
\mathbf{h}_{\mathbf{n}}(x)\;,  \label{eq:3.27} \\
\hat{L}_{\mathbf{n},0}\rho _{\mathbf{n}}(x,u) &=&-u\partial _{\tau }\varphi
(x)\;,  \label{eq:3.28}
\end{eqnarray}%
where
\be
\hat{L}_{\mathbf{n},u}=-\partial _{\tau }+iv_{F}\mathbf{n}\nabla +2iu\hat{h}%
_{\mathbf{n}}\;\label{eq:3.29}.
\ee
$\hat{h}_{\mathbf{n}}(x)$ is a matrix in the spin space with components $%
\hat{h}_{ij}=-\varepsilon _{ijk}\mathbf{h}_{k}$, so that $\hat{h}\mathbf{S}=%
\mathbf{h}\times \mathbf{S}$.

This result was obtained with the help of a generalized Schwinger ansatz\cite%
{schwinger} for $g_{\mathbf{n}}$
\be
g_{\mathbf{n}}\left( \mathbf{r},\tau ,\tau ^{\prime }|\phi \right) =\mathcal{%
T}_{\mathbf{n}}\left( \mathbf{r},\tau \right) g_{0}(\tau -\tau ^{\prime })%
\mathcal{T}_{\mathbf{n}}^{-1}\left( \mathbf{r},\tau ^{\prime }\right) ,
\ee%
where $g_{0}$ is the Green's function for free fermions $(\phi =0)$. It was
further assumed that $\phi $ varied smoothly on the scale of the Fermi
wavelength $\lambda _{F}=p_{F}^{-1}$.

Finally, substituting the decomposition Eq.~(\ref{eq:3.26}) into Eq.~(\ref%
{eq:3.25}), one finds
\be
\mathcal{Z}[\phi ]=\mathcal{Z}[0]\mathcal{Z}_{\rho }[\varphi ]\mathcal{Z}%
_{s}[\mathbf{h}]\;,
\ee%
where
\begin{eqnarray}
\mathcal{Z}_{\rho }[\varphi ] &=&\exp \left( 2\nu \int dudxd\hat{\mathbf{n}}%
\;\varphi _{\mathbf{n}}(x)\rho _{\mathbf{n}}(x)\right) \;,\quad \\
\mathcal{Z}_{s}[\mathbf{h}] &=&\exp \left( -2\nu \int dudxd\hat{\mathbf{n}}\;%
\mathbf{h}_{\mathbf{n}}(x)\mathbf{S}_{\mathbf{n}}(x)\right) \;,\quad
\label{eq:3.33}
\end{eqnarray}%
and $\mathbf{S}$ and $\rho $ fulfill the differential equations Eqs.~(\ref%
{eq:3.27}), (\ref{eq:3.28}). The equation for $\rho $ is readily solved
using a Fourier transform but we concentrate in the following on the spin
sector here.

Generally, one finds
\be
\mathcal{Z}=\mathcal{Z}[0]\mathcal{Z}_{\rho }\mathcal{Z}_{s}\,
\ee%
where
\begin{eqnarray}
\mathcal{Z}_{\rho } &=&\int \mathcal{D}\varphi W_{s}[\varphi ]\mathcal{Z}%
_{\rho }[\varphi ]\;, \\
\mathcal{Z}_{s} &=&\int \mathcal{D}\mathbf{h}W_{t}[\mathbf{h}-\mathbf{b}]%
\mathcal{Z}_{s}[\mathbf{h}]\;.  \label{eq:3.36}
\end{eqnarray}%
Our main task is to calculate the partition function $\mathcal{Z}_{s}.$

\subsubsection{Weight functions}

\label{subsubsec:weight} As is well known, the quasiclassical approximation
does not capture contributions that originate from frequencies of the order
of the Fermi energy $\varepsilon _{F}$. In the limit of weak interactions
considered here we may incorporate such contributions into the model by
replacing weights $W_{s}$, $W_{t}$ of Eqs.~(\ref{eq:Ws})--(\ref{eq:Wt}) by
weights $\mathcal{W}_{s}$, $\mathcal{W}_{t}$
\begin{eqnarray}
\mathcal{W}_{s}[\varphi ] &=&\exp \Big[-\frac{\nu }{2}\int dxd\hat{\mathbf{n}%
}\;\varphi _{\mathbf{n}}(x)\;\left[ \hat{\Gamma}_{s}^{-1}\varphi \right] (x,%
\mathbf{n})\Big],\;\quad \\
\mathcal{W}_{t}[\mathbf{h}] &=&\exp \Big[-\frac{\nu }{2}\int dxd\hat{\mathbf{%
n}}\;\mathbf{h}_{\mathbf{n}}(x)\;\left[ \hat{\Gamma}_{t}^{-1}\mathbf{h}%
\right] (x,\mathbf{n})\Big],\;\quad
\end{eqnarray}%
where
\be
\hat{\Gamma}_{s}=\hat{f}\frac{\nu \hat{V}_{s}}{1+2\nu \hat{V}_{s}},\qquad
\hat{\Gamma}_{t}=\hat{f}\frac{\nu \hat{V}_{t}}{1-2\nu \hat{V}_{t}}
\ee%
and we adopted the convention
\be
\left[ \hat{f}p\right] (x,\mathbf{n})=\int d\mathbf{r}_{1}\overline{f}(%
\mathbf{r}-\mathbf{r}_{1})p_{\mathbf{n}}(\mathbf{r}_{1},\tau )\;,
\ee%
\be
\left[ \hat{V}_{t,s}b\right] (x,\mathbf{n}_{1})=\int d\hat{\mathbf{n}}%
_{2}V_{t,s}(\theta _{12})b_{\mathbf{n}_{2}}(x)\;.
\ee%
The above argument is valid, however, only for the quadratic in $\mathbf{h}$
part of the weight function $W_{t}$ entering Eq.~(\ref{eq:3.36}). It does
not hold for the part containing $\mathbf{b}$, since the Green's function,
for which the quasiclassical approximation was used, depends only on the
field $\mathbf{h}$. Therefore, writing Eq.~(\ref{eq:3.36}) we should use the
following form for the weight $W_{t}[\mathbf{h}-\mathbf{b}]$
\begin{eqnarray}
&&W_{t}[\mathbf{h}-\mathbf{b}]\rightarrow \mathcal{W}_{t}[\mathbf{h},\mathbf{%
b}]=\mathcal{W}_{t}[\mathbf{h}]  \nonumber \\
&&\qquad \times \exp \Big[-\frac{1}{2}\int dx\;\mathbf{b}(x)\;\left[ \hat{V}%
_{t}^{-1}\mathbf{b}\right] (x)\Big]  \nonumber \\
&&\qquad \times \exp \Big[\int dxd\hat{\mathbf{n}}\;\mathbf{b}(x)\;\left[
\hat{V}_{t}^{-1}\mathbf{h}\right] (x,\mathbf{n})\Big].  \label{eq:3.42}
\end{eqnarray}%
Here and in what follows we omit the cutoff function $f$, whenever the
momentum is determined by the external field $\mathbf{b}$. This cannot
change results since we are interested only in small external momenta, $%
|q|\ll r_{0}^{-1}$, which allows us to put $f(\mathbf{q})=1$.

It is clear from Eq.~(\ref{eq:3.36}) that the magnetic field couples only to
the spin degrees of freedom and the charge sector does not play any role for
the spin susceptibility. Therefore, we can concentrate on the spin sector,
described by Eq.~(\ref{eq:3.36}) with the weight $W_{t}[\mathbf{h}-\mathbf{b}%
]$ determined by Eq.~(\ref{eq:3.42}).

\section{Supersymmetric representation}

\label{sec:susyrepresentation} In Sec.~\ref{subsec:Zs} we derive a
representation for $\mathcal{Z}_{s}$, Eq.~(\ref{eq:3.36}), in terms of a
functional integral over superfields. A detailed derivation has been
presented in Ref.\onlinecite{aleiner}. Therefore we only highlight the main
ideas here and relegate more technical details of the construction of the
model to Appendix \ref{app:derivation}. In Sec.~\ref{subsec:supervectors} we
collect the relevant definitions of supervectors and supermatrices that
enter the final model. This model is then presented in Sec.~\ref%
{subsec:effective}, rules of the perturbation theory are formulated in Sec.~%
\ref{subsec:perturbation} and a convenient diagrammatic representation is
introduced in Sec.~\ref{subsec:diagrams}.

\subsection{ $\mathcal{Z}_s$ as an integral over supervectors}

\label{subsec:Zs}

Using Eq.~(\ref{eq:3.33}) and Eqs.~(\ref{eq:3.27}), (\ref{eq:3.29}) one
arrives at the following form for the partition function $\mathcal{Z}_{s}[%
\mathbf{h}]$
\begin{eqnarray}
&&\mathcal{Z}_{s}[\mathbf{h}]=  \label{eq:4.1} \\
&&\qquad \exp \left( 2\nu \int dudxd\hat{\mathbf{n}}\;\mathbf{h}_{\mathbf{n}%
}(x)\left[ u\hat{L}_{\mathbf{n},u}^{-1}\partial _{\tau }\mathbf{h}_{\mathbf{n%
}}\right] (x)\right) .  \nonumber
\end{eqnarray}%
If one could find $\hat{L}_{\mathbf{n},u}^{-1}$ exactly for all $u\neq 0$,
the problem would be solved. However, since this is hardly possible for an
arbitrary $u$ and $\mathbf{h}$, we have to resort to some approximation
scheme. For this purpose it is advantageous to reexpress $\hat{L}_{\mathbf{n}%
,u}^{-1}$ in terms of a Gaussian functional integral. Using either bosonic
(complex) or fermionic (Grassmann) fields separately one would have to deal
with an $\mathbf{h}$-dependent normalization factor of the Gaussian
integral, which is inconvenient. This complication can be avoided by
introducing an integral that includes both bosonic and fermionic variables
on equal footing, as it has been used for a long time in the theory of
disordered systems, where the technique is known as the supersymmetry method%
\cite{review,book}. In the context of the present problem this approach has
been introduced in Ref. \onlinecite{aleiner}.

When using the Gaussian functional integration one should be careful,
however, since the operator $\hat{L}$ is not hermitian. In particular, the
sign of the $\mathbf{h}$-dependent term is not fixed and thus the
convergence of the Gaussian integral over bosonic variables, for which one
requires a positive real part of the kernel, is not easily insured.
Fortunately, it is known how to overcome this difficulty\cite{hermit}. One
can construct from the operator $\hat{L}$ Hermitian operators $\hat{L}%
^{\prime }=(\hat{L}+\hat{L}^{\dagger })/2$ and $\hat{L}^{\prime \prime }=-i(%
\hat{L}-\hat{L}^{\dagger })/2$ and arrange them into a new \emph{hermitian}
matrix operator
\be
\hat{M}=\left(
\begin{array}{cc}
\hat{L}^{\prime } & i\hat{L}^{\prime \prime } \\
-i\hat{L}^{\prime \prime } & -\hat{L}^{\prime }%
\end{array}%
\right) _{H}\label{eq:4.2}
\ee
The corresponding vector space will be called "Hermitized" or $H$- space.
One can reconstruct $\hat{L}^{-1}$ by summing certain matrix elements of the
inverse of $\hat{M}^{-1}$ as was shown in Ref. \onlinecite{hermit}.

The implementation of the ideas presented above leads to the following
identity
\be
\mathcal{Z}_{s}[\mathbf{h}]=\exp \left( \nu \int_{XX^{\prime }}\;\overline{%
\mathcal{F}_{\mathbf{h}}}(X)\mathcal{H}_{X,X^{\prime }}^{-1}\mathcal{F}_{%
\mathbf{h}}(X^{\prime })\right) \;.\quad\label{eq:4.3}
\ee
Here we use the collective variables
\be
X=(x,z),\;x=(\mathbf{r},\tau ),\;z=(u,\mathbf{n})\label{eq:4.4}
\ee
and the integration measure is specified as
\be
\int dX=\int_{X}=\int dxdz,\quad \int dz=\int_{0}^{1}du\int d\mathbf{n}%
\;.\qquad
\ee%
where $\int dx$ has been introduced in Eq.~(\ref{eq:2.6}).

To make contact with the previous discussion, we note that $(\Lambda
\mathcal{H})^{-1}$ corresponds to $\hat{M}^{-1}$ of Eq.~(\ref{eq:4.2}),
where $\Lambda $ is some constant matrix introduced below. $\mathcal{%
\overline{F}}_{\mathbf{h}}$ and $\mathcal{F}_{\mathbf{h}}$ are linear in $%
\mathbf{h}$ and their role is merely to select relevant components of $%
\mathcal{H}^{-1}$, the sum of which gives $\hat{L}_{\mathbf{n}}^{-1}$. The
operator $\mathcal{H}_{X,X^{\prime }}^{-1}$ can be written in terms of a
Gaussian functional integral as follows
\begin{eqnarray}
&&-\frac{1}{4i\nu }\mathcal{H}_{X,X^{\prime }}^{-1}=\left\langle \psi _{X}%
\overline{\psi }_{X^{\prime }}\right\rangle  \label{eq:4.6} \\
&=&\int \mathcal{D}(\psi ,\bar{\psi})\;\psi _{X}\overline{\psi }_{X^{\prime
}}\;\exp \left( 2i\nu \int_{X}\overline{\psi }_{X}\left( \mathcal{H}+i\delta
\Lambda \right) \psi _{X}\right) ,  \nonumber
\end{eqnarray}%
where $\psi $ and $\overline{\psi }=\psi ^{\dagger }\Lambda $ are
supervectors (see below). They contain both complex and Grassmann variables
on equal footing, which leads to the simple normalization of the Gaussian
integral
\be
1=\int \mathcal{D}(\psi ,\bar{\psi})\exp \left( 2i\nu
\int_{X}\overline{\psi _{X}}\left( \mathcal{H}+i\delta \Lambda
\right) \psi _{X}\right) \;.\label{eq:4.7}
\ee
The identities, Eqs.~(\ref{eq:4.3}), (\ref{eq:4.6}), are the basic
building blocks for the derivation of the model we want to work
with. In the next section we will define all quantities involved
in more detail, making it
possible to verify Eq.~(\ref{eq:4.3}) by explicit computation. \emph{%
Construction} of the theory that follows Ref.\onlinecite{aleiner} and adopts
the notations used in this paper is included in Appendix \ref{app:derivation}
for the interested reader.

It may be worth making a comment concerning the angular integration $\int d%
\mathbf{n}$. It will turn out later that for our purposes the most important
scattering processes are forward and backward scattering. By forward
scattering we mean scattering processes, in which both the incoming and
outgoing fermions have almost parallel momenta. Backward scattering refers
to a process, in which both incoming and outgoing momenta for each fermion
are almost anti-parallel to each other, while incoming and outgoing momenta
of different fermions are almost parallel.

It is then convenient to split the angular integration $\int d\hat{\mathbf{n}%
}$ into two half-spheres, one half-sphere contains "left-movers" the other
one "right movers". The arbitrariness involved in fixing the boundary in
dimension $d>1$ will not become important due to the quasi-one-dimensional
character of the relevant scattering processes. When separating sectors of
left and right moving particles it is then only necessary to integrate over
one half sphere and we denote this angular integration as $\int d\mathbf{n}$
with normalization $\int d\mathbf{n}=1/2$. As an example, the angular
integration in Eq.~(\ref{eq:4.1}) is now written as
\begin{eqnarray}
&&\int d\hat{\mathbf{n}}\mathbf{h}_{\mathbf{n}}\hat{L}_{\mathbf{n}%
}^{-1}\partial _{\tau }\mathbf{h}_{\mathbf{n}} \\
&=&\int d\mathbf{n}\;\mathbf{h}_{\mathbf{n}}\hat{L}_{\mathbf{n}%
}^{-1}\partial _{\tau }\mathbf{h}_{\mathbf{n}}+\mathbf{h}_{-\mathbf{n}}\hat{L%
}_{-\mathbf{n}}^{-1}\partial _{\tau }\mathbf{h}_{-\mathbf{n}}\;.  \nonumber
\end{eqnarray}

Let us perform two more manipulations to arrive at a form, where
only the averaging with weight $\mathcal{W}_{t}$ remains to be
done. Starting from Eq.~(\ref{eq:4.7}) one may verify by shifting
fields $\psi $, $\bar{\psi}$, that
\begin{eqnarray}
\mathcal{Z}_{s}[\mathbf{h}] &=&\exp \left( \nu \int_{XX^{\prime }}\;%
\overline{F_{\mathbf{h}}}(X)\mathcal{H}_{X,X^{\prime }}F_{\mathbf{h}%
}(X^{\prime })\right)  \label{eq:4.9} \\
&=&\int \mathcal{D}(\psi ,\bar{\psi})\;\exp \left( 2i\nu \int_{X}\overline{%
\psi }_{X}\left( \mathcal{H}+i\delta \Lambda \right) \psi _{X}\right)
\nonumber \\
&&\times \exp \left( \sqrt{-2i}\nu \int_{X}\left( \overline{F}_{\mathbf{h}%
}(X)\psi _{X}+\overline{\psi }_{X}F_{\mathbf{h}}(X)\right) \right) .
\nonumber
\end{eqnarray}%
Just as $\hat{L}$, $\mathcal{H}$ contains a part $\mathcal{H}_{\mathbf{h}}$,
that is linear in the field $\mathbf{h}$. We split off this part by writing
\be
\mathcal{H}=\mathcal{H}_{0}+\mathcal{H}_{\mathbf{h}}
\ee%
The final form of the model is then obtained by averaging the $\mathbf{h}$%
-dependent part of $\mathcal{Z}_{s}[\mathbf{h}]$ with the weight $\mathcal{W}%
_{t}[\mathbf{h}-\mathbf{b}]$ (compare with Eq.~(\ref{eq:3.36})).
\begin{eqnarray}
&&\mathcal{Z}_{s}=  \label{eq:4.11} \\
&&\int \mathcal{D}(\psi ,\bar{\psi})\exp \left( 2i\nu \int_{X}\overline{\psi
}_{X}\left( \mathcal{H}_{0}+i\delta \Lambda \right) \psi _{X}\right)
\mathcal{B}[\psi ,\bar{\psi},\mathbf{b}]  \nonumber
\end{eqnarray}%
\begin{eqnarray}
\mathcal{B}[\psi ,\bar{\psi},\mathbf{b}] =\int \mathcal{D}\mathbf{h}\;%
\mathcal{W}_{t}[\mathbf{h}-\mathbf{b}]\;\exp \left[ 2i\nu \int_{X}\overline{%
\psi }_{X}\mathcal{H}_{\mathbf{h}}\psi _{X}\right]  \nonumber \\
\times \exp \left( \sqrt{-2i}\nu \int_{X}\left( \overline{\mathcal{F}}_{%
\mathbf{h}}(X)\psi _{X}+\overline{\psi }_{X}\mathcal{F}_{\mathbf{h}%
}(X)\right) \right) \qquad  \label{eq:4.12}
\end{eqnarray}%
In the next subsection we will give the precise definition for $\psi ,\bar{%
\psi},\mathcal{F},\bar{\mathcal{F}}$ and $\mathcal{H}$. Equations~(\ref%
{eq:4.3}) and Eq.~(\ref{eq:4.6}) are discussed in Appendix \ref%
{app:derivation}. What remains then is to obtain the final model by
explicitly computing $\mathcal{B}$, Eq.~(\ref{eq:4.12}).

\subsection{Relevant supervectors and supermatrices}

\label{subsec:supervectors}

\subsubsection{Supervector $\protect\psi $ and its conjugation}

Let us first introduce the supervector ${\bm\psi }$ depending on
coordinates $X=(x,z)$ (cf. Eq.~(\ref{eq:4.4})). It has components
in the sectors of left
and right-moving particles labelled as $n$, the graded space of bosonic $%
\mathbf{S}$ and fermionic ${\bm\chi }$ variables labelled as $g$, the
Hermitized space labelled by $H$ and the spin space labelled by $s$. An
additional sector is introduced, which simplifies calculations with the
model. It has been termed \textquotedblleft electron-hole\textquotedblright\
$eh$ space in Ref. \onlinecite{aleiner} and plays a similar role as the
time-reversal sector in the $\sigma $-model description of disordered systems%
\cite{book}
\be
{\bm\psi }=\frac{1}{\sqrt{2}}\left(
\begin{array}{c}
{\bm\phi }^{\ast } \\
{\bm\phi }
\end{array}
\right) _{eh}\;\qquad {\bm\phi }(\mathbf{n})=\left(
\begin{array}{c}
{\bm\varphi }(\mathbf{n}) \\
{\bm\varphi }(-\mathbf{n})
\end{array}
\right) _{n},
\ee
where
\be
{\bm\varphi }=\left(
\begin{array}{cc}
{\bm\chi } &  \\
\mathbf{S} &
\end{array}%
\right) _{g},\quad {\bm\chi }=\left(
\begin{array}{c}
{\bm\chi }^{1} \\  {\bm\chi }^{2}
\end{array}%
\right) _{H}\quad \mathbf{S}=\left(
\begin{array}{c}
\mathbf{S}^{1}  \\
\mathbf{S}^{2}
\end{array}%
\right) _{H}\label{eq:4.14}
\ee%
Both $\mathbf{S}^{i}$ and ${\bm\chi }^{i}$ are vectors in the spin space
\be
\mathbf{S}^{i}=\left(
\begin{array}{c}
S_{x}^{i} \\
S_{y}^{i}  \\
S_{z}^{i}
\end{array}%
\right) _{s}\qquad {\bm\chi }^{i}=\left(
\begin{array}{ccc}
\chi _{x}^{i}   \\
\chi _{y}^{i}  \\
\chi _{z}^{i}
\end{array}%
\right) _{s}\label{eq:4.15}
\ee%
The components $\chi _{x}^{i},\chi _{y}^{i},\chi _{z}^{i}$ are anticommuting
(Grassmann) fields.

The conjugate vector $\overline{\psi }$ is defined as
\be
\overline{\bm\psi }={\bm\psi }^{\dagger }\Lambda ,
\ee%
where
\be
\Lambda =\left(
\begin{array}{cc}
1 & 0 \\
0 & -1%
\end{array}%
\right) _{H}
\ee%
is the third Pauli matrix in the Hermitized space. An important symmetry
that arises due to introducing the $(eh)$ sector is
\be
\bar{\psi}=(C\psi )^{T}
\ee%
where $C$ is the following matrix
\be
C=\left(
\begin{array}{cc}
C_{0} & 0 \\
0 & -C_{0}%
\end{array}%
\right) _{H},\quad C_{0}=\left(
\begin{array}{cc}
c_{1} & 0 \\
0 & c_{2}%
\end{array}%
\right) _{g}
\ee%
and matrices $c_{i}$ have structure in the $eh$ sector.
\be
c_{1}=\left(
\begin{array}{cc}
0 & -1 \\
1 & 0%
\end{array}%
\right) _{eh},\quad c_{2}=\left(
\begin{array}{cc}
0 & 1 \\
1 & 0%
\end{array}%
\right) _{eh}
\ee%
For Grassmann variables the convention $(\chi ^{\ast })^{\ast }=-\chi $ is
used. The conjugation of matrices is introduced as
\be
\overline{A}=CA^{T}C^{T}\label{eq:4.20}
\ee
where the special transposition appropriate for supermatrices
\cite{book} should be used. The important property
\be
\overline{\psi }A\phi =\overline{\phi }\,\overline{A}\,\psi
\;,\label{eq:4.21}
\ee
where $\psi $, $\phi $ are supervectors, is one of the main motivations for
introducing the $(eh)$ sector. When calculating higher cumulants later using
Wick's theorem the number of contractions can be reduced considerably with
the help of relation Eq.~(\ref{eq:4.21}).

\subsubsection{The matrix $\mathcal{H}$}

The matrix $\mathcal{H}$ is split into a $\mathbf{h}$-dependent and a $%
\mathbf{h}$-independent part
\begin{eqnarray}
\mathcal{H} &=&\mathcal{H}_{0}+\mathcal{H}_{\mathbf{h}} \\
\mathcal{H}_{0} &=&-iv_{0}\tau _{3}\Sigma _{3}\mathbf{n}\nabla -\Lambda
_{1}\partial _{\tau },\quad \mathcal{H}_{\mathbf{h}}=-2i\tau _{3}\hat{%
\mathbbm{H}}_{\mathbf{n}}  \nonumber
\end{eqnarray}%
Different constant matrices in this expression are
\be
\Lambda _{1}=\left(
\begin{array}{cc}
0 & 1 \\
1 & 0%
\end{array}%
\right) _{H},\; \Sigma _{3}=\left(
\begin{array}{cc}
1 & 0 \\
0 & -1%
\end{array}%
\right) _{n},\; \tau _{3}=\left(
\begin{array}{cc}
1 & 0 \\
0 & -1%
\end{array}%
\right) _{eh}\label{eq:4.24}
\ee%
and
\be
\mathbbm{H}_{\mathbf{n}}(x)=\left(
\begin{array}{cc}
\hat{h}_{\mathbf{n}}(x) & 0 \\
0 & \hat{h}_{-\mathbf{n}}(x)%
\end{array}%
\right) _{n}\;.\label{eq:4.25}
\ee

\subsubsection{Vector $\mathcal{F}_{\mathbf{h}}$}

Vector $\mathcal{F}$ does not have the full symmetry in supersymmetric $g$%
-space. Instead, it projects onto the bosonic sector. The role of the
fermionic fields in Eq.~(\ref{eq:4.9}) is only to provide the normalization.
We present $\mathcal{F}_{\mathbf{h}}$ as a product of an $\mathbf{h}$%
-dependent and an $\mathbf{h}$-independent part. The latter one is
\be
\mathcal{F}_{0}=\frac{1}{\sqrt{2}}\left(
\begin{array}{cc}
0 &  \\
1 &
\end{array}%
\right) _{g}\otimes \left(
\begin{array}{cc}
1 &  \\
1 &
\end{array}%
\right) _{n}\otimes \left(
\begin{array}{cc}
\left(
\begin{array}{cc}
1 &  \\
1 &
\end{array}%
\right) _{eh} &  \\
\left(
\begin{array}{cc}
-1 &  \\
1 &
\end{array}%
\right) _{eh} &
\end{array}%
\right) _{H}\;.
\ee%
The charge conjugated vector is $\overline{\mathcal{F}_{0}}=\left( C\mathcal{%
F}_{0}\right) ^{T}$, where only $c_{2}$ is effective when evaluating the
right hand side. $\mathcal{F}_{\mathbf{h}}$ and $\overline{\mathcal{F}}_{%
\mathbf{h}}$ are then given as
\be
\mathcal{F}_{\mathbf{h},\alpha }(X)=\partial _{X}(\alpha )\mathbbm{H}_{%
\mathbf{n}}(x)\mathcal{F}_{0},\;
\ee%
and $\overline{\mathcal{F}}_{\mathbf{h}}=\left( C\mathcal{F}_{\mathbf{h}%
}\right) ^{T}$, where
\be
\partial _{X}(\alpha )=\left(
\begin{array}{cc}
1 & 0 \\
0 & u\Big(\alpha \partial _{\tau }+(1-\alpha )iv_{0}\mathbf{n}\nabla \Sigma
_{3}\Big)%
\end{array}%
\right) _{eh}\label{eq:4.27}
\ee
Here an additional parameter $\alpha $ has been introduced into the model
and we will comment on it in the following subsection.

\subsubsection{Parameter $\protect\alpha$ and the weight $\mathcal{W}_t$.}

In view of Eq.~(\ref{eq:4.1}) one would expect only $\partial _{\tau }$ to
enter Eq.~(\ref{eq:4.27}), corresponding to $\alpha =1$. Choosing different
values of $\alpha $, however, may be convenient as we will see below when
studying the renormalization of the model. We will set $\alpha =1/2$ there,
treating temporal and spatial derivatives in a symmetric way. For $\alpha
\neq 1$ relation Eq.~(\ref{eq:4.4}) needs to be modified and this
modification eventually changes the weight $\mathcal{W}_{t}$. Any physical
quantity calculated with the model is of course independent of the choice of
$\alpha $. As is shown in Appendix \ref{app:derivation}, when introducing
parameter $\alpha \neq 1$, relation Eq.~(\ref{eq:4.4}) takes the form
\begin{eqnarray}
&&\mathcal{Z}_{s}[\mathbf{h}]=\exp \left( -\nu (1-\alpha )\int_{\hat{\mathbf{%
n}},x}\mathbf{h}_{\mathbf{n}}^{2}(x)\right)  \label{eq:4.28} \\
&&\times \exp \left( -4i\nu ^{2}\int_{XX^{\prime }}\;\overline{F_{\mathbf{h}%
,\alpha }}(X)\left\langle {\bm\psi }_{X}\overline{{\bm\psi }}_{X^{\prime
}}\right\rangle F_{\mathbf{h},\alpha }(X^{\prime })\right) .  \nonumber
\end{eqnarray}%
For $\alpha =1$ it coincides with Eq.~(\ref{eq:4.4}) and one can use the
form for $\mathcal{B}$ given in Eq.~(\ref{eq:4.12}). For general $\alpha $
it seems natural to absorb the exponential in the first line of Eq.~(\ref%
{eq:4.28}), into the weight $\mathcal{W}_{t}$ and thus change $\mathcal{B}$
to
\begin{eqnarray}
&&\mathcal{B}[\psi ,\bar{\psi},\mathbf{b}]=\int \mathcal{D}\mathbf{h}\;%
\mathcal{W}_{t}[\mathbf{h},\mathbf{b},\alpha ]\;\exp \left[ 2i\nu \int_{X}%
\overline{\psi }_{X}\mathcal{H}_{\mathbf{h}}\psi _{X}\right]  \nonumber \\
&&\times \exp \left[ \sqrt{-2i}\nu \int_{X}\left[ \overline{F_{\mathbf{h}%
,\alpha }(X)}\psi _{X}+\overline{\psi _{X}}F_{\mathbf{h},\alpha }(X)\right] %
\right] ,  \label{eq:4.29}
\end{eqnarray}%
where
\be
\mathcal{W}_{t}[\mathbf{h},\mathbf{b},\alpha ]=\mathcal{W}_{t}[\mathbf{h},%
\mathbf{b}]\;\mbox{e}^{-\nu (1-\alpha )\int_{\hat{\mathbf{n}},x}\mathbf{h}_{%
\mathbf{n}}^{2}(x)}.
\ee%
Clearly, this change in $\mathcal{W}_{t}$ only affects the quadratic form in
$\mathbf{h}$ but not the part containing $\mathbf{b}$. Therefore, to make
the change explicit we may write here
\begin{eqnarray}
&&\mathcal{W}_{t}[\mathbf{h},\mathbf{b}=0,\alpha ] \\
&=&\exp \Big[-\frac{\nu }{2}\int dx(d\mathbf{n})\mathbf{h}_{\mathbf{n}}(x)\;%
\left[ \hat{\Gamma}_{t}^{-1}(\alpha )\mathbf{h}\right] (x,\mathbf{n})\Big],
\nonumber
\end{eqnarray}
where
\be
2\hat{\Gamma}_{t}(\alpha )=\hat{f}\frac{2\nu \hat{V}_{t}}{1-2\nu \alpha \hat{%
V}_{t}}.
\ee%
The final step in the derivation of the model is the calculation of $%
\mathcal{B}$ in Eq.~(\ref{eq:4.29}).

\subsection{Effective low energy theory}

\label{subsec:effective} From Eq.~(\ref{eq:4.11}) together with $\mathcal{B}$
given in Eq.~(\ref{eq:4.29}) we find
\begin{eqnarray}
\mathcal{Z}_s=\int \mathcal{D}({\bm \psi},\overline{{\bm \psi}}%
)\;\exp\left(-\sum_{i}\mathcal{S}_{i}\right)  \label{eq:4.33}
\end{eqnarray}
Next we specify the different parts $\mathcal{S}_{i}$ of the effective
action.

The interaction-independent part is
\be
\mathcal{S}_{0}=-2i\nu \int dX\;\overline{{\bm\psi }}_{X}\big(\mathcal{H}%
_{0}+i\delta \Lambda \big){\bm\psi }_{X}.\label{eq:4.34}
\ee
There are three different interaction vertices present in the theory
\begin{eqnarray}
&&\mathcal{S}_{2}=-2i\nu \sum_{ij}\lambda _{ij}\int dXdX_{1}  \label{eq:4.35}
\\
&&\left( \overline{\psi }_{X,\delta }\tau _{3}\Pi _{j}\partial _{X}\mathcal{F%
}_{0}\right) \;{\Gamma }_{X,X_{1}}^{i}\;\left( \overline{\mathcal{F}_{0}}%
\overline{\partial _{X_{1}}}\Pi _{j}\tau _{3}\psi _{X_{1},\delta }\right)
\nonumber \\
&&\mathcal{S}_{3}=-4\sqrt{-2i}\nu \sum_{ij}\lambda _{ij}\;\varepsilon
_{\delta \beta \gamma }\;\int dXdX_{1}  \label{eq:4.36} \\
&&\left( \overline{\psi }_{X,\delta }u\tau _{3}\Pi _{j}\psi _{X,\beta
}\right) \;{\Gamma }_{X,X_{1}}^{i}\left( \overline{\mathcal{F}_{0}}\overline{%
\partial _{X_{1}}}\tau _{3}\psi _{X_{1},\gamma }\right)  \nonumber \\
&&\mathcal{S}_{4}=-4\nu \sum_{ij}\lambda _{ij}\;\varepsilon _{\delta \beta
\gamma }\varepsilon _{\delta \beta _{1}\gamma _{1}}\int dXdX_{1}
\label{eq:4.37} \\
&&\left( \overline{\psi }_{X,\beta }u\tau _{3}\Pi _{j}\psi _{X,\gamma
}\right) {\Gamma }_{X,X_{1}}^{i}\left( \overline{\psi }_{X_{1},\beta
_{1}}u_{1}\tau _{3}\Pi _{j}\psi _{X_{1},\gamma _{1}}\right)  \nonumber
\end{eqnarray}%
Summation over spin indices is implied and we use the totally
antisymmetric tensor $\varepsilon _{\alpha \beta \gamma }$ with
$\varepsilon _{123}=1$. This part of the action would be
sufficient for a calculation of the thermodynamic potential in the
absence of a magnetic field and it coincides with the action
written in Ref.\onlinecite{aleiner}. Here we used the notation
\be
{\Gamma }_{X,X^{\prime }}^{i}=\gamma _{i}\left( \widehat{\mathbf{n}\mathbf{n}%
^{\prime }}\right) \;f(\mathbf{r}-\mathbf{r}^{\prime })\;\delta (\tau -\tau
^{\prime })
\ee%
and
\begin{eqnarray}
\gamma _{1}(\widehat{\mathbf{n}\mathbf{n}_{1}}) &=&\left( \frac{\nu \hat{V}%
_{t}}{1-2\nu \alpha \hat{V}_{t}}\right) (\widehat{\mathbf{n},\mathbf{n}_{1}}%
)\equiv \gamma _{f}^{0}  \label{eq:4.39} \\
\gamma _{2}(\widehat{\mathbf{n}\mathbf{n}_{1}}) &=&\left( \frac{\nu \hat{V}%
_{t}}{1-2\nu \alpha \hat{V}_{t}}\right) (\widehat{\mathbf{n},-\mathbf{n}_{1}}%
)\equiv \gamma _{b}^{0}  \label{eq:4.40}
\end{eqnarray}%
Matrices $\Pi _{i}$ are
\be
\Pi _{1}=\mathbbm{1},\quad \Pi _{2}=\Sigma _{3}
\ee%
The form of
\be
\lambda _{ij}=\left(
\begin{array}{cc}
1 & 1 \\
1 & -1%
\end{array}%
\right)
\ee%
was determined from the following identities
\begin{eqnarray}
\Sigma _{+}A\Sigma _{+}+\Sigma _{-}A\Sigma _{-} &=&\frac{1}{2}\left( \Pi
_{1}A\Pi _{1}+\Pi _{2}A\Pi _{2}\right)  \label{eq:4.43} \\
\Sigma _{+}A\Sigma _{-}+\Sigma _{-}A\Sigma _{+} &=&\frac{1}{2}\left( \Pi
_{1}A\Pi _{1}-\Pi _{2}A\Pi _{2}\right)  \label{eq:4.44}
\end{eqnarray}%
Taking into account that relevant scattering events are quasi
one-dimensional, as will be seen later, Eqs.~(\ref{eq:4.43}), (\ref{eq:4.44}%
) also explain the labelling in Eqs.~(\ref{eq:4.39}), (\ref{eq:4.40}), where
$\gamma _{1}$ is classified as forward scattering and $\gamma _{2}$ as
backward scattering.

The presence of a magnetic field introduces three more terms, namely
\begin{eqnarray}
&&\mathcal{S}_{b0}=-\nu \;\eta \;\int dx\;\mathbf{b}^{2}(x) \label{eq:4.45}\\
&&\mathcal{S}_{b1}=-2\nu \sqrt{-2i}\;\eta \;\int dX\;b_{\delta }(x)\left(
\overline{\psi }_{X,\delta }\tau _{3}\partial _{X}\mathcal{F}_{0}\right)
\qquad  \label{eq:4.46} \\
&&\mathcal{S}_{b2}=4\nu \;\varepsilon _{\delta \beta \gamma
}\;\eta \;\int dX\;b_{\delta }(x)\left( \overline{\psi }_{X,\beta
}u\tau _{3}\psi _{X,\gamma }\right)\label{eq:4.47}
\end{eqnarray}%
In these expressions
\begin{equation}
\eta =\frac{1}{1-2\alpha \nu \overline{V_{t}}},  \label{eq:4.48}
\end{equation}%
where the bar in $\overline{V_{t}}$ means averaging over the full solid
angle. We remind that $\eta $, the interaction amplitudes $\Gamma _{i}$ as
well as $\partial _{X}$, $\overline{\partial }_{X}$ depend on parameter $%
\alpha $ introduced in \ref{eq:4.27}. Here we suppressed the label for the
sake of brevity.

\subsection{Rules of Perturbation Theory}

\label{subsec:perturbation}

\subsubsection{Gaussian averages}

A perturbation theory can be set up in a standard way using a cumulant
expansion and Wick's theorem. Gaussian averages are taken with respect to $%
\mathcal{S}_{0}$ and it is therefore convenient to work with the matrix
Green's function
\begin{eqnarray}
&&\hat{\mathcal{G}}(X_{1},X_{2})=-4i\nu \left\langle {\bm\psi }%
_{X_{1}}\otimes \overline{{\bm\psi }}_{X_{2}}\right\rangle _{0}
\label{eq:4.49} \\
&&\left\langle (\dots )\right\rangle _{0}=\int \mathcal{D}{\bm\psi }\;(\dots
)\exp (-\mathcal{S}_{0}[{\bm\psi }])
\end{eqnarray}%
Due to supersymmetry no normalization factor arises.
$\hat{\mathcal{G}}$ is a matrix in spin space but its spin
structure is trivial and we denote
\be
\hat{\mathcal{G}}_{\alpha \beta }(X_{1},X_{2})=\mathcal{G}_{\mathbf{n}%
_{1}}(x_{1}-x_{2})\delta _{\alpha \beta }\delta _{\mathbf{n}_{1},\mathbf{n}%
_{2}}\delta (u_{1}-u_{2}).\quad
\ee%
The Fourier transform of $\mathcal{G}$ is introduced as
\begin{eqnarray}
&&\mathcal{G}_{\mathbf{n}_{1}}(x_{1}-x_{2})= \\
&&\qquad T\sum_{\omega _{n}}\int d\mathbf{p}\;\mathcal{G}_{\mathbf{n}_{1}}(%
\mathbf{p},\omega _{n})\;\mbox{e}^{i\mathbf{p}(\mathbf{r}_{1}-\mathbf{r}%
_{2})-i\omega _{n}(\tau _{1}-\tau _{2})}.  \nonumber
\end{eqnarray}%
Here $\omega _{n}=2\pi Tn$ are bosonic Matsubara frequencies and
\be
\mathcal{G}_{\mathbf{n}}(\mathbf{p},\omega _{n})=\frac{1}{v_{F}\mathbf{n}%
\mathbf{p}\tau _{3}\Sigma _{3}+i\omega \Lambda _{1}-i\delta \Lambda _{1}}.
\ee%
Similar to the specific heat \cite{aleiner}, the temperature dependence of
the susceptibility is determined by non-zero Matsubara frequencies only.
Therefore, the term containing the infinitesimal $\delta $ in the Green
functions will not become important in our calculations and will not be
written from now on. The matrix Green function $\mathcal{G}_{\mathbf{n}}$ is
diagonal in spin space.

In addition to averages of the type written in Eq.~(\ref{eq:4.49}) one has
to account for non-standard averages of the type $\left\langle \psi
_{X_{1},\alpha }\psi _{X_{2},\beta }\right\rangle _{0}\neq 0$ and $%
\left\langle \overline{\psi }_{X_{1},\alpha }\overline{\psi }_{X_{2},\beta
}\right\rangle _{0}\neq 0$, which arise due to the $eh$ sector. It is,
however, sufficient to work with Eq.~(\ref{eq:4.49}), since expressions
involving such non-standard averages can easily be transformed to a more
standard form with the help of the relation $\overline{{\bm\psi }}_{X_{1}}A{%
\bm\psi }_{X_{2}}=\overline{{\bm\psi }}_{X_{2}}\overline{A}{\bm\psi }%
_{X_{1}} $, which is valid for any supermatrix $A$. The charge conjugation
operation $A\rightarrow \overline{A}$ has been defined in Eq.~(\ref{eq:4.20}%
) .

It is often convenient to work with a generalization of the trace operation $%
\mbox{tr}$, used for conventional matrices, to the so-called supertrace $%
\mbox{str}$ \cite{book}. It is defined as
\be
\mbox{str}\left(
\begin{array}{cc}
a & \sigma \\
\rho & b%
\end{array}%
\right) _{g}=\mbox{tr}a-\mbox{tr}b\;.
\ee%
Two more useful relations are
\begin{eqnarray}
&&\mbox{str}(AB)=\mbox{str}(BA)\;, \\
&&\overline{\psi }_{1,\alpha }A\psi _{2,\beta }=-\mbox{str}\Big(A\left( \psi
_{2,\alpha }\overline{\psi }_{1,\beta }\right) \Big)\;.
\end{eqnarray}%
$A,B$ are arbitrary supermatrices. It follows from the definition of the
supertrace that for matrices $A_{i}$ that have the full symmetry in $g$%
-space, $\mbox{str}(A_{i})=0$. The matrix Green's function $\mathcal{G}$ is
a particularly important example. One immediately concludes that the
following relations (and straightforward generalizations thereof) hold for
such matrices
\be
\left\langle \left( \overline{{\bm\psi }}A_{1}{\bm\psi }\right)
\right\rangle _{0}=\left\langle \left( \overline{{\bm\psi }}A_{1}{\bm\psi }%
\right) \left( \overline{\psi }A_{2}{\bm\psi }\right) \right\rangle _{0}=0.
\ee%
These important relations considerably reduce the number of diagrams to be
considered in the perturbation theory.

\subsection{Diagrammatic representation}

\label{subsec:diagrams} 
\begin{figure}[t]
\setlength{\unitlength}{2.3em}
\begin{picture}(12,10)

\put(1,9){\includegraphics[width=3\unitlength]{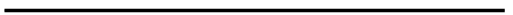}}
\put(2.5,9.5){$\mathcal{G}$}

\put(1,7){\includegraphics[width=3\unitlength]{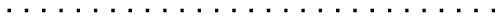}}
\put(2,7.5){$\mathcal{F}_0\overline{\mathcal{F}}_0$}

\put(1,3.8){\includegraphics[width=3\unitlength]{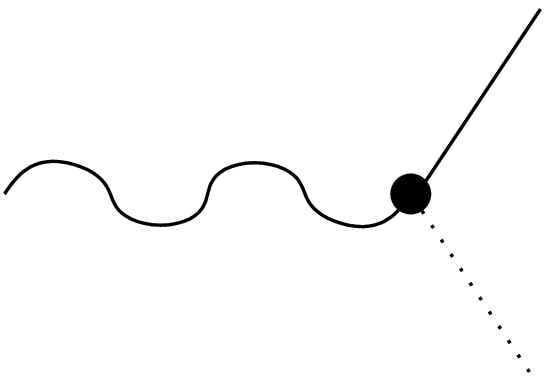}}
\put(2.2,5.4){$\mathcal{S}_{b1}$}

\put(1,1){\includegraphics[width=3\unitlength]{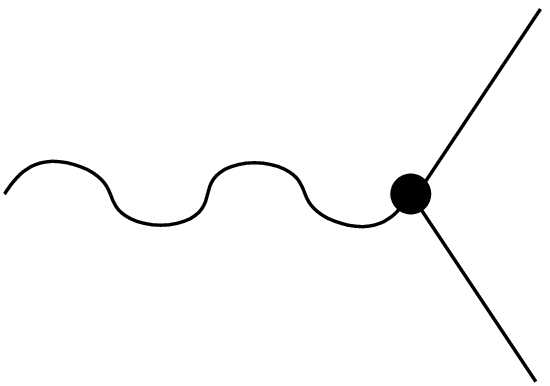}}
\put(2.2,2.6){$\mathcal{S}_{b2}$}

\put(6,5.8){\includegraphics[width=2.3\unitlength]{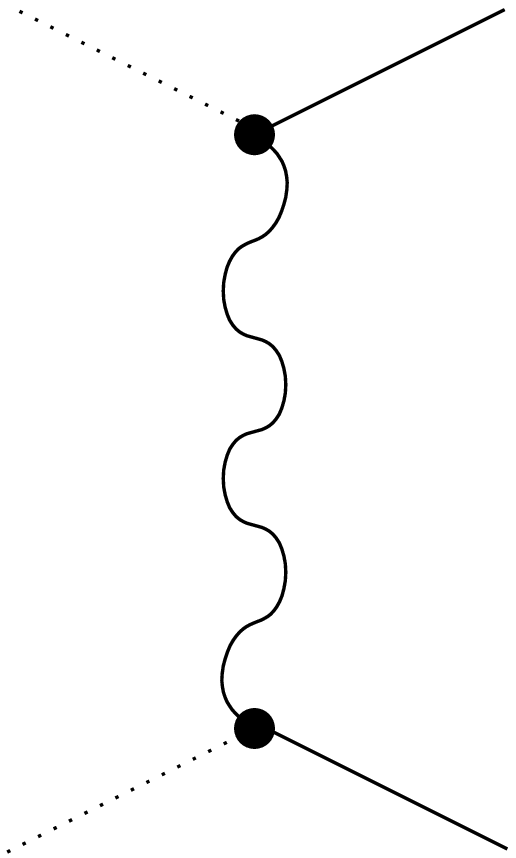}}
\put(7.5,7.6){$\mathcal{S}_{2}$}

\put(6,0.8){\includegraphics[width=2.3\unitlength]{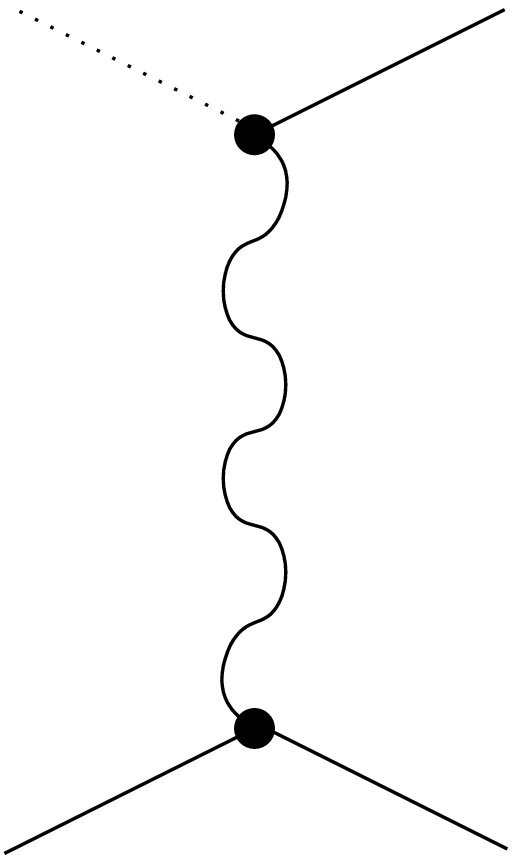}}
\put(7.5,2.6){$\mathcal{S}_{3}$}

\put(9,3.8){\includegraphics[width=2.3\unitlength]{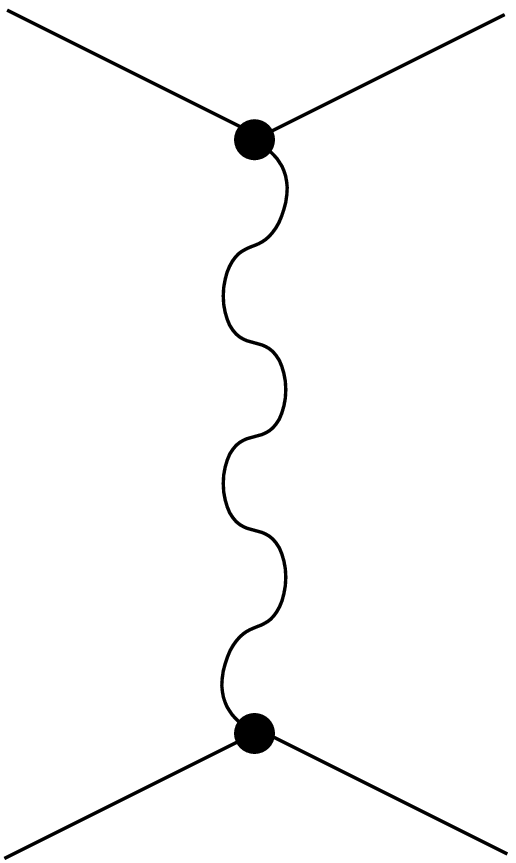}}
\put(10.5,5.6){$\mathcal{S}_{4}$}

\end{picture}
\caption{Basic building blocks of the perturbation theory.}
\label{fig:blocks}
\end{figure}

Figure \ref{fig:blocks} displays the building blocks
that we will use for the diagrammatic representation of the perturbation
theory. The three interaction vertices reflect the structure of $\mathcal{S}%
_{2}$, $\mathcal{S}_3$ and $\mathcal{S}_4$. They are plotted in such a way
that \emph{small} momenta flow \emph{along} the interaction line. The dotted
lines symbolize the structure $\mathcal{F}_0\overline{\mathcal{F}}_0$. These
lines do carry neither momentum nor frequency. They are however convenient
to make contact with conventional diagrams formulated in terms of electron
Green's functions as we will discuss now.

To this end it is instructive to make comparison with the original model.
Since the charge channel has been separated, we need to consider fermions
interacting in the triplet channel only, i.e. the free action Eq.~(\ref%
{eq:2.3}) and an interaction part of the form
\be
\mathcal{S}_{int,t}=-\frac{1}{2}\int_{p_{1}p_{2}q}\;\mathbf{S}%
(p_{1},-q)\Gamma _{t}(\theta
_{12})\mathbf{S}(p_{2},q),\label{eq:4.58}
\ee
where we remind that $\mathbf{S}(p,q)=\chi ^{\dagger }\left( p-\frac{q}{2}%
\right) {\bm\sigma }\chi \left( p+\frac{q}{2}\right) $ and we used
four dimensional notation for momenta and energies.

In contrast, the effective low energy model of Eq.~(\ref{eq:4.33}) is
formulated in terms of spin modes. The structure of the terms appearing
after expansion of $\int \mathbf{h}\hat{L}_{\mathbf{h}}\partial _{\tau }%
\mathbf{h}$ in Eq.~(\ref{eq:4.1}) in the field $\mathbf{h}$ is the
translation to the spin mode language of a closed Fermion loop with $n\geq 2$
fields $\mathbf{h}$ coupled to it. The expansion of $\hat{L}_{\mathbf{h}}$
in powers of $\mathbf{h}$ is performed here assuming that $\hat{L}_{\mathbf{h%
}=0}\propto (v_{0}\mathbf{n}\mathbf{p}-i\omega )^{-1}$ describes free
propagation of spin modes (compare to $\mathcal{S}_{0}$ of Eq.~(\ref{eq:4.34}%
)). After integration over $\mathbf{h}$, which reduces to contracting pairs
of fields $\langle \mathbf{h}\mathbf{h}\rangle $, one obtains a theory of
\emph{interacting} spin modes. Not all fields $\mathbf{h}$ in $\int \mathbf{h%
}\hat{L}_{\mathbf{h}}\partial _{\tau }\mathbf{h}$ enter in an equivalent
way, however, and this explains the presence of three different interaction
terms in the model. When contracting two fields $\mathbf{h}$ that appear due
to an expansion of $\hat{L}_{\mathbf{h}}$ in $\mathbf{h}$, one finds an
interaction vertex of the type represented by $\mathcal{S}_{4}$ (see Fig.~%
\ref{fig:blocks}). If only one such field is involved one comes to $\mathcal{%
S}_{3}$, otherwise to $\mathcal{S}_{2}$.

\begin{figure}[tb]
\setlength{\unitlength}{2.3em}
\begin{picture}(12,5)

\put(1,1.5){\includegraphics[width=2.5\unitlength]{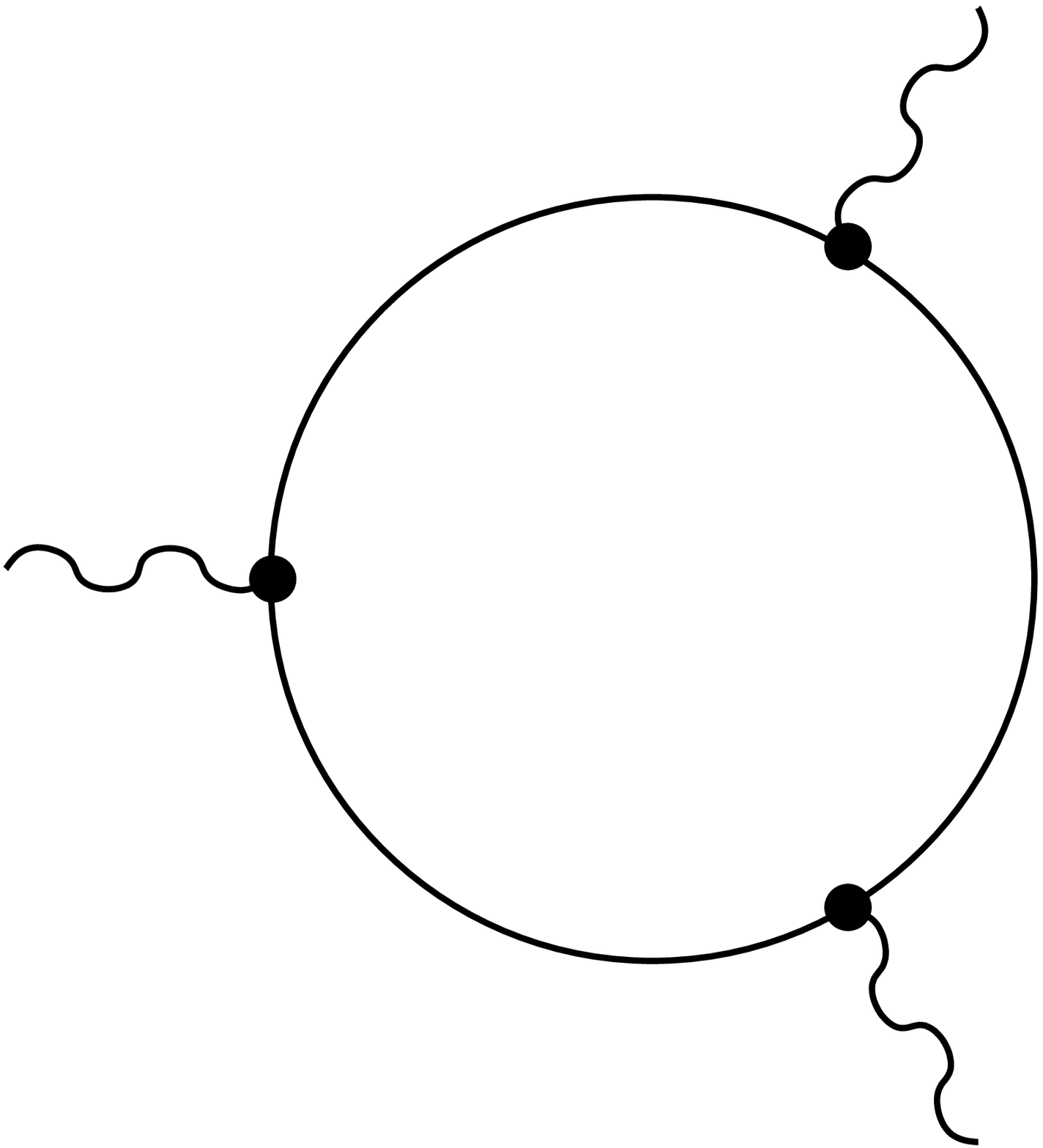}}
\put(4,2.8 ){$\leftrightarrow$}

\put(5,2.3){\includegraphics[width=2.1\unitlength]{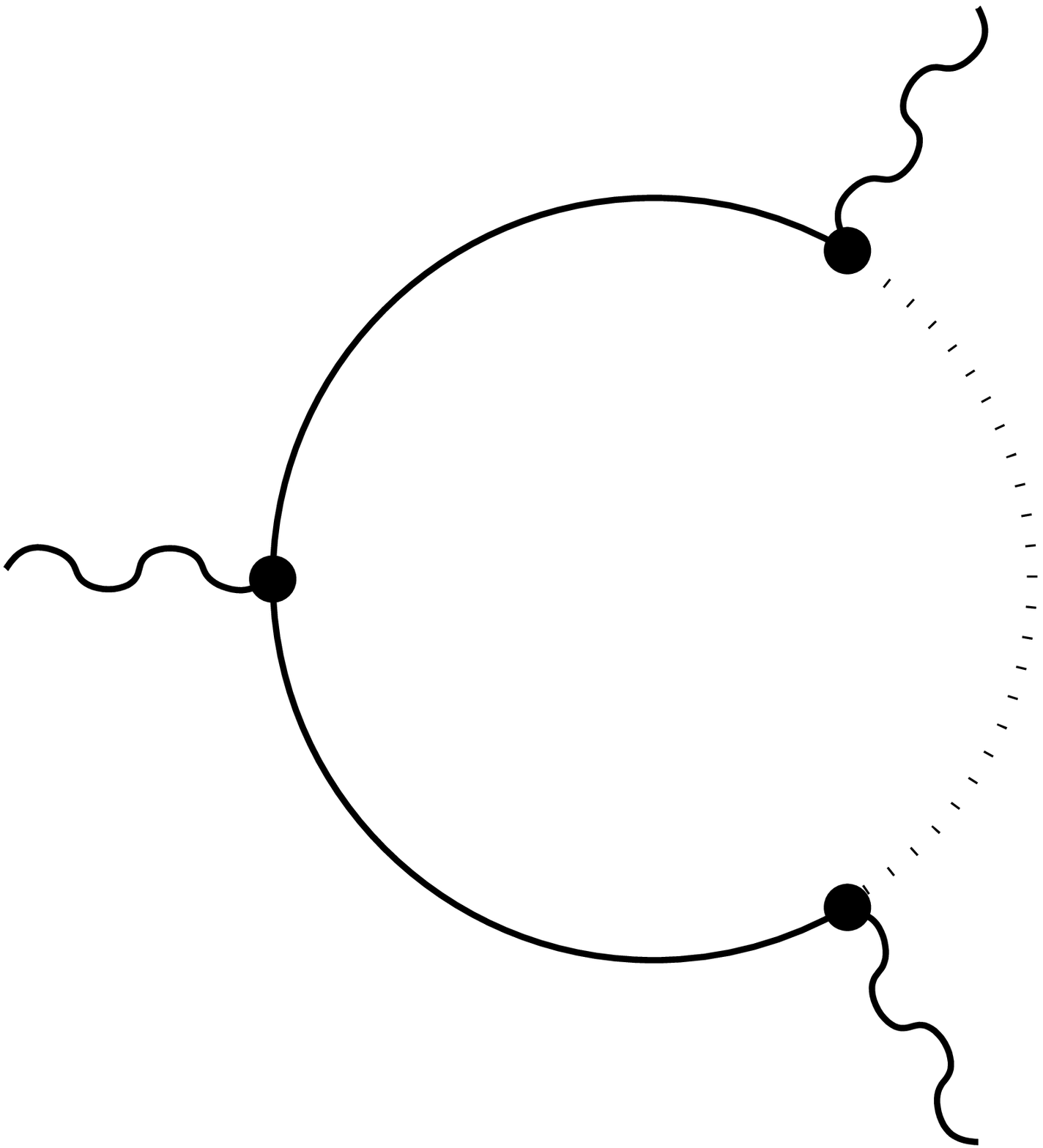}}

\put(8.4,2.3){\includegraphics[width=2.1\unitlength]{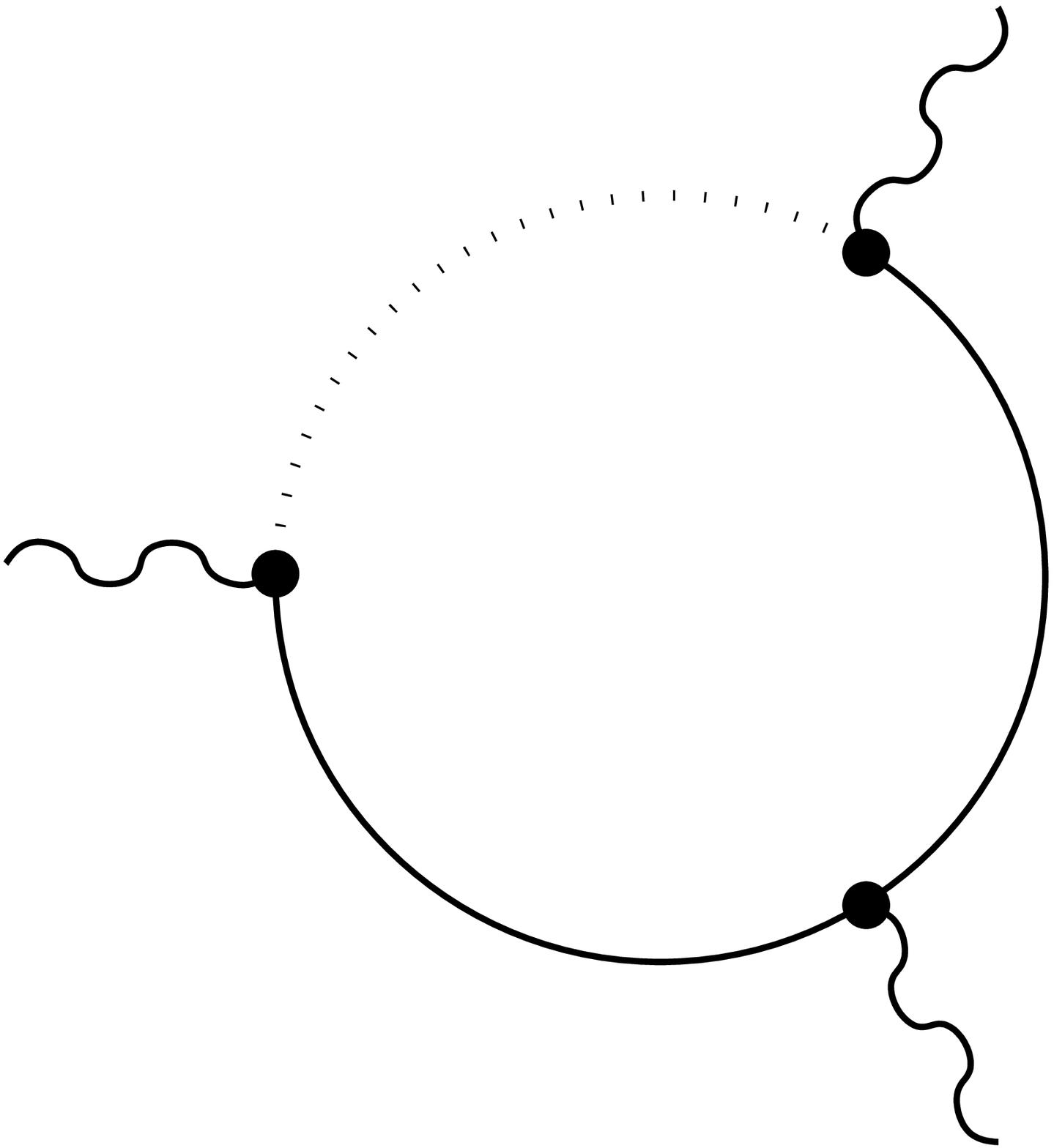}}
\put(7,0.2){\includegraphics[width=2.1\unitlength]{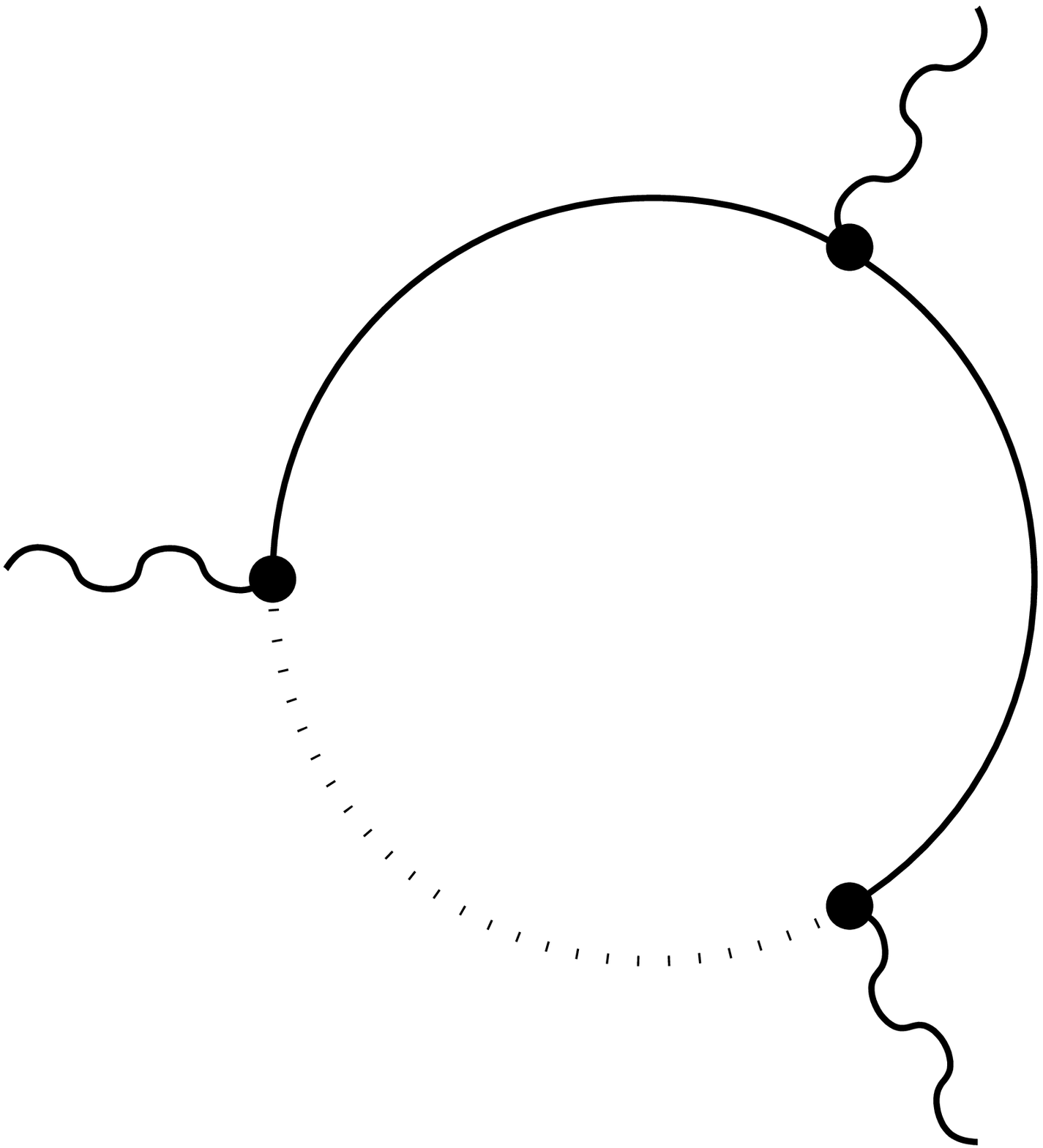}}
\end{picture}
\caption{A closed loop of fermionic Green's function contrasted with the
spin mode representation.}
\label{fig:translation1}
\end{figure}

\begin{figure}[tb]
\setlength{\unitlength}{2.3em}
\begin{picture}(12,8)

\put(2.5,6){\includegraphics[width=5.5\unitlength]{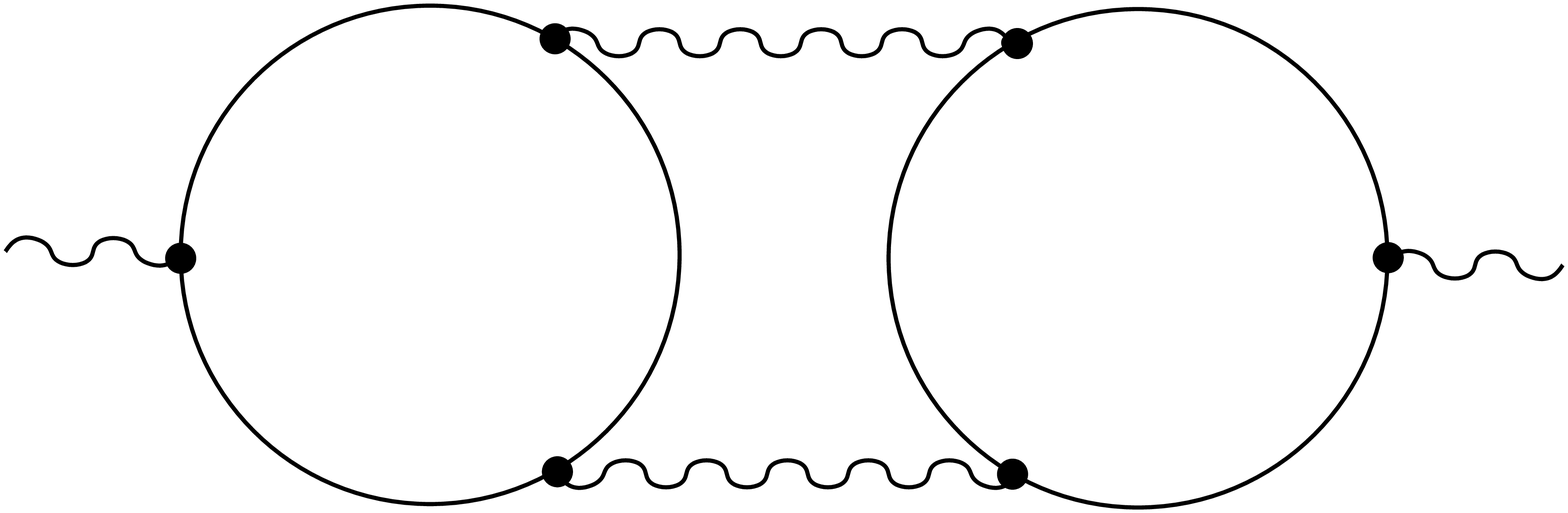}}
\put(8.3,6.7){$\leftrightarrow$}

\put(0.2,3.5){\includegraphics[width=5.5\unitlength]{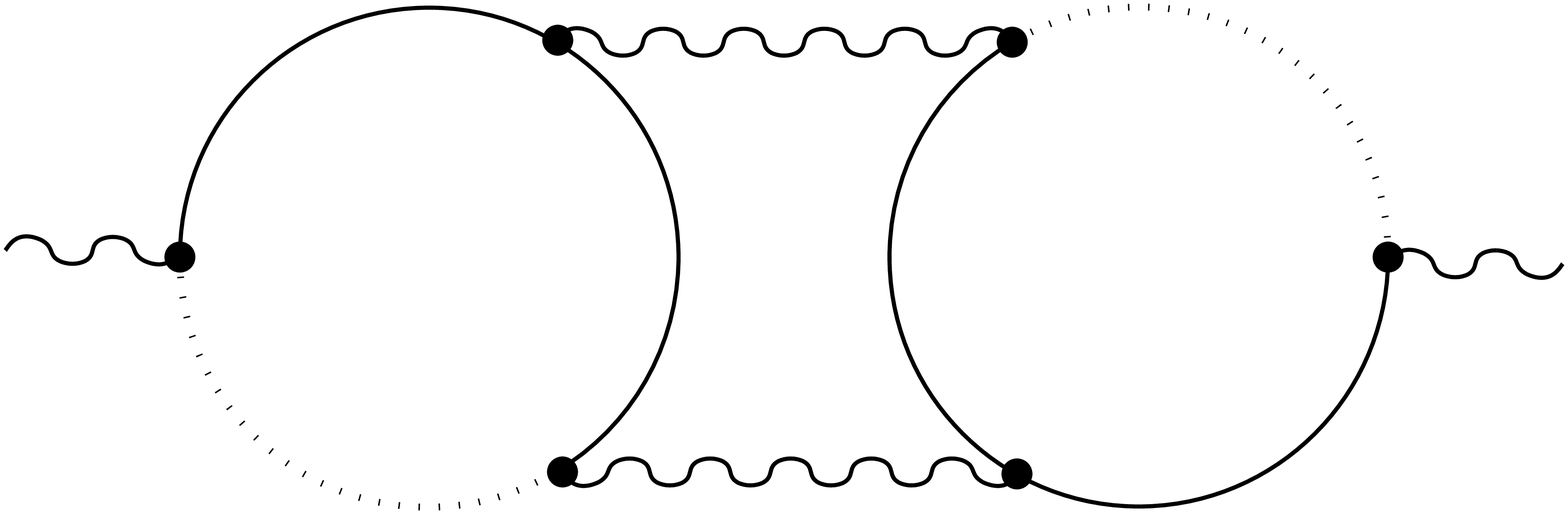}}
\put(2.8,3.8){\scriptsize{$\mathcal{S}_3$}}
\put(2.8,5.3){\scriptsize{$\mathcal{S}_3$}}

\put(5.9,3.5){\includegraphics[width=5.5\unitlength]{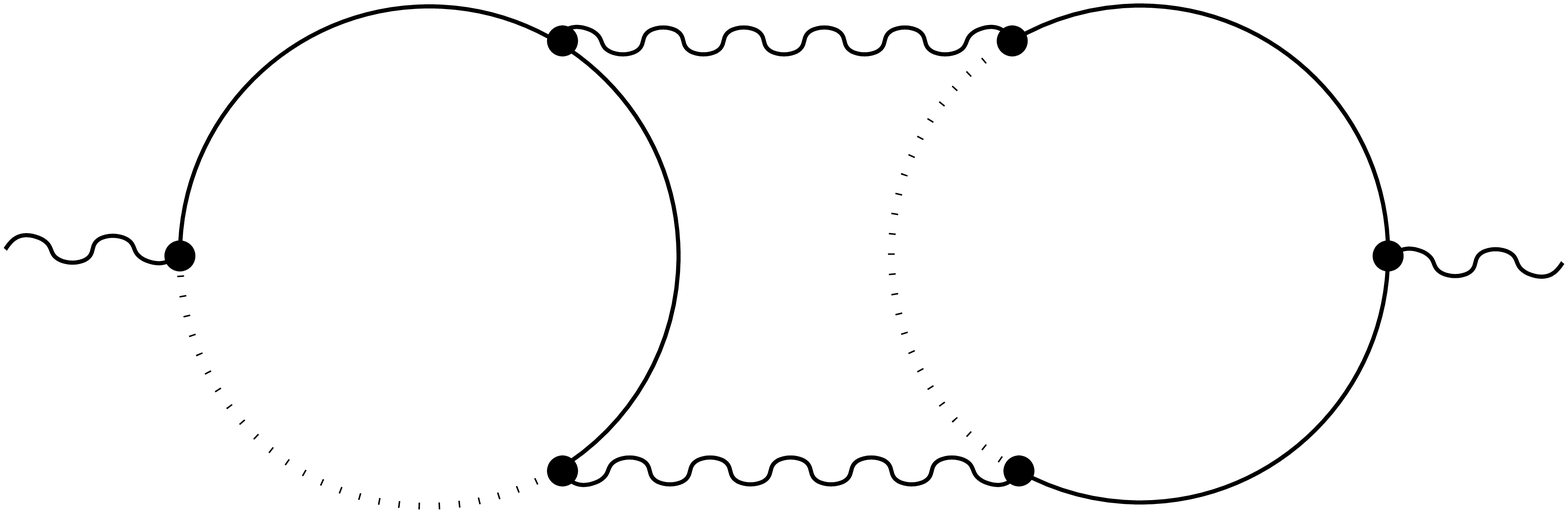}}
\put(8.5,3.8){\scriptsize{$\mathcal{S}_2$}}
\put(8.5,5.3){\scriptsize{$\mathcal{S}_3$}}

\put(0.2,0.7){\includegraphics[width=5.5\unitlength]{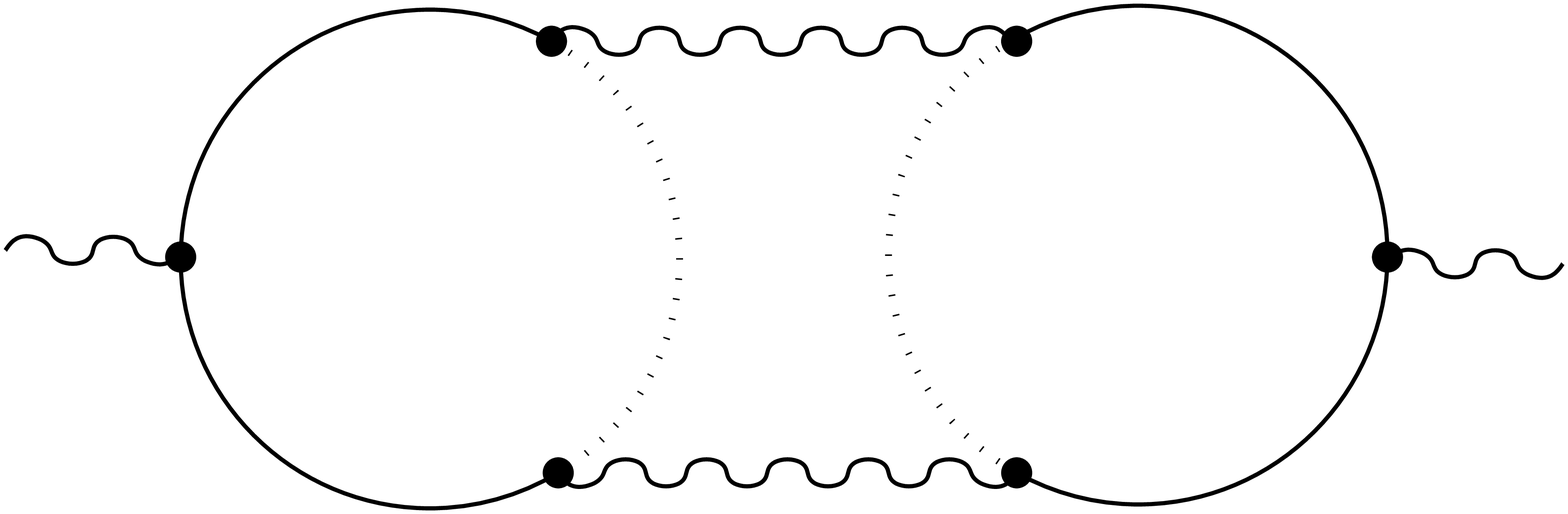}}
\put(2.8,1.0){\scriptsize{$\mathcal{S}_2$}}
\put(2.8,2.5){\scriptsize{$\mathcal{S}_2$}}

\put(5.9,0.7){\includegraphics[width=5.5\unitlength]{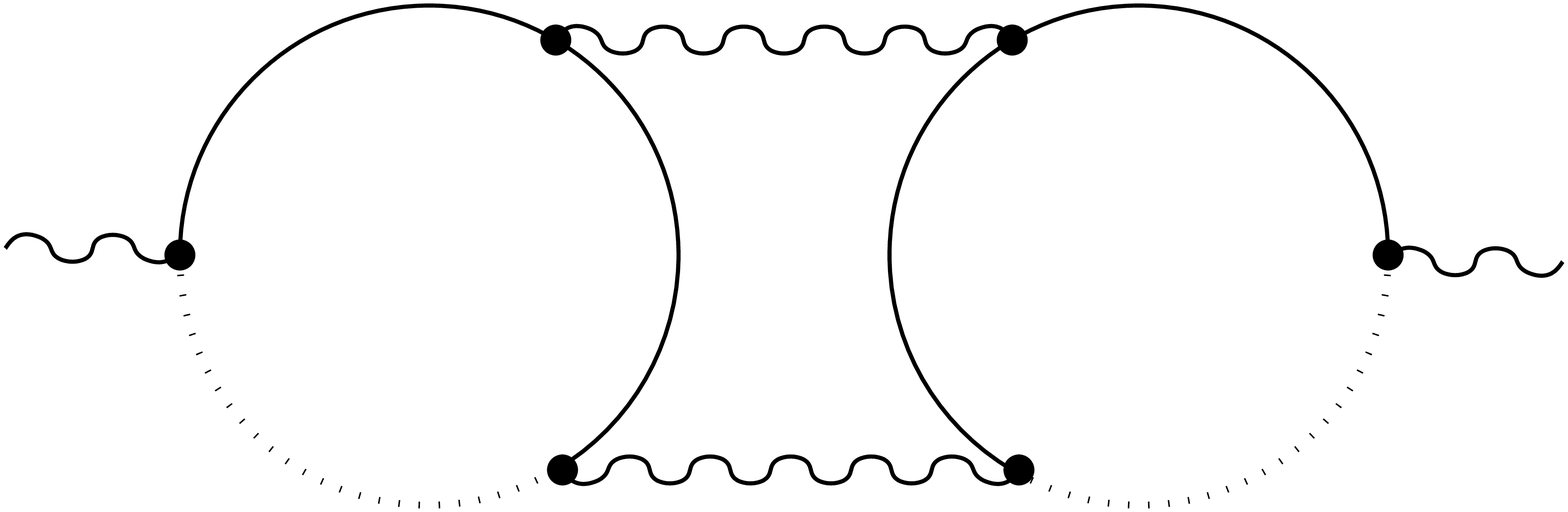}}
\put(8.5,1.0){\scriptsize{$\mathcal{S}_2$}}
\put(8.5,2.5){\scriptsize{$\mathcal{S}_4$}}
\end{picture}
\caption{A particular diagram for the susceptibility formulated in the
fermionic and the spin mode representation.}
\label{fig:translation2}
\end{figure}

Let us summarize the discussion with the help of Figs.~\ref{fig:translation1}%
, \ref{fig:translation2}. Here a closed loop of fermionic Green's
functions (Fig.~\ref{fig:translation1}) as well as a particular
diagram for the susceptibility (Fig.~\ref{fig:translation2}) are
shown formulated first in terms of fermionic Green's functions and
then also as a diagram for interacting spin modes in the effective
theory. We see that one diagram of the conventional perturbation
theory produces several diagrams of the expansion in the spin
modes (we marked the corresponding interactions vertices by
$\mathcal{S}_{i}$, $i=2,3,4$).

At first glance it looks as if our effective perturbation theory has become
even more complicated than the original one. However, it is not so because
its not sufficient to just write the diagrams. One should calculate them
singling out the most interesting low energy contributions. This singling
out has already been performed when deriving the effective theory for the
spin excitations. A larger number of the diagrams in the new theory
corresponds to different possibilities of obtaining low energy contributions
when integrating in the diagrams of the conventional perturbation theory.

The perturbative expansion obtained from the low energy effective action is
equivalent to expressions obtained from Eqs.~(\ref{eq:2.3}) and (\ref%
{eq:4.58}) after expanding $\xi _{\mathbf{p}+\mathbf{q}}\sim \xi _{\mathbf{p}%
}+v_{F}\mathbf{n}\mathbf{q}$ in the vicinity of the Fermi surface in each
loop and subsequent integration in $\xi _{\mathbf{p}}$. The present model,
however, organizes the terms in a way, that is much more convenient for
identifying the most important contributions.

Let us remark that the field $\mathbf{b}$ enters the diagrams in the same
way as $\mathbf{h}$ and that for each loop there is one fixed angle $\mathbf{%
n}$, which is as a direct consequence of the integration over $\xi _{\mathbf{%
p}}$.

\subsection{Bare spin susceptibility}

\label{subsec:bare} The static spin susceptibility $\chi $ at $T=0$,
\be
\chi =2\nu \eta _{\alpha =1},\label{eq:4.59}
\ee
where $\eta $ is given by Eq. (\ref{eq:4.48}), can be obtained
straightforwardly from $\mathcal{S}_{b0},$ Eq. (\ref{eq:4.45}), for $\alpha
=1$. For representations with an arbitrary $\alpha \neq 1$, the term $%
\mathcal{S}_{b0}$ alone does not provide the full answer. For general $%
\alpha $ one should consider the combination $\mathcal{S}_{b0}-1/2\left%
\langle \!\left\langle \mathcal{S}_{b1}^{2}\right\rangle \!\right\rangle $
with $\mathcal{S}_{b1}$ from Eq. (\ref{eq:4.46}). The average in this
formula is to be taken with the full quadratic form $\mathcal{S}_{0}+%
\mathcal{S}_{2}$. This procedure effectively amounts to a ladder summation.
In particular, we will later choose $\alpha =1/2$. In this case one arrives
at Eq.~(\ref{eq:4.59}) with the help of the identity $2\nu \eta _{1}=2\nu
\eta _{1/2}+2\nu \eta _{1/2}^{2}(1-\nu \overline{V}_{t}\eta _{1/2})^{-1}$,
where the first term is obtained from $\mathcal{S}_{b0}$ and the second one
from $-1/2\left\langle \!\left\langle \mathcal{S}_{b1}^{2}\right\rangle
\!\right\rangle $.

\section{\label{sec:Renormalization and Perturbation Theory}Renormalization}

As sketched in the introduction, logarithmic corrections appear in the model
and they can be summed in a renormalization scheme as it has been done for
the interaction amplitudes in Ref. \onlinecite{aleiner}. The appearing
logarithms and thus the renormalized amplitudes generally depend on the
deviation in angles from the ideal forward or backward scattering case. It
is therefore immediately clear that this renormalization scheme cannot
include as effective charges physical quantities like the susceptibility
that do not depend on such angles. This is why we consider the
renormalization of interaction amplitudes and external vertices first and
then include the renormalized values into a perturbation theory for the
susceptibility.

\subsection{Generalization}

During the process of the renormalization new terms may appear in the
action. To consistently take such terms into account, they should be
included into the model from the beginning. To this end a generalization of
Eqs.~(\ref{eq:4.35})--(\ref{eq:4.37}) is introduced in Sec.~\ref%
{subsubsec:genaction}. This step is prepared in Sec.\ref%
{subsubsec:decomposition}.

\subsubsection{Decomposition}

\label{subsubsec:decomposition}

In Eqs.~(\ref{eq:4.35})--(\ref{eq:4.37}) the following decomposition of
supermatrix $P$ was used
\be
P=\sum_{i=1}^{2}P^{i},\qquad P^{i}=\frac{1}{2}\sum_{k=1}^{2}\lambda _{ik}\Pi
_{k}P\Pi _{k},
\ee%
where $[A^{1},\Sigma _{3}]=0$ and $\{A^{2},\Sigma _{3}\}=0$ for arbitrary
matrices $A$.

For a generalization we consider $4$ supermatrices $\Pi _{i}$ with $[\Pi
_{i},\Pi _{j}]=0$ and $\Pi _{i}^{2}=1$, namely
\be
\Pi _{1}=1,\Pi _{2}=\Sigma _{3},\Pi _{3}=\Lambda _{1}\tau _{3},\Pi
_{4}=\Lambda _{1}\Sigma _{3}\tau _{3}
\ee

We decompose a supermatrix $P$ in such a way that $P=\sum_{i=1}^{4}P^{i}$
and
\begin{eqnarray}
&&[P^{1},\Pi _{2}]=0,\quad \{P^{1},\Pi _{3}\}=0 \\
&&[P^{2},\Pi _{2}]=0,\quad \lbrack P^{2},\Pi _{3}]=0 \\
&&\{P^{3},\Pi _{2}\}=0,\quad \lbrack P^{3},\Pi _{3}]=0 \\
&&\{P^{4},\Pi _{2}\}=0\quad \{P^{4},\Pi _{3}\}=0
\end{eqnarray}%
An explicit formula for $P^{i}$ can be given
\be
P^{i}=\frac{1}{4}\sum_{k=1}^{4}\lambda _{ik}\Pi _{k}P\Pi _{k},\;\lambda
_{ik}=\left(
\begin{array}{rrrr}
1 & 1 & -1 & -1 \\
1 & 1 & 1 & 1 \\
1 & -1 & 1 & -1 \\
1 & -1 & -1 & 1%
\end{array}%
\right)
\ee%
It is easily seen that
\be
\mbox{str}\lbrack P_{i}P_{j}]=\delta _{ij}\mbox{str}\lbrack P_{i}^{2}],\quad %
\mbox{str}\lbrack PQ]=\sum_{k=1}^{4}\mbox{str}\lbrack P_{k}Q_{k}].
\ee%
The following useful relation can be checked by direct computation
\begin{eqnarray}
&&\sum_{k_{1},k_{2}=1}^{4}\;\lambda _{i_{1}k_{1}}\lambda _{i_{2}k_{2}}\;\big[%
AL_{k_{1}k_{2}}BL_{k_{1}k_{2}}\big]  \nonumber \\
&=&4\delta _{i_{1},i_{2}}\sum_{k=1}^{4}\;\lambda _{i_{1}k}\;\big[A\Pi
_{k}B\Pi _{k}\big],
\end{eqnarray}%
where $L_{k_{1}k_{2}}=\Pi _{k_{1}}\Pi _{k_{2}}$.

\subsubsection{Generalized action}

\label{subsubsec:genaction} After this preparation we introduce the
generalized model. We start with $S_{4}$.
\begin{eqnarray}
\mathcal{S}_{4} &=&-2\nu \sum_{ij=1}^{4}\lambda _{ij}\;\varepsilon _{\delta
\beta \gamma }\varepsilon _{\delta \beta _{1}\gamma _{1}}\int dXdX_{1} \\
&&\left( \overline{\psi }_{X,\beta }u\tau _{3}\Pi _{j}\psi _{X,\gamma
}\right) \hat{\Gamma}_{i}(X,X_{1})\left( \overline{\psi }_{X_{1},\beta
_{1}}u_{1}\tau _{3}\Pi _{j}\psi _{X_{1},\gamma _{1}}\right)  \nonumber
\end{eqnarray}%
The amplitudes $\hat{\Gamma}_{i}$ are
\begin{eqnarray}
&&\hat{\Gamma}_{i}(X,X_{1})  \label{opgamma} \\
&&\qquad =\Gamma _{i}(\widehat{\mathbf{n}\mathbf{n}_{1}};u,u_{1},(\mathbf{r}-%
\mathbf{r}_{1})^{\perp })\;f(\mathbf{r}-\mathbf{r}_{1})\;\delta (\tau -\tau
^{\prime })  \nonumber
\end{eqnarray}%
and by comparison with Eq.~(\ref{eq:4.36}) one finds their bare values
\be
\Gamma _{i}(\theta ,u,u_{1};\mathbf{r}^{\perp })=\left\{
\begin{array}{cc}
\gamma _{f}^{0}(\theta ) & i=1,2 \\
\gamma _{b}^{0}(\theta ) & i=3,4%
\end{array}%
\right. ,
\ee%
where $\mathbf{r}^{\perp }=\mathbf{r}-(\mathbf{n}\mathbf{r})$ is the
component of $\mathbf{r}$ transverse to $\mathbf{n}$ and one should keep in
mind that important initial and final angles $\mathbf{n}$, $\mathbf{n}_{1}$
are almost parallel or almost antiparallel to each other. The initial values
for $\Gamma _{i}$ do not depend on $u$, $u_{1}$ and $\mathbf{r}^{\perp }$
but develop such a dependence under renormalization.

The generalization of Eq.~(\ref{eq:4.35}) reads
\begin{eqnarray}
&&\mathcal{S}_{2}=-i\nu \sum_{ij}\sum_{\sigma _{1},\sigma _{2}=\pm }\lambda
_{ij}\int dXdX_{1} \\
&&\left( \overline{\psi }_{X,\delta }\tau _{3}\Pi _{j}\partial _{X}\mathcal{F%
}_{\sigma _{1}}\right) \;\hat{{\bm\Delta }}_{i}^{\sigma _{1}\sigma
_{2}}(X,X_{1})\left( \overline{\mathcal{F}_{\sigma _{2}}}\overline{\partial
_{X_{1}}}\Pi _{j}\tau _{3}\psi _{X_{1},\delta }\right)   \nonumber
\end{eqnarray}%
In this formula $\mathcal{F}_{\pm }=\tau _{\pm }\mathcal{F}_{0}$, $\overline{%
\mathcal{F}_{\pm }}=\overline{\mathcal{F}}_{0}\tau _{\mp }$ and $\tau _{\pm
}=(1\pm \tau _{3})/2$ are projection operators that change under charge
conjugation as $\overline{\tau _{\pm }}=\tau _{\mp }$. In analogy to Eq.~(%
\ref{opgamma}) we defined
\begin{eqnarray}
&&\hat{{\bm\Delta }}_{i}^{\sigma _{1}\sigma _{2}}(X,X_{1}) \\
&&\quad ={\bm\Delta }_{i}^{\sigma _{1}\sigma _{2}}\left( \widehat{\mathbf{n}%
\mathbf{n}_{1}};u,u_{1},(\mathbf{r}-\mathbf{r}_{1})^{\perp }\right) \;f(%
\mathbf{r}-\mathbf{r}_{1})\;\delta (\tau -\tau ^{\prime })  \nonumber
\end{eqnarray}%
Due to the relation $\left( \mathcal{F}_{0}\overline{\mathcal{F}_{0}}\right)
^{i}=0,(i=1,4)$ only $(2,3)$- components enter the action. The bare values
of the vertices are equal to
\be
{\bm\Delta }_{i}^{\sigma _{1}\sigma _{2}}(\theta ,u,u_{1};\mathbf{r}_{\perp
})=\left\{
\begin{array}{cc}
\gamma _{f}^{0}(\theta ) & i=2 \\
\gamma _{b}^{0}(\theta ) & i=3%
\end{array}%
\right.
\ee

Finally, we write the cubic term in the form
\begin{eqnarray}
&&\mathcal{S}_{3}=-2\sqrt{-2i}\varepsilon _{\delta \beta \gamma
}\sum_{ij}\sum_{\sigma =\pm }\lambda _{ij}\int dXdX_{1} \\
&&\left( \overline{\psi }_{X,\delta }u\tau _{3}\Pi _{j}\psi _{X,\beta
}\right) \hat{\mathcal{B}}_{i}^{\sigma }(X,X_{1})\left( \overline{\mathcal{F}%
_{\sigma }}\overline{\partial _{X_{1}}}\tau _{3}\Pi _{j}\psi _{X_{1},\gamma
}\right)  \nonumber
\end{eqnarray}%
with
\begin{eqnarray}
&&\hat{\mathcal{B}}_{i}^{\sigma }(X,X_{1}) \\
&&\quad =\mathcal{B}_{i}^{\sigma }\left( \widehat{\mathbf{n}\mathbf{n}_{1}}%
;u,u_{1},(\mathbf{r}-\mathbf{r}_{1})^{\perp }\right) \;f(\mathbf{r}-\mathbf{r%
}_{1})\;\delta (\tau -\tau ^{\prime }),  \nonumber
\end{eqnarray}%
where the bare values of this vertex are
\be
\mathcal{B}_{i}^{\sigma }(\theta ,u,u_{1};\mathbf{r}_{\perp })=\left\{
\begin{array}{cc}
\gamma _{f}^{0}(\theta ) & i=1,2 \\
\gamma _{b}^{0}(\theta ) & i=3,4%
\end{array}%
\right.
\ee

As we will see, in the approximation we consider, a generalization for the
terms $\mathcal{S}_{0},\mathcal{S}_{b1},\mathcal{S}_{b2}$ will be necessary
only in the $1d$ case.

\subsection{Renormalization scheme}

We use a standard momentum shell renormalization group scheme. Separating
fast and slow fields in the action we integrate over the fast fields and
determine in this way the flow of coupling constants as a function of a
running cutoff. In our case this amounts to a resummation of the
perturbation theory in the leading logarithmic approximation. A quantity $y$
is expanded in a series of the form
\be
y=\sum_{n}\left[ \gamma \ln (\dots )\right] ^{n}\;a_{n}(\gamma ),
\ee%
and one attempts to find a Taylor expansion of $a_{n}(\gamma )$. We assume
during the renormalization that the coupling constants $\gamma $ are small, $%
\gamma \ll 1$.

In our case it convenient to define fast fields ${\bm\phi }$ and slow fields
$\bm\Psi $ with respect to the frequency only. The reason is the anisotropy
in momentum. As one can see, relevant momenta $p_{\parallel }$ are of the
order of $\omega /v_{F}$, while momenta $\mathbf{p}_{\perp }$ do not
contribute to the logarithm and enter as parameters. Thus we write
\be
{\bm\psi }(X)=\bm\Psi (X)+{\bm\phi }(X),
\ee%
where the fast fields ${\bm\phi }$ have the frequencies $\omega $ in the
interval,
\be
\kappa \omega _{c}<|\omega |<\omega _{c}
\ee%
while the slow ones $\bm\Psi $ carry frequencies
\be
|\omega |<\kappa \omega _{c},
\ee%
where $\omega _{c}$ is the running cut-off and $\kappa <1$. Fast modes are
integrated over in the Gaussian approximation using averages of the form
\be
\left\langle \dots \right\rangle _{0}=\int d{\bm\phi }\;(\dots )\exp \left( -%
\mathcal{S}_{0}[{\bm\phi }]\right) .
\ee%
This results in a change in $\mathcal{S}$
\be
\delta \mathcal{S}[\bm\Psi ]=-\ln\left\langle \exp \Big(-\mathcal{S}[\bm\Psi +{%
\bm\phi }]\Big)\right\rangle _{0}-\mathcal{S}[\bm\Psi ],
\ee%
that will now be determined explicitly. In diagrams the Green's function of
the fast modes will be denoted by a thick solid line in order to
discriminated it from the Green's function of slow modes.

\subsection{Renormalization of interaction amplitudes}

\label{subsec:renintamplitudes}

\begin{figure}[tb]
\setlength{\unitlength}{2.3em}
\begin{picture}(12,8.5)

\put(0.5,5){\includegraphics[width=4\unitlength]{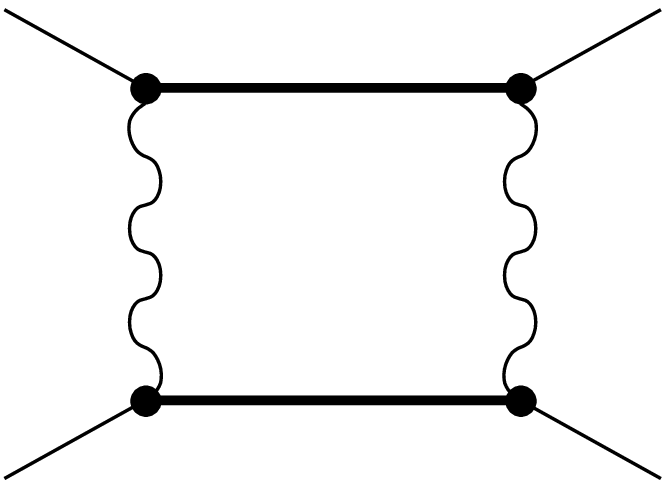}}
\put(2.2,4.6){$(\delta\Gamma)$}\put(0.9,6.3){$\Gamma$}
\put(3.2,6.3){$\Gamma$}

\put(6,5){\includegraphics[width=4\unitlength]{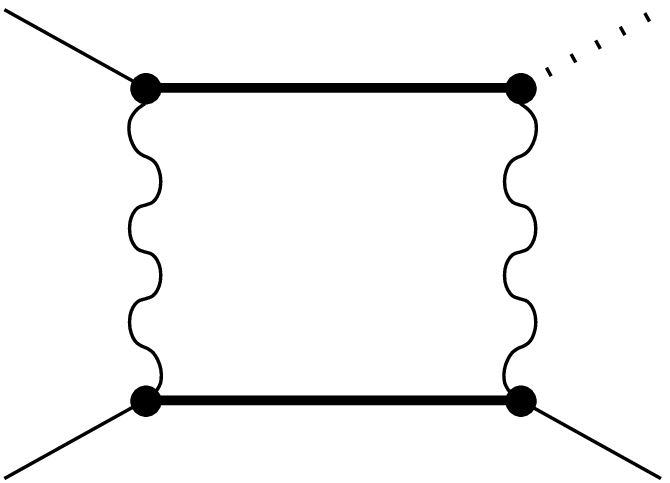}}
\put(7.6,4.6){$(\delta\mathcal{B})$}\put(6.3,6.3){$\Gamma$}
\put(8.6,6.3){$\mathcal{B}$}

\put(0.5,1){\includegraphics[width=4\unitlength]{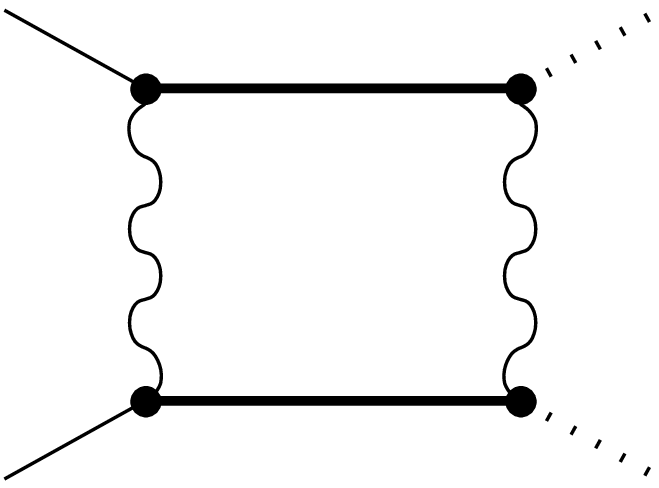}}
\put(2.1,0.6){$(\delta\bfDelta)$}\put(0.85,2.3){$\Gamma$}
\put(3.05,2.3){$\bfDelta$}

\put(6,1){\includegraphics[width=4\unitlength]{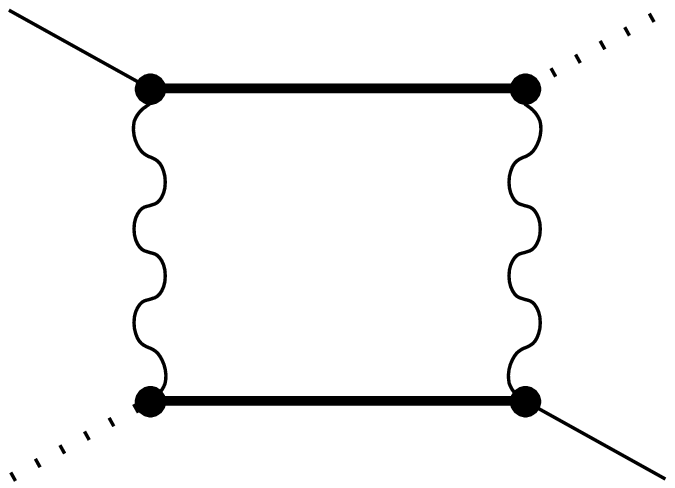}}
\put(7.5,0.6){$(\delta\bfDelta)$} \put(6.35,2.3){$\mathcal{B}$}
\put(8.55,2.3){$\mathcal{B}$}

\end{picture}
\caption{Relevant diagrams for the renormalization of the interaction
amplitudes as found in Ref.~\onlinecite{aleiner}}
\label{fig:intamplitudes}
\end{figure}

The renormalization of the interaction amplitudes was considered in Ref. %
\onlinecite{aleiner}. Here we merely summarize the results, since we will
use them later on. Let us note that for the renormalization group the
symmetric choice $\alpha =1/2$ is most convenient. Relevant diagrams are
shown in Fig.~ \ref{fig:intamplitudes}.

The result of the analysis in Ref.\onlinecite{aleiner} was that the model is
reproduced under renormalization and the changes in $\Gamma_i, {\bm \Delta}%
_i $ and $\mathcal{B}_i$ can conveniently be written in the form
\begin{eqnarray}
&&\delta\hat{\Gamma}_i=\mathfrak{B}_{\Gamma_i}\left(\hat{\Gamma}_j;\hat{%
\mathcal{B}}_j;\hat{{\bm \Delta}}_j\right)\delta\xi \\
&&\delta\hat{\mathcal{B}}_i=\mathfrak{B}_{\mathcal{B}_i}\left(\hat{\Gamma}_j;%
\hat{\mathcal{B}}_j;\hat{{\bm \Delta}}_j\right)\delta\xi \\
&&\delta\hat{{\bm \Delta}}_i=\mathfrak{B}_{{\bm \Delta}_i}\left(\hat{\Gamma}%
_j;\hat{\mathcal{B}}_j;\hat{{\bm \Delta}}_j\right)\delta\xi
\end{eqnarray}
where
\begin{eqnarray}
\delta\xi=uu_1\mu_d\overline{f}_\perp\left(\frac{\mathbf{r}_\perp}{r_0}%
\right)\ln\kappa^{-1}
\end{eqnarray}
Here $\mu_d$ in $d$ dimensions is given as
\begin{eqnarray}
\mu_d=\frac{2}{\pi\nu v_Fr_0^{d-1}}=\left\{%
\begin{array}{ccc}
2\;, & d=1 &  \\
4(p_Fr_0)^{-1}\;, & d=2 &  \\
4\pi(p_Fr_0)^{-2}\;, & d=3 &
\end{array}%
\right.  \label{eq:5.29}
\end{eqnarray}
and
\begin{eqnarray}
\overline{f}_\perp\left(\frac{\mathbf{r}_\perp}{r_0}\right)=r_0^{d-1}\int
\frac{d^{d-1}p_\perp}{(2\pi)^{d-1}}\;\mbox{e}^{i\mathbf{p}_\perp\mathbf{r}%
_\perp} f(\mathbf{p}_\perp)  \label{eq:5.30}
\end{eqnarray}
Therefore the amplitudes can be written in a scaling form
\begin{eqnarray}
&&\Gamma_i(\theta;u,u_1;\mathbf{r}_1)=\gamma_i[\xi(\theta;u,u_1;\mathbf{r}%
_\perp);\gamma_i^0(\theta)] \\
&&\mathcal{B}_i^{\sigma}(\theta;u,u_1;\mathbf{r}_\perp)=\beta_i^\sigma[%
\xi(\theta;u,u_1;\mathbf{r}_\perp);\gamma_i^0(\theta)]  \nonumber \\
&&{\bm \Delta}_i^{\sigma_1\sigma_2}(\theta;u,u_1;\mathbf{r}%
_\perp)=\Delta_i^{\sigma_1\sigma_2}[\xi(\theta;u,u_1;\mathbf{r}%
_\perp);\gamma_i^0(\theta)]  \nonumber
\end{eqnarray}
where $\gamma_1^0=\gamma_2^0=\gamma_f$, $\gamma_3^0=\gamma_4^0=\gamma_b$
give the initial conditions for the flow. The flow stops at $%
\max(\theta,T/\varepsilon_\infty)$, so that
\begin{eqnarray}
\xi(\theta;u,u_1;\mathbf{r}_\perp)=-uu_1\mu_d\overline{f}_\perp\left(\frac{%
\mathbf{r}_\perp}{r_0}\right)\ln\left(\max\left[\theta,\frac{T}{%
\varepsilon_\infty}\right]\right).  \nonumber \\
\end{eqnarray}

For any perturbative calculation of the spin susceptibility the amplitudes $%
\gamma _{1},\beta _{1}^{\pm },\Delta _{1}^{\pm ,\pm }$ cannot enter. This
immediately follows from the relation $\Lambda _{1}\tau _{3}\mathcal{F}_{0}=-%
\mathcal{F}_{0}$, which means that the matrix structure of the Green's
function in $H$-space becomes trivial for every closed loop in perturbation
theory. Since in this paper we are interested in the perturbative sector of
the model only we give here the relevant RG equations for the backward
scattering components
\begin{eqnarray}
&&\frac{d{\gamma }_{3}\left( \xi \right) }{d\xi }=-\left[ {\gamma }_{3}(\xi )%
\right] ^{2}; \\
&&\frac{d{\beta }_{3}^{+}\left( \xi \right) }{d\xi }=-2{\gamma }_{3}\left(
\xi \right) {\beta }_{3}^{+}\left( \xi \right) ;\;\frac{d{\beta }%
_{3}^{-}\left( \xi \right) }{d\xi }=-{\gamma }_{3}\left( \xi \right) {\beta }%
_{3}^{-}\left( \xi \right) ;  \nonumber \\
&& \\
&&\frac{d\Delta _{3}^{++}\left( \xi \right) }{d\xi }=-2\Delta
_{3}^{++}\left( \xi \right) \gamma _{3}\left( \xi \right) -2\left[ \beta
_{3}^{+}\left( \xi \right) \right] ^{2}; \\
&&\frac{d\Delta _{3}^{-+}\left( \xi \right) }{d\xi }=\frac{d\Delta
_{3}^{+-}\left( \xi \right) }{d\xi }=-2\beta _{3}^{-}\left( \xi \right)
\beta _{3}^{+}\left( \xi \right) .
\end{eqnarray}%
There is a subtle point related to the amplitude $\Delta _{3}^{--}$. Instead
of a flow equation the relation
\be
\Delta _{3}^{--}\left( \xi \right) \gamma _{3}(\xi )=\left[ \beta
_{3}^{-}\left( \xi \right) \right] ^{2}
\ee%
was fixed in Ref. \onlinecite{aleiner} to cancel ultraviolet divergencies
that would otherwise develop under a change in the cut-off. We will come
back to this point in Sec.~\ref{sec:sus1d} below.

Appropriate boundary conditions have already been specified when introducing
the model above. The solutions of the flow equations are
\begin{eqnarray}
&&\gamma_{3}\left( \xi\right) =\beta_{3}^{-}\left(
\xi\right)=\Delta_{3}^{--}\left( \xi\right)= \frac{1}{\xi_{b}^{\ast}+\xi}; \label{eq:5.38}\\
&&\beta_{3}^{+}\left( \xi\right) =\Delta_{3}^{+-}\left( \xi\right)
=\Delta_{3}^{-+}\left( \xi\right)=\frac{\xi_{b}^{\ast}}{\left( \xi_{b}^{\ast
}+\xi\right) ^{2}}; \label{eq:5.39}\\
&&\Delta_{3}^{++}\left( \xi\right) =\frac{2\xi_{b}^{\ast2}}{\left(
\xi _{b}^{\ast}+\xi\right) ^{3}}-\frac{\xi_{b}^{\ast}}{\left(
\xi_{b}^{\ast}+\xi\right) ^{2}},\label{eq:5.40}
\end{eqnarray}
where we introduced the notation
\begin{eqnarray}
\xi_{b}^{\ast}(\theta) \equiv \frac{1}{\gamma_b(\theta)}>0.
\end{eqnarray}
and the backscattering amplitude $\gamma_b^0$ is defined in Eq.~(\ref%
{eq:4.40}).

\subsection{Renormalization of $\mathcal{S}_0$, $\mathcal{S}_{b0}$, $%
\mathcal{S}_{b1}$ and $\mathcal{S}_{b2}$}

\label{subsec:rens0}

In this section we consider the renormalization of the terms $\mathcal{S}%
_{0} $, $\mathcal{S}_{b0}$, $\mathcal{S}_{b1}$ and $\mathcal{S}_{b2},$ Eqs. (%
\ref{eq:4.34}, \ref{eq:4.45}-\ref{eq:4.47}). It will be shown that for the
one-loop RG considered in this paper vertex corrections cancel in dimensions
$d=2,3$ and, as a result, these terms are not renormalized. This is no
longer true for $d=1$. Unlike in higher dimensions no angular integration is
performed in one spatial dimension and this fact is responsible for the
appearing of additional logarithmic corrections as will be shown below. When
selecting the relevant corrections in $1d$, we have in mind a comparison to
the well known result of Dzyaloshinskii and Larkin\cite{dl}.

\subsubsection{Corrections to $\mathcal{S}_0$}

\label{subsubsec:s0corrections} 
\begin{figure}[tb]
\setlength{\unitlength}{2.3em}
\begin{picture}(12,5)

\put(4,0.5){\includegraphics[width=3.5\unitlength]{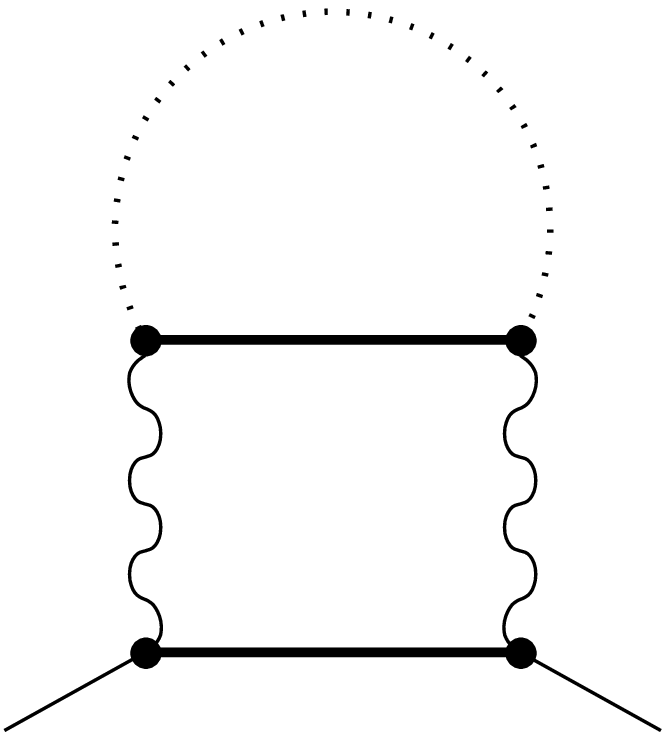}}

\put(4.2,1.7){$\mathcal{B}$} \put(6.2,1.7){$\mathcal{B}$}

\put(4.0,0.8){$x$} \put(5.3,0.2){$(\delta\mathcal{S}_0)$}
\put(7.3,0.8){$x_1$}
\put(5.3,1.1){$\bfn_1$}\put(5.3,2.75){$\bfn_2$}
\end{picture}
\caption{Logarithmic correction to $\mathcal{S}_0$ in $d=1$. }
\label{fig:rens0}
\end{figure}

This contribution has been noticed before%
\cite{aleiner} but was discarded, since for the renormalization of the
interaction amplitudes this term was already beyond the desired accuracy.
The relevant diagram is shown in Fig.~\ref{fig:rens0}. In the presence of
external vertices this term should be considered.

After expanding field $\Psi (x_{1})$ around a point $x$ one obtains an
expression containing the following integral
\be
T\sum_{\omega }\;\int d\mathbf{p}\;\frac{v_{F}\mathbf{n}_{1}\mathbf{p}%
+i\omega }{v_{F}\mathbf{n}_{1}\mathbf{p}-i\omega }\frac{1}{(v_{F}\mathbf{n}%
_{2}\mathbf{p}+i\omega )^{2}}A(\mathbf{p})
\ee%
where $A$ is a product of the vertex parts $\mathcal{B}$ and cut-off
functions $f$ appearing in the expression. The crucial point is that there
is a free integration over the vector $\mathbf{n}_{2}$ and unlike the
contributions from diagrams shown in Fig.~\ref{fig:intamplitudes} a
logarithm can be obtained only in $d=1$. The result can be written as
\be
\delta \mathcal{S}_{0}=-2i\nu \int dX\;\overline{{\bm\psi }}_{X}\mathcal{H}%
_{0}\;\delta \mathcal{R}\;{\bm\psi }_{X}\;,
\ee%
where
\begin{eqnarray}
&&\delta \mathcal{R}=u\int du_{1}\;\delta \xi (u,u_{1}) \\
&&\times \left[ (1+\Pi _{3})\beta _{1}^{+}\beta _{1}^{-}+(1-\Pi _{3})\beta
_{3}^{+}\beta _{3}^{-}\right] \;.  \nonumber
\end{eqnarray}%
It seems natural to interpret this term as a correction to $\mathcal{S}_{0}$%
. For our purposes it is more convenient, however, not to allow $\mathcal{S}%
_{0}$ to change. This can be achieved by rescaling fields ${\bm\psi }$ after
each renormalization step in such a way $\delta \mathcal{R}$ is removed from
$\mathcal{S}_{0}$. This is why we do not write additional RG equations here.
In turn, this rescaling of the fields can lead to additional corrections in
the flow equations for the interaction amplitudes or external vertices. For
the interaction amplitudes, it is in fact easily seen that taking these
corrections into account would be an overstepping of accuracy. This is no
longer true for the external field vertices as will be discussed below.

\subsubsection{Corrections to $\mathcal{S}_{b2}$}

\begin{figure}[tb]
\setlength{\unitlength}{2.3em}
\begin{picture}(12,5.5)

\put(0.5,1.5){\includegraphics[width=3\unitlength]{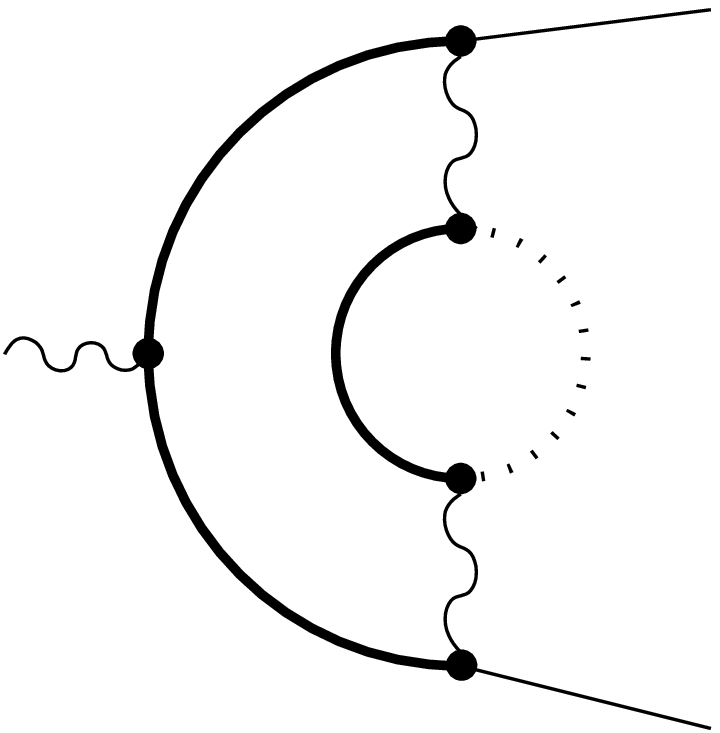}}
\put(2.65,2.1){$\mathcal{B}$} \put(2.65,3.9){$\mathcal{B}$}

\put(4,1.5){\includegraphics[width=3\unitlength]{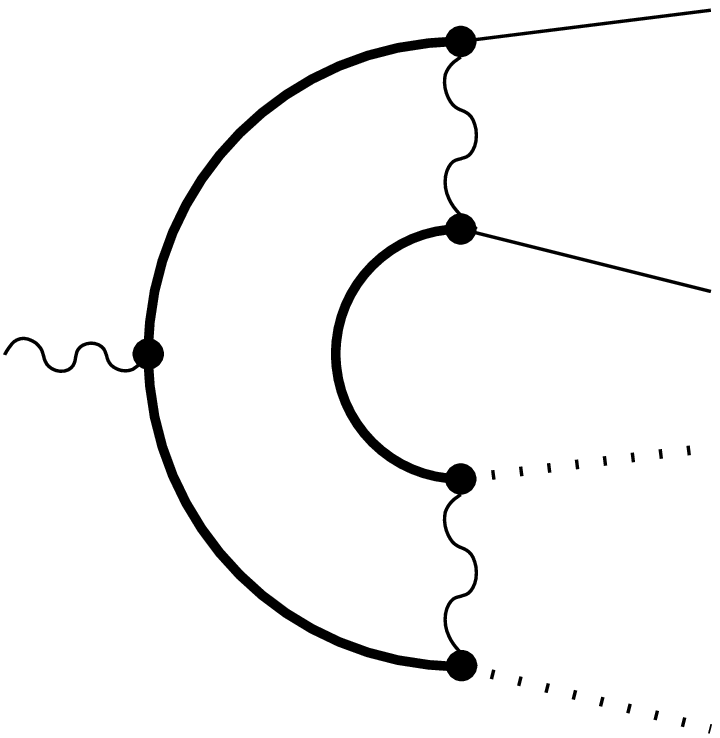}}
\put(6.15,2.1){$\Gamma$} \put(6.15,3.9){$\bfDelta$}

\put(7.5,1.5){\includegraphics[width=3\unitlength]{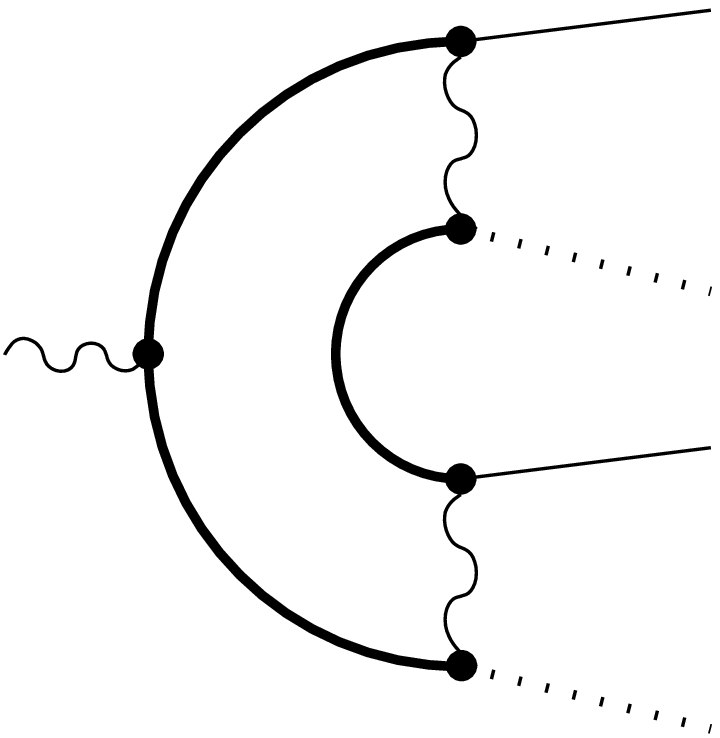}}
\put(9.65,2.1){$\mathcal{B}$} \put(9.65,3.9){$\mathcal{B}$}

\put(2,0.5){$\delta\mathcal{S}_{b2}^{(1)}$}
\put(5.5,0.5){$\delta\mathcal{S}_{b2}^{(2)}$}
\put(9,0.5){$\delta\mathcal{S}_{b2}^{(3)}$}
\end{picture}
\caption{Diagrams for the corrections to $\mathcal{S}_{b2}$. Logarithmic
corrections from $\protect\delta\mathcal{S}_{b2}^{(1)}$ and $\protect\delta%
\mathcal{S}_{b2}^{(2)}$ cancel each other in any dimension. $\protect\delta%
\mathcal{S}_{b2}^{(3)}$ gives a logarithmic correction in $d=1$, but for the
susceptibility it gives corrections beyond our accuracy.}
\label{fig:rensb2}
\end{figure}

Relevant contributions to the term $\mathcal{S}_{b2}$ are represented in
Fig.~\ref{fig:rensb2}. The correction $\delta \mathcal{S}_{b2}^{(2)}$ is
determined by the vertices $\Delta $ and $\Gamma $ and can be written in the
form
\begin{eqnarray}
\delta \mathcal{S}_{b2}^{(2)} &=&8\nu \eta \;\varepsilon _{\alpha \beta
\gamma }\;\int dXdX_{1}\;uu_{1}b_{\alpha }(x)\; \\
&&\times \gamma _{3}\Delta _{3}^{+-}\;\delta \xi \;\overline{\psi }_{X,\beta
}\mathbbm{K}_{3}^{++}\psi _{X_{1},\gamma }\;\tilde{f}(x-x_{1})\;,  \nonumber
\end{eqnarray}%
where
\be
\mathbbm{K}=\mathcal{F}_{0}\overline{\mathcal{F}}_{0},\quad \mathbbm{K}%
^{++}=\tau _{+}\mathbbm{K}\tau _{+}
\ee%
and
\be
\tilde{f}(x-x_{1})=\overline{f}(\mathbf{r}-\mathbf{r}_{1})\delta (\tau -\tau
_{1})\;.
\ee%
The correction is logarithmic in any dimension. However, the form of $%
\mathcal{S}_{b2}^{\left( 2\right) }$ is different from that of $\mathcal{S}%
_{b2}$ because it contains integration over both $\mathbf{n}$ and $\mathbf{n}%
_{1}$, which contrast the bare form $\mathcal{S}_{b2}$, Eq. (\ref{eq:4.47})
Moreover, the matrix $\mathbbm{K}$ breaks the symmetry in $g$-space
(superspace) and, at first glance, one should introduce additional
renormalization coupling constants.

Fortunately there is another diagram that exactly cancels the previous one.
It is also shown in Fig.~\ref{fig:rensb2} and its contribution equals%
\begin{eqnarray}
\delta \mathcal{S}_{b2}^{(1)} &=&-8\nu \eta \;\varepsilon _{\alpha \beta
\gamma }\;\int dXdX_{1}\;uu_{1}b_{\alpha }(x)\; \\
&&\times \beta _{3}^{+}\beta _{3}^{-}\;\delta \xi \;\overline{\psi }%
_{X,\beta }\mathbbm{K}_{3}^{++}\psi _{X_{1},\gamma }\;\tilde{f}(x-x_{1})
\nonumber
\end{eqnarray}%
In fact, one comes to the exact cancellation
\be
\delta \mathcal{S}_{b2}^{(2)}=-\delta \mathcal{S}_{b2}^{(1)}
\ee%
by virtue of the relation
\be
\gamma _{3}\Delta _{3}^{+-}=\beta _{3}^{+}\beta _{3}^{-}
\ee%
that follows immediately from Eqs. (\ref{eq:5.38}-\ref{eq:5.40}).

Finally there is an additional logarithmic contribution in $1d$, $\delta
\mathcal{S}_{b2}^{(3)}$, represented in Fig.~{\ref{fig:rensb2}}. It has a
similar form as $\delta S_{0}$, but taking this correction into account
would mean overstepping the accuracy for our problem. The reason is that due
to the supersymmetry no diagram for the susceptibility can be formed with
the help of the vertex $\mathcal{S}_{b2}$ without including additional
interaction amplitudes. As a consequence, the leading correction to the spin
susceptibility in $d=1$ resulting from this contribution would be $\delta
\chi \propto \gamma ^{3}\ln (\dots )$, which is beyond our accuracy. For the
same reason the rescaling of the fields, which is necessary to bring $%
\mathcal{S}_{0}$ to its bare form after the renormalization, need not be
considered here. This result relies on the symmetry in the $g$-space and
this is why it was important to check the cancellation of the terms
violating the supersymmetry.

\subsubsection{Corrections to $\mathcal{S}_{b0}$}

\begin{figure}[tb]
\setlength{\unitlength}{2.3em}
\begin{picture}(12,5.5)

\put(2,1.5){\includegraphics[width=8\unitlength]{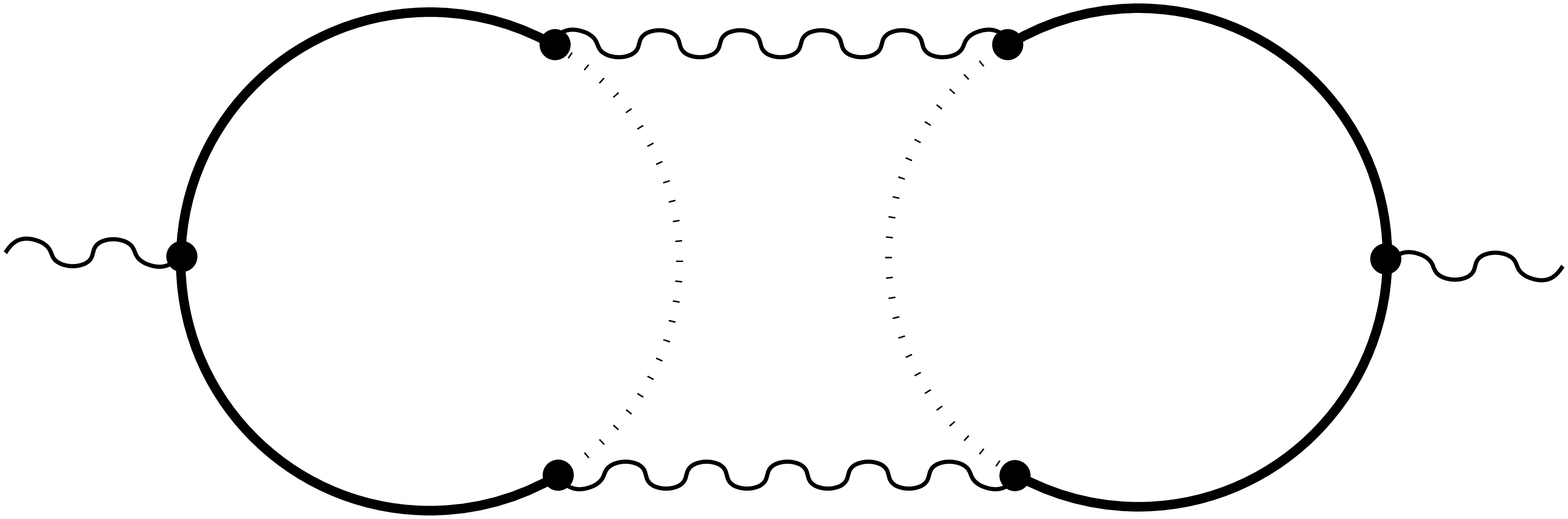}}

\put(5.5,0.5){$\delta\mathcal{S}_{b0}$}
\put(5.8,2){$\bfDelta$}\put(5.8,4.2){$\bfDelta$}

\end{picture}
\caption{Logarithmic correction to $\mathcal{S}_{b0}$ in $d=1$.}
\label{ren:sb0}
\end{figure}

The analytic expression corresponding to the diagram shown in Fig.~\ref%
{ren:sb0} is quadratic in $\mathbf{b}$ but does not contain any
slow field $\bm\Psi $. Therefore, we attribute the corresponding
contribution to the renormalization of $\mathcal{S}_{b0}$, Eq.
(\ref{eq:4.45}). Clearly, in this case the rescaling of the fields
is not important. Since there is a free integration over both the
vectors $\mathbf{n}_{1}$ and $\mathbf{n}_{2}$ for $d>1$, the
correction is logarithmic only in $d=1$. This is similar to what
happens when calculating the correction to $\mathcal{S}_{0}$.

In $d=1$, we write $\mathcal{S}_{b0}$ in the form
\be
S_{b0}=-\frac{1}{2}\nu \eta \int
dxdu_{1}du_{2}\;\mathbf{b}^{2}(x)\;\sigma (\xi
,u_{1},u_{2})\label{eq:5.51}
\ee
and set
\be
\sigma (\xi =0,u_{1},u_{2})=1.
\ee%
Then the correction $\delta \sigma $ to this quantity takes the form
\begin{eqnarray}
\delta \sigma &=&-\frac{1}{2}\eta \;u_{1}u_{2}\;\delta \xi
\Big(\left( \Delta ^{+-}\right) ^{2}+\left( \Delta
_{3}^{-+}\right) ^{2}+2\Delta _{3}^{++}\Delta _{3}^{--}\Big).
\nonumber \\\label{eq:5.53} &&
\end{eqnarray}

\subsubsection{Correction to $\mathcal{S}_{b1}$}

\begin{figure}[tb]
\setlength{\unitlength}{2.3em}
\begin{picture}(12,5.5)

\put(0.5,1.7){\includegraphics[width=5.5\unitlength]{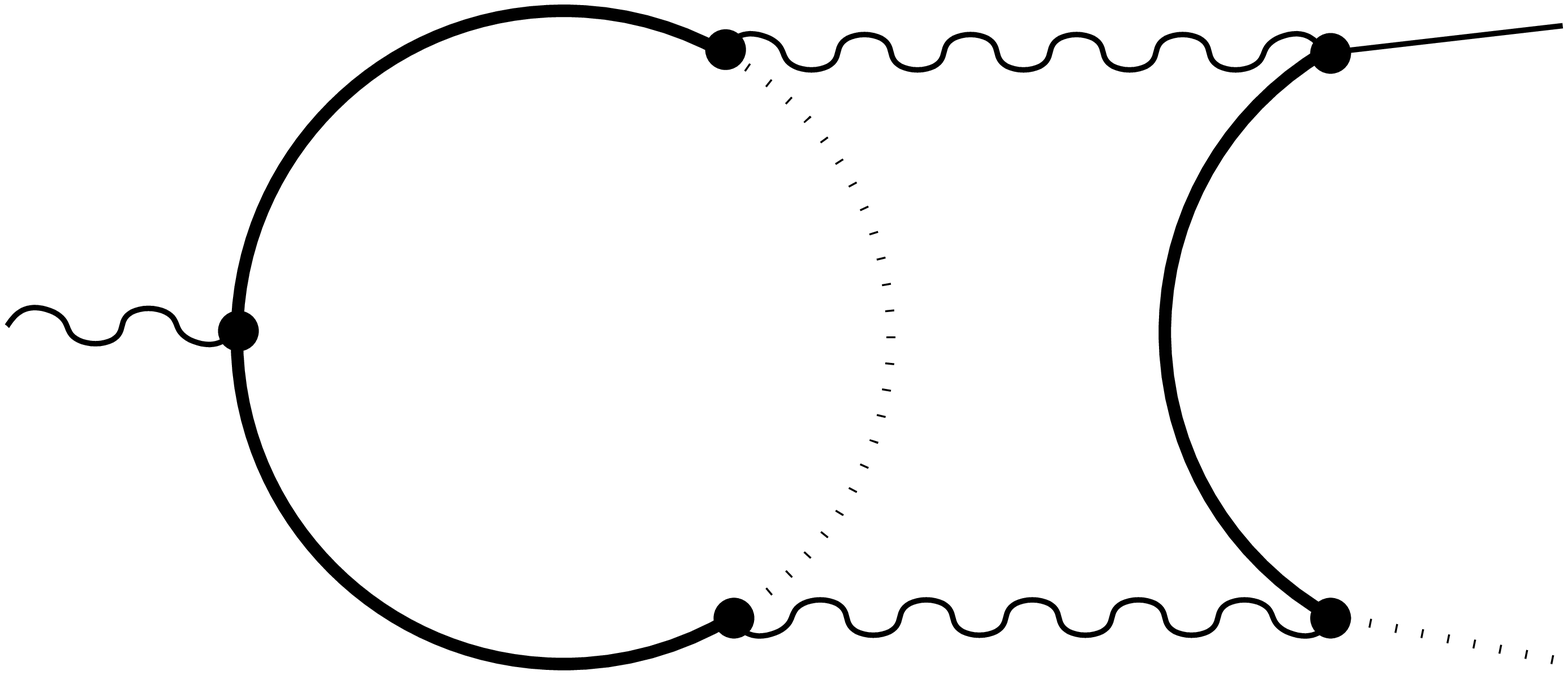}}
\put(3.8,2.1){$\bfDelta$}\put(3.9,4.1){$\mathcal{B}$}

\put(7,1.3){\includegraphics[width=3\unitlength]{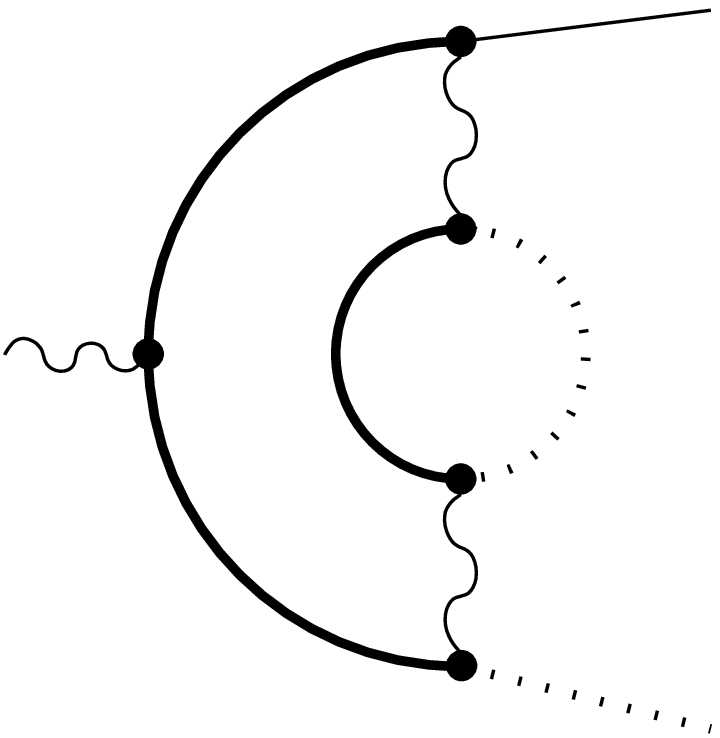}}
\put(9.1,1.9){$\bfDelta$}\put(9.1,3.6){$\mathcal{B}$}

\put(2.5,0.5){$\delta\mathcal{S}_{b1}:$ $\delta\mu_a+\delta\mu_b$}
\put(8,0.5){$\delta\mathcal{S}_{b1}$: $\delta\mu_c^{(1)}$}
\end{picture}
\caption{Logarithmic corrections to $\protect\delta\mathcal{S}_{b1}$ in $d=1$%
.}
\label{fig:rensb1}
\end{figure}

There are two separate contributions to $\mathcal{S}_{b1}$, Eq. (\ref%
{eq:4.46}), represented in Fig.~\ref{fig:rensb1}. In $d>1$ the slow field $%
\bm\Psi $ fixes the vector $\mathbf{n}$ in one of the Green's functions $%
\mathcal{G}$ only, while the other vector $\mathbf{n}^{\prime }$ is
integrated over. As a consequence, a logarithmic correction is obtained only
in $d=1$. In one dimension the rescaling of the fields (cf. Sec.~\ref%
{subsubsec:s0corrections}) is also important and gives an additional
contribution.

Considering the case $d=1$ we present $\mathcal{S}_{b1}$ in the form
\begin{eqnarray}
\mathcal{S}_{b1} &=&\sqrt{-2i}\nu \eta \int dxdz_{1}dz_{2}\;b_{\alpha }(x)
\nonumber \\
&&\qquad \times \overline{\mathcal{F}}_{0}\tau _{3}\overline{\mathcal{D}}%
(\xi ,u_{1},u_{2})\psi _{\alpha }(x,u_{2})\;.
\end{eqnarray}%
Trivial "angular" integration in $1d$ (weighted summation over the
directions) has been performed. The operator
$\overline{\mathcal{D}}$ is defined as
\be
\overline{\mathcal{D}}(\xi ,u_{1},u_{2})=\left(
\begin{array}{cc}
\frac{u_{2}}{2}\Big(\mu _{a}iv_{F}\Sigma _{3}\nabla +\mu _{b}\partial _{\tau
}\Big) & 0 \\
0 & \mu _{c}%
\end{array}%
\right) .\label{eq:5.55}
\ee
Here $\mu _{i}=\mu _{i}(\xi ,u_{1},u_{2})$, and initially $\mu _{i}(\xi
=0,u_{1},u_{2})=1$, $i=a,b,c$. In this case
\be
\overline{\mathcal{D}}(\xi =0,u_{1},u_{2})=\overline{\partial }(x,u_{2})
\ee%
and we come back to the original form displayed in Eq.~(\ref{eq:4.46}).

The diagrams in Fig.~{\ref{fig:rensb1}} represent corrections to $\mu _{i}$.
The left diagram determines corrections $\delta \mu _{a}$ and $\delta \mu
_{b}$
\be
\delta \mu _{a}=-\frac{u_{1}}{2}\left( \beta _{3}^{-}\Delta
_{3}^{+-}+\beta _{3}^{+}\Delta _{3}^{--}\right) \delta \xi
=-\delta \mu _{b}\label{eq:5.57}
\ee
The contribution $\delta \mu _{c}$ consists of two parts
\be
\delta \mu _{c}=\delta \mu _{c}^{(1)}+\delta \mu
_{c}^{(2)}.\label{eq:5.58}
\ee
The correction $\delta \mu _{c}^{(1)}$ is represented by the right diagram
in Fig.~\ref{fig:rensb1} and reads
\be
\delta \mu _{c}^{(1)}=-u_{2}\left( \beta _{3}^{+}\Delta
^{-+}+\beta _{3}^{-}\Delta _{3}^{+-}\right) \delta \xi
.\label{eq:5.59}
\ee
The correction $\delta \mu _{c}^{(2)}$ is due to the rescaling of the fields
has to be performed at each RG step to keep the form of $\mathcal{S}_{0}$
fixed
\be
\delta \mu _{c}^{(2)}=-2u_{2}\beta _{3}^{+}\beta _{3}^{-}\delta
\xi\label{eq:5.60}
\ee
Note that the forward scattering components drop out as could be expected.

\subsubsection{RG equations and their solution}

We found logarithmic corrections to the vertices $\mathcal{S}_{b0}$, $%
\mathcal{S}_{b1}$ and $\mathcal{S}_{b2}$ in dimensionality $d=1$ only. This
means that these terms are not renormalized in the first order in the
dimensionalities $d=2,3$ and the vertices $\gamma ,$ $\beta $ and $\Delta $
given by Eqs. (\ref{eq:5.38}-\ref{eq:5.40}) are sufficient to determine the
susceptibitlity.

At the same time, the renormalization of the vertices $\mathcal{S}_{b0}$, $%
\mathcal{S}_{b1}$ and $\mathcal{S}_{b2}$ is very important in $d=1$. Both
functions $\mu _{i}$ and $\sigma $ from Eqs. (\ref{eq:5.51}, \ref{eq:5.55})
do not have a simple form and one should write and solve proper RG
equations. For the function $\sigma $ related to $\delta \mathcal{S}_{b0}$,
we write $\sigma =\sigma (\xi ,u_{1},u_{2})$ and using the correction $%
\delta \sigma $, Eq. (\ref{eq:5.53}), obtain the following differential
equation
\be
\frac{\partial \sigma }{\partial \xi }=-\frac{1}{2}\eta \;u_{1}u_{2}\left(
\frac{6{\xi _{b}^{\ast }}^{2}}{(\xi +\xi _{b}^{\ast })^{4}}+\frac{2\xi
_{b}^{\ast }}{(\xi +\xi _{b}^{\ast })^{3}}\right) ,
\ee%
With the boundary condition $\sigma (\xi =0,u_{1},u_{2})=1$ we obtain
\begin{eqnarray}
&&\sigma (\xi ,u_{1},u_{2})= \\
&&\qquad 1+\frac{1}{2}u_{1}u_{2}\;\eta \left( \frac{2{\xi _{b}^{\ast }}^{2}}{%
(\xi +\xi _{b}^{\ast })^{3}}-\frac{\xi _{b}^{\ast }}{(\xi +\xi _{b}^{\ast
})^{2}}-\frac{1}{\xi _{b}^{\ast }}\right) \;.  \nonumber
\end{eqnarray}%
The corresponding differential equations for $\mu _{i}$ are to be
obtained from the forms of the corrections, Eqs.
(\ref{eq:5.57}-\ref{eq:5.60}), and can be written as
\begin{eqnarray}
\frac{\partial \mu _{a}}{\partial \xi } &=&-u_{1}\frac{\xi _{b}^{\ast }}{%
(x+\xi _{b}^{\ast })^{3}}=-\frac{\partial \mu _{b}}{\partial \xi } \\
\frac{\partial \mu _{c}}{\partial \xi } &=&-u_{2}\left( \frac{{3\xi
_{b}^{\ast }}^{2}}{(\xi +\xi _{b}^{\ast })^{4}}+\frac{\xi _{b}^{\ast }}{(\xi
+\xi _{b}^{\ast })^{3}}\right)
\end{eqnarray}%
with the boundary conditions $\mu _{i}(\xi =0)=1$. Integrating
these equations we obtain (only $\mu _{a}$ and $\mu _{c}$ will
enter our results)
\begin{eqnarray}
\mu _{a} &=&1+\frac{u_{1}}{2}\left( \frac{\xi _{b}^{\ast }}{(\xi +\xi
_{b}^{\ast })^{2}}-\frac{1}{\xi _{b}^{\ast }}\right) \\
\mu _{c} &=&1+\frac{u_{2}}{2}\left( \frac{2{\xi _{b}^{\ast }}^{2}}{(\xi +\xi
_{b}^{\ast })^{3}}+\frac{\xi _{b}^{\ast }}{(\xi +\xi _{b}^{\ast })^{2}}-%
\frac{3}{\xi _{b}^{\ast }}\right) .
\end{eqnarray}

The calculations presented in this subsection allowed us to obtain all
effective vertices entering the RG scheme. This gives us a possibility to
calculate the susceptibility for all dimensions $d=1,2,3.$ The result for $%
d=1$ is well known\cite{dl} from a renormalization group treatment for the
initial electron model. We will reproduce now this result using the derived
equations in order to check the formalism of the bosonization used here.
Only after that we will concentrate on calculating the susceptibility in the
higher dimensionalities $d=2,3$.

\section{Spin susceptibility in $d=1$}

\label{sec:sus1d} We can now determine the temperature dependent correction
to the static spin susceptibility in $d=1$.

\begin{figure}[tb]
\setlength{\unitlength}{2.3em}
\begin{picture}(12,4.5)
\put(7.9,1.5){$\tilde{\Delta}$}
\put(0.5,1.3){\includegraphics[width=3\unitlength]{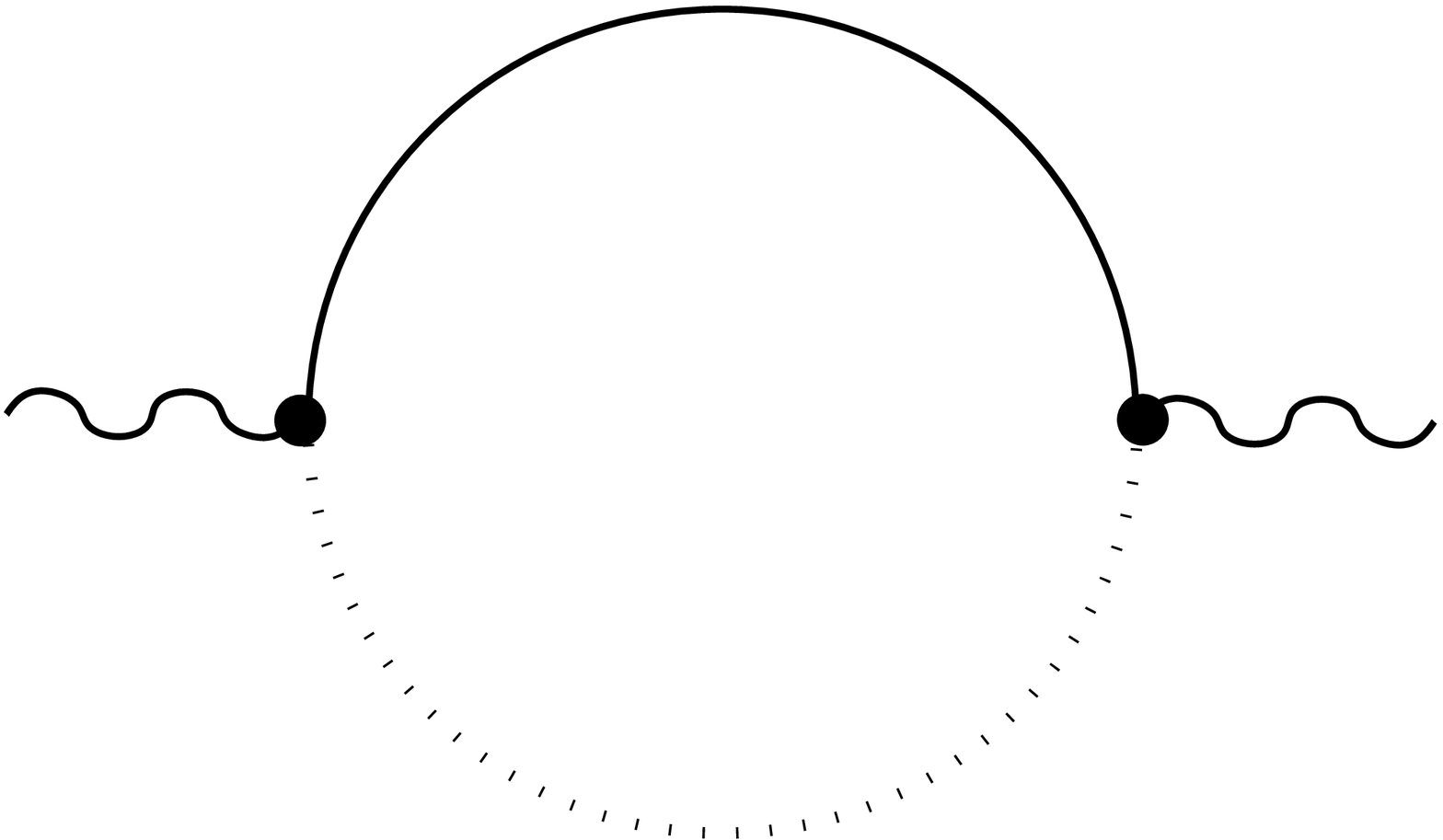}}
\put(5,1.3){\includegraphics[width=6\unitlength]{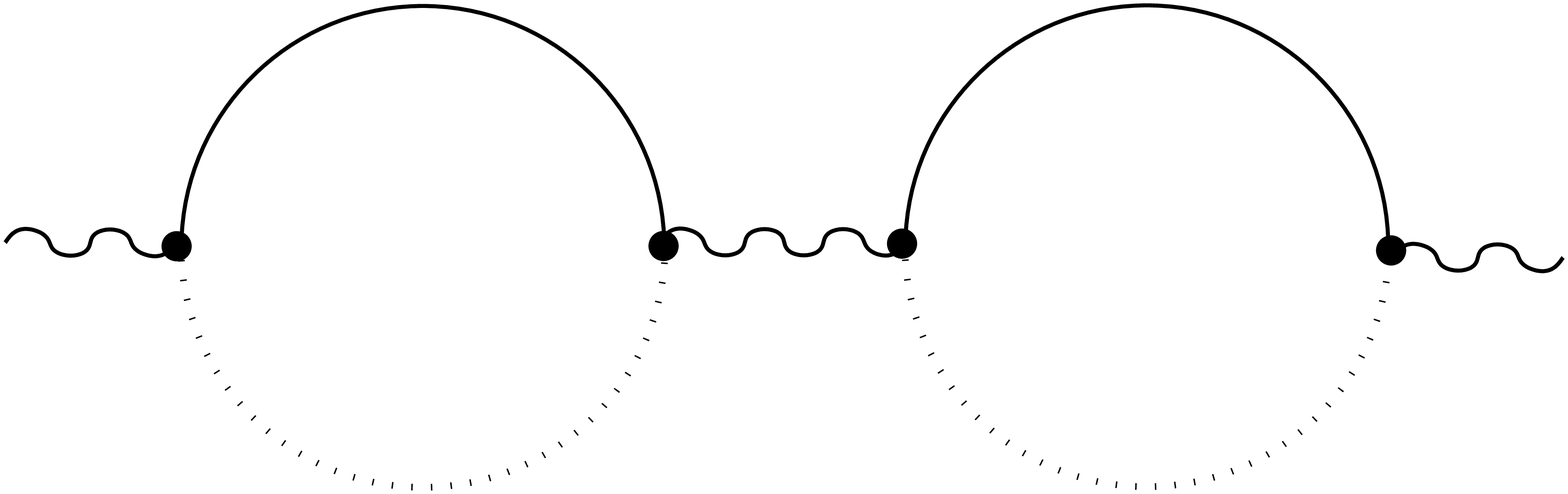}}

\put(1.6,0.5){$\delta\chi_2$} \put(7.7,0.5){$\delta\chi_3$}

\end{picture}
\caption{These two diagrams with renormalized external vertices ($\protect%
\delta\protect\chi_2$) and interaction amplitude ($\protect\delta\protect\chi%
_3$) determine the correction to the susceptibility in $d=1$ together with
the diagram of Fig.~\protect\ref{ren:sb0}. }
\label{fig:sus1d}
\end{figure}

We write the correction to the susceptibility as
\begin{equation}
\delta \chi =\delta \chi _{1}+\delta \chi _{2}+\delta \chi _{3}  \label{a3}
\end{equation}%
and denote the contribution from $\delta \mathcal{S}_{b0}$, Fig.~\ref%
{ren:sb0}, as $\delta \chi _{1}$. The second term $\delta \chi _{2}$ in Eq. (%
\ref{a3}) is the contribution that corresponds to the diagram shown in Fig.~%
\ref{fig:sus1d} on the left hand side.

Here only the renormalized vertex of $\mathcal{S}_{b1}$ enters but no
additional interaction amplitude. A diagram with this property does not
exist for vertex $\mathcal{S}_{b2}$. Finally, the diagram shown on the right
hand side of Fig.~\ref{fig:sus1d} gives a correction termed $\delta \chi
_{3} $. It involves a renormalized interaction amplitude. The corresponding
expressions take the form
\begin{eqnarray}
\delta \chi _{1} &=&\nu \int du_{1}du_{2}\;\sigma (\xi ,u_{1},u_{2}) \\
\delta \chi _{2} &=&2\nu \int du_{1}du_{2}du_{3}\;u_{2}\;\mu
_{c}(\xi ,u_{3},u_{2})\mu _{a}(\xi ,u_{1},u_{2})  \nonumber \\
&& \\
\delta \chi _{3} &=&\frac{1}{2}\nu\int
du_{1}du_{2}\;u_{1}u_{2}\;\sum_{\alpha ,\beta =\pm }\tilde{\Delta}%
_{3}^{\alpha \beta }(\xi ,u_{1},u_{2})  \label{eq:6.3}
\end{eqnarray}

We do not write factors of $\eta$ in $d=1$, since we want avoid
unnecessary complications while focusing on the leading
temperature dependent corrections.
In Eq.~(\ref{eq:6.3}) we introduced the interaction amplitudes $\tilde{\Delta%
}_{3}^{\alpha \beta }$. Naively one would expect amplitudes $\Delta
_{3}^{\alpha \beta }$ Eqs. (\ref{eq:5.38}-\ref{eq:5.40}) to enter here but
this would not be correct. In fact, this question is intimately related a
subtle point related to the renormalization of $\Delta _{3}^{--}$ already
alluded to in Sec.~\ref{subsec:renintamplitudes}.

When calculating corrections to $\Delta _{3}^{--}$ within the
renormalization scheme, the authors of Ref.\onlinecite{aleiner} found
ultraviolet divergencies that could be cancelled only provided the condition
$\Delta _{3}^{--}\gamma _{3}=\left( \beta _{3}^{-}\right) ^{2}$ is imposed.
Since $\beta _{3}^{-}$ and $\gamma _{3}$ can be determined independently,
this condition fixes $\Delta _{3}^{--}$. For large temperatures, where one
can use the bare values of these amplitudes, this relation is automatically
fulfilled. It is crucial to note now that it was necessary to fix $\Delta
_{3}^{--}$ only because this amplitude itself enters $\mathcal{S}_{2}[{\bm%
\phi }]$, where fields ${\bm\phi }$ are the fast modes.

Returning to the diagram for $\delta \chi _{3}$, Fig.~\ref{fig:sus1d} we see
that the frequencies flowing through the Green's functions are determined by
the external vertices and therefore are vanishingly small. In particular,
they are smaller than any frequency considered in the renormalization
scheme. We argue that the part of $\mathcal{S}_{2}$ that contains the fields
at vanishingly small frequencies should be split off from the beginning and
when separating fast and slow modes, it must always contain slow fields
only. Correspondingly, the interaction amplitudes, termed $\tilde{\Delta}%
_{3}^{\alpha \beta }$, are renormalized but do not play any role when
calculating corrections to the interaction vertices. In such a situation,
there is no reason to fix $\tilde{\Delta}_{3}^{--}$ as was done previously
for $\Delta _{3}^{--}$. Instead, one should follow the renormalization group
scheme and derive a proper RG equation for $\tilde{\Delta}_{3}^{--}$.

The relevant diagrams have been already presented in Fig.\ref%
{fig:intamplitudes} and the result of the RG procedure can be expressed by
the equation
\be
\frac{\partial \tilde{\Delta}_{3}^{--}}{\partial \xi }=-2(\beta
_{3}^{-})^{2}=-\frac{2}{(\xi +\xi _{b}^{\ast })^{2}}\label{eq:6.4}
\ee
The solution of Eq. (\ref{eq:6.4}) with the initial condition $\tilde{\Delta}%
_{3}^{--}(\xi =0)=1/\xi _{b}^{\ast }$ takes the form
\be
\tilde{\Delta}_{3}^{--}=\frac{2}{\xi +\xi _{b}^{\ast }}-\frac{1}{\xi
_{b}^{\ast }}\;.
\ee%
This should be contrasted with
\begin{equation}
\Delta _{3}^{--}=\frac{1}{\xi +\xi _{b}^{\ast }}  \label{eq:6.6}
\end{equation}

We checked our reasoning by a perturbative calculation at order $\gamma ^{2}$%
, where the difference between $\tilde{\Delta}_{3}^{--}$ and $\Delta
_{3}^{--}$ is already noticeable.

Finally, we use the identities
\begin{eqnarray}
\frac{1}{1+X} &=&\int_{0}^{1}du_{1}du_{2}\;u_{1}u_{2}\;\left(
z_{12}+z_{12}^{2}+2z_{12}^{3}\right)  \\
&=&\int_{0}^{1}du_{1}du_{2}\;u_{1}^{2}\left( z_{12}^{2}+2z_{12}^{3}\right)
\end{eqnarray}%
where $z_{12}=1/(1+u_{1}u_{2}X),$ that can be checked by a direct
computation of the integrals. Then, recalling that $X=2\gamma _{b}\ln
(\varepsilon _{F}/T)$ we come to the following temperature dependent
correction to the spin susceptibility.
\begin{equation}
\delta \chi (T)=\frac{2\nu \gamma _{b}}{1+2\gamma _{b}\ln
(\varepsilon _{F}/T)}  \label{eq:6.9}
\end{equation}%
This result has first been obtained by Dzyaloshinskii and Larkin\cite{dl}.
Eq. (\ref{eq:6.9}) serves as a good check of the bosonization approach used
here. Actually, the calculations within the framework of the bosonization
method of Ref. \cite{aleiner} are most difficult in $d=1$. It is clear that
this method is less convenient for calculations in $1d$ than the other well
developed ones \cite{gntbook}. However, calculations in $d=2,3$ are somewhat
less involved and the present approach is the most convenient tool for
calculations in these dimensionalities. In the next Section we concentrate
on such calculations.

\section{\label{sec:Non-analytic corrections to spin susceptibility in
$d=2,3$}Non-analytic corrections to spin susceptibility in $d=2,3$}

Non-analytic corrections to the spin susceptibility have been considered in
several works before \cite{coffey, belitz,
chubukov1,sarma,Catelani,baranov,chitov,chubukov2,chubukov3}. A linear in $T$
behavior at order $\gamma ^{2}$ was obtained in $2d$, while the potential
analog in $3d$, a $T^{2}\ln T$ behavior, was found to be absent and the
first correction in $3d$ was proportional to $T^{2}$. We will show now that
there are logarithmic corrections to these results and sum up the leading
logarithms.

Let us repeat that, as it has been demonstrated in
Sec.~\ref{subsec:rens0}, the terms $\mathcal{S}_{b0}$,
$\mathcal{S}_{b1}$ and $\mathcal{S}_{b2}$, Eqs.
(\ref{eq:4.45}-\ref{eq:4.47}), are not renormalized in dimensions
$d>1$ . Therefore we can perform a perturbative analysis with the
renormalized
interaction vertices $\gamma ,$ $\beta $, and $\Delta ,$ Eqs. (\ref{eq:5.38}-%
\ref{eq:5.40}), keeping the bare values of $\mathcal{S}_{b0}$, $\mathcal{S}%
_{b1}$ and $\mathcal{S}_{b2}$.

\begin{figure}[tb]
\setlength{\unitlength}{2.3em}
\begin{picture}(12,13)

\put(0.5,10.5){\includegraphics[width=5\unitlength,]{nmodsus3}}
\put(2.8,10){$(1)$} \put(2.9,12.2){\scriptsize{$\bfDelta$}}
\put(2.9,10.8){\scriptsize{$\bfDelta$}}

\put(6.2,10){\includegraphics[width=2.9\unitlength,height=2.9\unitlength]{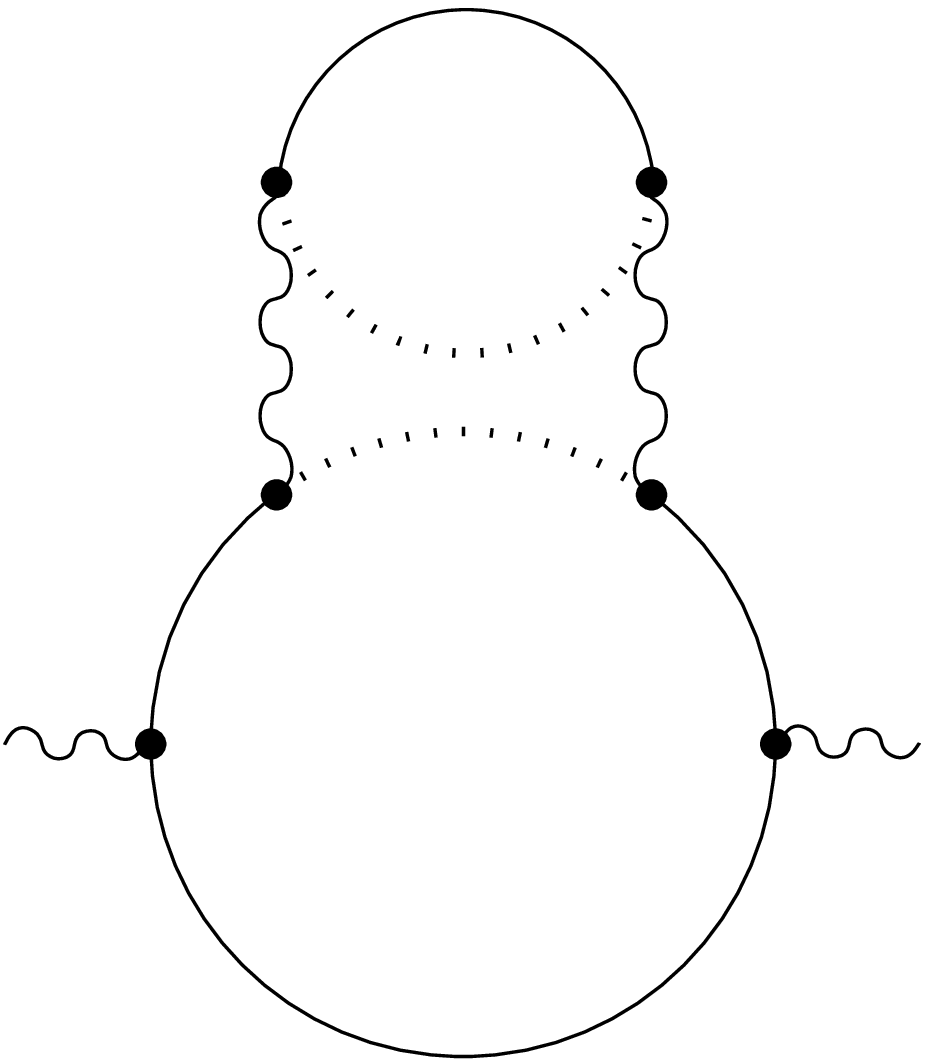}}
\put(8.7,10){$(2)$}
\put(6.5,11.9){\scriptsize{$\bfDelta$}}\put(8.35,11.9){\scriptsize{$\bfDelta$}}

\put(0.5,7.5){\includegraphics[width=5\unitlength]{nmodsus4}}
\put(2.8,7){$(3)$} \put(2.9,7.8){\scriptsize{$\bfDelta$}}
\put(3.0,9.2){\scriptsize{$\Gamma$}}

\put(6,7.5){\includegraphics[width=5\unitlength]{nmodsus1}}
\put(8.3,7){$(4)$} \put(8.4,7.8){\scriptsize{$\mathcal{B}$}}
\put(8.4,9.2){\scriptsize{$\mathcal{B}$}}

\put(0.5,4.5){\includegraphics[width=5\unitlength]{nmodsus2}}
\put(2.8,4){$(5)$} \put(2.9,4.8){\scriptsize{$\bfDelta$}}
\put(2.9,6.2){\scriptsize{$\mathcal{B}$}}

\put(6.2,4.3){\includegraphics[width=2.9\unitlength,height=2\unitlength]{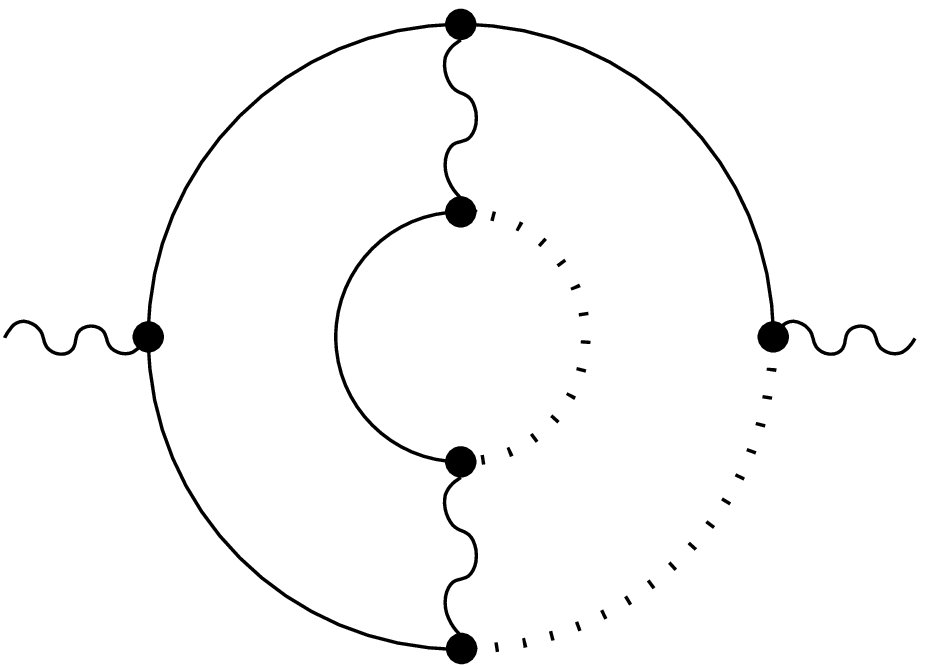}}
\put(8.7,4.2){$(6)$}\put(7.8,4.6){\scriptsize{$\bfDelta$}}
\put(7.8,5.8){\scriptsize{$\mathcal{B}$}}

\put(3.55,0.9){\includegraphics[width=2.9\unitlength]{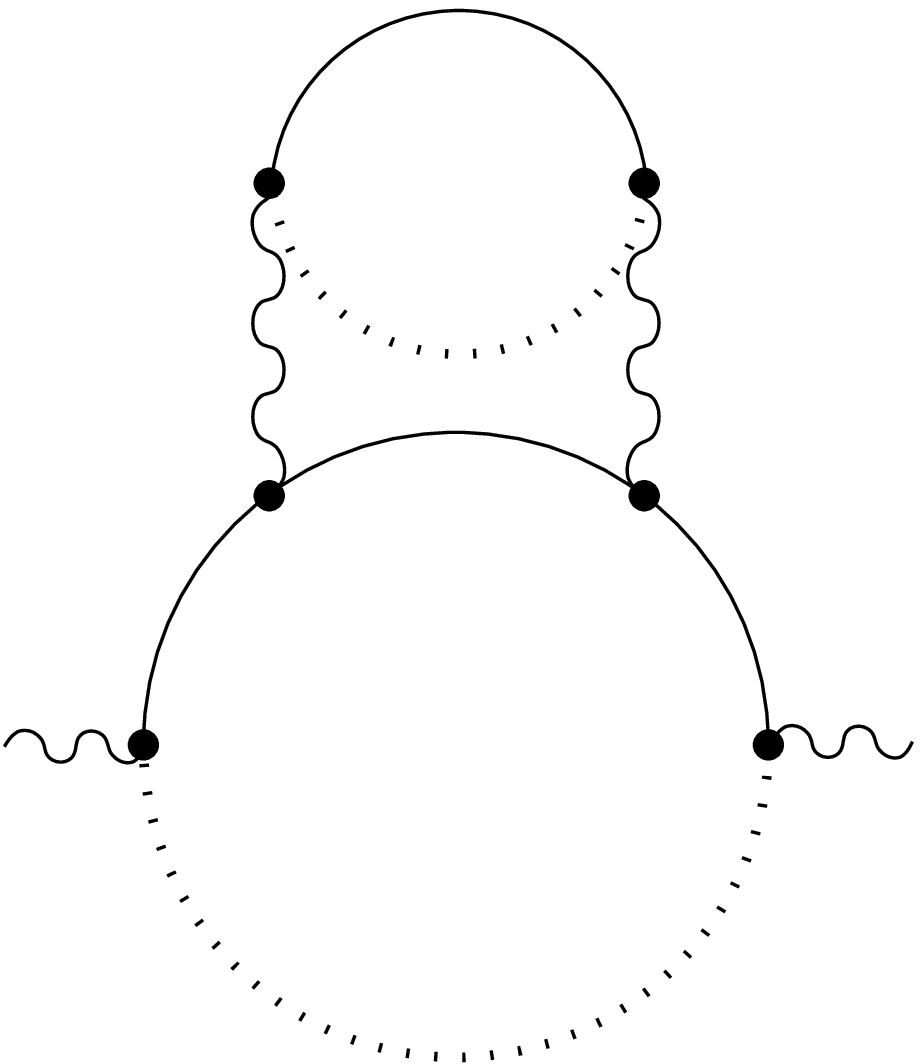}}
\put(6,0.8){$(7)$}

\put(3.95,3){\scriptsize{$\mathcal{B}$}}
\put(5.75,3){\scriptsize{$\mathcal{B}$}}

\end{picture}
\caption{These are the diagrams responsible for the non-analytic temperature
dependence of the spin susceptibility.}
\label{highdsus}
\end{figure}

The relevant diagrams leading to $T^{d-1}$ corrections are displayed in Fig.%
\ref{highdsus}. The solid lines carry the frequencies $\omega $
and the momenta $k$ of the order of $T$ and $T/v_{F}$,
respectively. They are smaller than characteristic energies in the
Green function entering the vertices because the latter are
responsible for the logarithmic contributions. This means that the
vertices can be taken at zero external frequencies and momenta and
this is the reason why one may just take the values of the
vertices from Eqs. (\ref{eq:5.38}-\ref{eq:5.40}). The same
procedure has been used in Ref. \cite{aleiner} for calculation of
the specific heat. Putting the bare values for the vertices
$\gamma $, $\beta $ and $\Delta $ would give the peturbative
results of Refs. \cite{coffey,
chubukov1,sarma,Catelani} in $d=2,3$. In this limit, the diagrams of Fig. %
\ref{highdsus} correspond to the conventional diagrams considered in those
works.

As concerns diagrams containing the amplitude $\gamma _{f}$ of the forward
scattering, we did not find any logarithmic contributions. This is because
one obtains integrals of products of Green functions containing poles in the
same half plane of complex variables $\mathbf{kn.}$

Considering the contributions of the diagrams in Fig. \ref{highdsus} and
comparing them with self-energy and vertex corrections in $1d$, Fig. \ref%
{fig:sus1d} one can see that there is a close analogy between the terms
responsible for the logarithmic corrections in one dimension and those
responsible for the non-analytic behavior in higher dimensions. Within our
formalism, the main difference between the two cases is the additional
angular integration in dimensions $d>1$.

Let us now turn to the computation of the diagrams in two and three
dimensions. Calculating the terms of the perturbation theory corresponding
to the diagrams displayed in Fig. \ref{highdsus} one finds that some of them
show unphysical divergencies in the limit of vanishing momenta and
frequencies. Therefore one should sum up certain diagrams first before
taking the limit.

To demonstrate this feature explicitly, let us consider the backscattering
contribution for the diagram $3$ of Fig. \ref{highdsus}. When evaluating the
term $\chi _{3}(\mathbf{q},\varepsilon )$ corresponding to this diagram one
finds terms containing the product
\be
\mathcal{G}_{\mathbf{n}_{1}}(\mathbf{q},\varepsilon )\mathcal{G}_{\mathbf{n}%
_{2}}(\mathbf{q},-\varepsilon )
\ee%
where $\mathbf{q}$ and $\varepsilon $ are the external momentum and
frequency.

The integral over the internal momenta is ultraviolet divergent and must be
cut with the help of the function $f,$ Eq. (\ref{eq:3.6}). At the same time,
the limits $\mathbf{q}\rightarrow 0$, $\varepsilon \rightarrow 0$ in the
diagram $3$, Fig. \ref{highdsus}, cannot be taken unambiguously.

In order to get rid of such unphysical divergencies, we note that this term
contains the product of the interaction amplitudes $\Delta _{3}^{--}\Gamma
_{3}$. A closer inspection reveals that this term is intimately related to
the renormalization of $\Delta $ and the ultraviolet divergence encountered
during the renormalization (see Fig. \ref{fig:intamplitudes}). It follows
from the results of the renormalization that this divergence must be
cancelled with the help of diagram $4$ using the relation $\Delta
_{3}^{--}\gamma _{3}=(\beta _{3}^{-})^{2}$.

The relation to the conventional perturbation theory, which is made obvious
by including the dotted lines into the diagrams, in fact strongly suggests
first to group several different diagrams before evaluating them. These are
the groups
\begin{eqnarray}
\chi _{a} &=&\chi _{1}+\chi _{3}+\chi _{4}+\chi _{5} \\
\chi _{b} &=&\chi _{2}+\chi _{6}+\chi _{7}
\end{eqnarray}%
Strictly speaking, diagram $6$ differs topologically from diagrams $2$ and $%
7 $. It nevertheless turns out to be advantageous to combine them, since the
expressions are similar at low energies that are considered in the model we
use.

For convenience of the reader, and since the algebra is rather tedious,
explicit expressions for the diagrams are included in Appendix \ref%
{app:diagrams} . Using the mutual relations between the seven interaction
amplitudes $\Delta _{3}^{\pm ,\pm }$, $\beta _{3}^{\pm }$ and $\gamma _{3}$,
one finds rather simple expressions for $\chi _{a}$ and $\chi _{b}$, that
allow to take easily the limit $\mathbf{q\rightarrow }0\mathbf{,}$ $%
\varepsilon \rightarrow 0$. After taking this limit they read

\begin{eqnarray}
\chi_a&=&-32\eta^2 T\sum_\omega\int d\mathbf{p} \int d\mathbf{n}_1d\mathbf{n}%
_2  \label{eq:7.4} \\
&&\frac{\omega^2}{(v_F\mathbf{n}_1\mathbf{p}-i\omega)^2(v_F\mathbf{n}_2%
\mathbf{p}+i\omega)^2}\;Y(\mathbf{p},\widehat{\mathbf{n}_1\mathbf{n}_2})
\nonumber \\
\chi_b&=&32\eta^2 T\sum_\omega\int d\mathbf{p} \int d\mathbf{n}_1d\mathbf{n}%
_2  \label{eq:7.5} \\
&&\frac{i\omega(v_F\mathbf{n}_2\mathbf{p}-i\omega)}{(v_F\mathbf{n}_1\mathbf{p%
}-i\omega)^3(v_F\mathbf{n}_2\mathbf{p}+i\omega)}Y(\mathbf{p},\widehat{%
\mathbf{n}_1\mathbf{n}_2})  \nonumber
\end{eqnarray}
where
\begin{eqnarray}
Y(\mathbf{p},\theta)=\left[\int d^d\mathbf{r}\mbox{e}^{-i\mathbf{p}\mathbf{r}%
}\frac{\gamma_b(\theta)}{1+\overline{f}_\perp\left(\frac{\mathbf{r}_\perp}{%
r_0}\right)X(\theta)}\overline{f}(\mathbf{r})\right]^2  \label{eq:7.6}
\end{eqnarray}
and
\begin{eqnarray}
X(\theta)=-\mu_d\gamma_b(\theta)\ln\left(\max\{\theta,T/\varepsilon_0\}%
\right)
\end{eqnarray}
The numerical coefficient $\mu_d$ was introduced in Eq.~(\ref{eq:5.29}). The
integration over $u_1$, $u_2$ was performed with the help of the following
relations
\begin{eqnarray}
&&\int du_1du_2 u_1^2u_2^2\Big(z_1z_2+2(z_1^3z_2+z_1z_2^3) \\
&&\qquad+(z_1^2z_2+z_1z_2^2)+2z_1^2z_2^2\Big)=\frac{1}{(1+x_1)(1+x_2)}
\nonumber \\
&&\int du_1du_2 u_1^3u_2 \Big(2z_1^2z_2^2+2(z_1^3z_2+z_1z_2^3) \\
&&\qquad+(z_1^2z_2+z_1z_2^2)\Big)=\frac{1}{(1+x_1)(1+x_2)}  \nonumber
\end{eqnarray}
where
\begin{eqnarray}
z_i=\frac{1}{1+u_1u_2x_i}
\end{eqnarray}

The non-analytic contribution to the spin susceptibility is found from the
small region of phase space, for which the angles $\mathbf{n}_{1}$ and $%
\mathbf{n}_{2}$ are close to each other, $|\mathbf{n}_{1}-\mathbf{n}_{2}|\ll
1$. We therefore introduce
\begin{eqnarray}
&&\mathbf{n}=(\mathbf{n}_{1}+\mathbf{n}_{2})/2,\quad \mathbf{m}=\mathbf{n}%
_{1}-\mathbf{n}_{2} \\
&&p_{\parallel }=\mathbf{p}\mathbf{n},\quad \mathbf{p}_{\perp }=\mathbf{p}%
-p_{\parallel }\mathbf{n}
\end{eqnarray}%
and perform the integration in $p_{\parallel }$ in Eqs.~(\ref{eq:7.4}) and (%
\ref{eq:7.5}).

As a result, we obtain the following formula for the non-analytic correction
to the spin-susceptibility in dimension $d>1$
\begin{eqnarray}
\delta \tilde{\chi}(T) &=&\chi _{a}(T)+\chi _{b}(T)  \label{eq:7.13} \\
&=&\frac{256}{v_{F}}\eta ^{2}T\sum_{\omega }|\omega |^{3}\int \frac{d^{d-1}%
\mathbf{p}_{\perp }}{(2\pi )^{d-1}}\int d\mathbf{n}_{1}d\mathbf{n}_{2}
\nonumber \\
&&\times \frac{3(v_{F}\mathbf{m}\mathbf{p}_{\perp })^{2}-4\omega ^{2}}{%
((v_{F}\mathbf{m}\mathbf{p}_{\perp })^{2}+4\omega ^{2})^{3}}\;Y(p_{\parallel
}\sim 0,\mathbf{p}_{\perp },|\mathbf{m}|)  \nonumber
\end{eqnarray}%
The main contribution to the integrals in Eq. (\ref{eq:7.4}, \ref{eq:7.5})
comes from $p_{\parallel }$ of the order of $T/v_{F}$ and this is why we can
set $p_{\parallel }\sim 0$ in the argument of the function $Y$.

Equation \ref{eq:7.13} contains a sum over bosonic Matsubara frequencies and
we write this sum symbolically as
\be
\delta \tilde{\chi}(T)=T\sum_{\omega _{n}}\mathbbm{R}(\omega _{n})\;.
\ee%
Technically it is more convenient to calculate the deviation from the
zero-temperature limit instead of computing the sum, i.e. to calculate the
quantity
\be
\delta \chi (T)=\delta \tilde{\chi}(T)-\delta \tilde{\chi}(T=0)\;.
\ee%
Using the Poisson formula the temperature dependent correction to the
susceptibility $\delta \chi (T)$ can be represented as
\begin{eqnarray}
\delta \chi (T) &=&\left( T\sum_{\omega }-\int \frac{d\omega }{2\pi }\right) %
\mathbbm{R}(\omega )  \label{eq:7.16} \\
&=&\sum_{l\neq 0}\int \frac{d\omega }{2\pi }\;\mathbbm{R}(\omega )\;\exp
\;\left( -i\frac{l\omega }{T}\right)  \label{eq:7.17}
\end{eqnarray}

The further evaluation is slightly different in dimensions $d=2$ and $d=3$
and we discuss the two cases separately.

\subsection{Non-analytic correction in two dimensions}

Rescaling the momentum and integrating over the angle $\mathbf{n}$ we obtain

\begin{eqnarray}
&&\delta\chi(T)=\frac{32}{v_F^2}\eta^2\;\left(T\sum_\omega-\int\frac{d\omega%
}{2\pi}\right)\;\int_0^1\frac{d|\mathbf{m}|}{2\pi}\frac{1}{|\mathbf{m}|}%
\qquad \\
&&\qquad\times\int\frac{dk}{2\pi}\;\frac{3k^2-1}{(k^2+1)^3}%
\;Y\left(p_\parallel=0,p_\perp=\frac{2|\omega|k}{v_F|\mathbf{m}|},|\mathbf{m}%
|\right)  \nonumber
\end{eqnarray}

The integral over $|\mathbf{m}|$ is logarithmic and therefore not very
sensitive to the upper limit that can safely be set to $1$. One notices that
the momentum dependence of $Y$ is crucial here coupling the integrals in $|%
\mathbf{m}|$ and $k$. If $Y$ were independent of the momentum, the $k$
integral would be equal to zero, whereas, at the same time, the integral
over $\mathbf{m}$ would diverge at the lower limit.

Fortunately, this uncertainty can easily be avoided taking into the momentum
dependence of the function $Y$.

After introducing the Fourier transform of $Y$
\be
\bar{Y}(|r|;\theta )=\int \frac{dp_{\perp }}{2\pi }\;\mbox{e}^{ip_{\perp
}r}\;Y(p_{\perp },p_{\parallel }=0,\theta )
\ee%
the momentum integration can be performed with the help of the identity
\be
\int_{-\infty }^{\infty }\frac{dk}{(2\pi )}\;\frac{3k^{2}-1}{(1+k^{2})^{3}}%
\mbox{e}^{-ikb}=-\frac{1}{4}|b|^{2}\mbox{e}^{-|b|}.\label{eq:7.20}
\ee
where $b=2r|\omega |/(v_{F}|\mathbf{m}|)$.

Then, we use Eq.~(\ref{eq:7.17}) and obtain the following expression
\begin{eqnarray}
\left( T\sum_{\omega }-\int \frac{d\omega }{2\pi }\right) \beta ^{2}\omega
^{2}\mbox{e}^{-\beta |\omega |} &=&T\left( 2x^{2}\frac{\coth x}{\sinh ^{2}x}-%
\frac{2}{x}\right) ,  \nonumber  \label{eq:7.21} \\
&&
\end{eqnarray}%
where $x=2\pi T|r|/(v_{F}|\mathbf{m}|)=\pi T\beta $, and change the
integration variable from $|\mathbf{m}|$ to $x$. As a result, we find
\begin{eqnarray}
\delta \chi (T) &=&-\frac{4T}{\pi v_{F}^{2}}\eta ^{2}\int_{-\infty }^{\infty
}dr\mathcal{I}_{2}(a)\bar{Y}\left( |r|,\theta =\frac{2\pi T}{\varepsilon
_{\infty }}\frac{r}{r_{0}}\frac{1}{x}\right) ,  \nonumber \\
&&
\end{eqnarray}%
where
\be
\mathcal{I}_{2}(a)=\int_{a}^{\infty }dx\left( \frac{2x\coth x}{\sinh ^{2}x}-%
\frac{2}{x^{2}}\right)
\ee%
and $a=\frac{2\pi T}{\varepsilon _{\infty }}\frac{r}{r_{0}}\frac{1}{\Delta
\phi }$.

We reintroduced formally an upper cut-off $\Delta \phi $ for the integration
over $|\mathbf{m}|$ but it will drop out from the final result. One can see
that the essential $r$ as controlled by the function $Y$ (and thus $f$) are
small, $r\lesssim r_{0}$, while essential $x$ in the integral are large, $%
x>1 $. This means that the main contribution in the integral over the angles
$\theta $ comes from $\theta $ of the order of $T/\varepsilon _{\infty }$.
Therefore we can with logarithmic accuracy set $\theta =0$ in the argument
of $\bar{Y}$. In turn, it means that the integral over $x$ is rather
insensitive to the lower bound as long as $x<1$ and we may safely extend the
integration range in $x$ to the interval $(0,\infty )$. Then, the
integrations over $x$ and $r$ can be easily performed. To this end we note
that $\mathcal{I}_{2}(0)=-1$ and introduce the notation
\be
Y(\theta )=Y(\theta ,\mathbf{p}=0)
\ee%
to formulate our result for the susceptibility $\delta \chi $,
\be
\delta \chi ^{2d}(T)=2\eta ^{2}\frac{T}{\varepsilon _{F}}\chi
_{0}^{2d}\;Y(\theta =0)\label{eq:7.25}
\ee
In Eq. (\ref{eq:7.25}), $\chi _{0}^{2d}=m/\pi $. The vertex part $\eta $,
Eq. (\ref{eq:4.48}), should be understood as $\eta _{\alpha =1}$ and it is a
result of an additional summation of ladder diagrams including $\mathcal{S}%
_{2}$, in close analogy to the discussion in Sec.~\ref{subsec:bare}. The
limit $\theta =0$ corresponds to the backward scattering. Before further
discussing this result in Sec.~\ref{subsec:final} we turn to the three
dimensional case.

\subsection{Non-analytic correction in three dimensions}

In $3d$ one obtains from Eq.~(\ref{eq:7.13}) after rescaling of momenta and
integration over $|\mathbf{n}|$ the following expression
\begin{eqnarray}
\delta \chi (T) =\frac{32\pi }{v_{F}^{3}}\eta ^{2}\;\left(
T\sum_{\omega
}-\int \frac{d\omega }{2\pi }\right) |\omega |\;\int_{0}^{1}\frac{d|\mathbf{m%
}|}{2\pi }\frac{1}{|\mathbf{m}|}  \nonumber \\
\int \frac{d^{2}k}{(2\pi )^{2}}\;\frac{3(\mbox{e}_{\mathbf{m}}\mathbf{k}%
)^{2}-1}{((\mbox{e}_{\mathbf{m}}\mathbf{k})^{2}+1)^{3}}\;Y\Big(p_{\parallel
} =0,\mathbf{p}_{\perp }=\frac{2|\omega |\mathbf{k}}{v_{F}|\mathbf{m}|},|%
\mathbf{m}|\Big)  \nonumber \\
&&  \label{eq:7.26}
\end{eqnarray}%
where $\mbox{e}_{\mathbf{m}}=\mathbf{m}/|\mathbf{m}|$.

We see that the integral over $\mathbf{m}$ in Eq. (\ref{eq:7.26}) is
logarithmic. However, corrections of the form $\delta \chi \propto \gamma
^{2}T^{2}\ln T$ are absent and this is due to the fact that the integral
over $k$ vanishes provided the momentum dependence of $Y$ is neglected.
Nevertheless, if the function $Y$ depends on the momentum $\mathbf{p}_{\perp
}$ the entire integral is finite.

After introducing the Fourier transform of $Y$ as
\be
\bar{Y}(|\mathbf{r}|,\theta )=\int \frac{d^{2}p}{(2\pi )^{2}}\mbox{e}^{i%
\mathbf{p}_{\perp }\mathbf{r}}\;Y(p_{\parallel }=0,\mathbf{p}_{\perp
},\theta )
\ee%
it is convenient to decompose the vectors $\mathbf{k},\mathbf{p}_{\perp },%
\mathbf{r}$ into components parallel and perpendicular to $\mbox{e}_{\mathbf{%
m}}$, such that $\mathbf{r}=(\tilde{r}_{\parallel },\tilde{r}_{\perp }).$
Then, one can then proceed in close analogy to the calculation in $2d$.
Using Eq.~(\ref{eq:7.20}) and Eq.~(\ref{eq:7.21}) one arrives at
\begin{eqnarray}
\delta \chi (T) &=&-\frac{4T^{2}}{v_{F}^{3}}\eta ^{2}\int d^{2}r\mathcal{I}%
_{3}(a)\bar{Y}\Big(|\mathbf{r}|,\theta =\frac{2\pi T}{\varepsilon _{\infty }}%
\frac{|\mathbf{r}|}{r_{0}}\frac{1}{x}\Big),\no  \\
&&  \label{eq:7.28}
\end{eqnarray}%
where
\be
\mathcal{I}_{3}(a)=\int_{a}^{\infty }dx\left( 2\frac{\coth x}{\sinh ^{2}x}-%
\frac{2}{x^{3}}\right)
\ee%
and $a=\frac{2\pi T}{\varepsilon _{\infty }}\frac{|\mathbf{r}|}{r_{0}}\frac{1%
}{\Delta \phi }$. Again, we reintroduced the upper cut-off $\Delta \phi $
for the $\mathbf{m}$ integration.

The integral over $x$ in Eq. (\ref{eq:7.28}) shows a somewhat stronger
dependence on the lower cutoff $a$ as compared to the two-dimensional case
with $\mathcal{I}_{3}(0)=-1/3$, $\mathcal{I}_{3}(1)\sim -0.28$. This means
that angles larger than $T/\varepsilon _{\infty }$ start contributing more
significantly. Still, the dominant contribution to the integral comes from $%
x>1$, so that we can set $\theta =0$ in the argument of $\bar{Y}$ with
logarithmic accuracy. Then, we come to the following result for the
temperature dependent correction to the spin susceptibility
\be
\delta \chi ^{3d}(T)=\frac{\pi ^{2}}{3}\eta ^{2}\frac{T^{2}}{\varepsilon
_{F}^{2}}\;\chi _{0}^{3d}\;Y(\theta =0)
\ee%
where $\chi _{0}^{3d}=mp_{F}/\pi ^{2}$ and $Y(\theta )=Y(\mathbf{p}=0,\theta
)$ and $\eta =\eta _{\alpha =1}$, see the remarks below Eq.~(\ref{eq:7.25}).

We see that, as in the $2d$ case, this correction is determined completely
by the backward scattering ($\theta =0$).

\subsection{\label{subsec:final} Final results in $d=2,3$}

\subsubsection{General results}
The quantity $Y(\theta =0)$ can be considered as the square of an effective
temperature dependent backward scattering amplitude $\gamma _{b}(T)$, and we
write it in the form $Y(\theta =0)=\gamma _{b}^{2}(T)$, where
\begin{equation}
\gamma _{b}(T)=\gamma _{b}\;\int \frac{d^{d-1}\mathbf{r}_{\perp }}{%
r_{0}^{d-1}}\frac{\overline{f}_{\perp }\left( \frac{\mathbf{r}_{\perp }}{%
r_{0}}\right) }{1+\overline{f}_{\perp }\left( \frac{\mathbf{r}_{\perp }}{%
r_{0}}\right) X(T)}\;,  \label{b0}
\end{equation}%
$\gamma _{b}=\gamma _{b}\left( \theta =0\right) $, $X(T)=\mu _{d}\gamma
_{b}\ln (\varepsilon _{\infty }/T)$ and $\overline{f}_{\perp }(\mathbf{r}%
_{\perp }/r_{0})$ was defined in Eq.~(\ref{eq:5.30}).

So, our results can be written in the most general form as
\begin{eqnarray}
&&\delta \chi ^{2d}(T)=2\eta ^{2}\gamma _{b}^{2}(T)\frac{T}{\varepsilon _{F}}%
\chi _{0}^{2d}\;  \label{eq:7.32} \\
&&\delta \chi ^{3d}(T)=\frac{\pi ^{2}}{3}\eta ^{2}\gamma _{b}^{2}(T)\frac{%
T^{2}}{\varepsilon _{F}^{2}}\;\chi _{0}^{3d}  \label{eq:7.33}
\end{eqnarray}%
and we remind the reader that $\eta =\eta _{\alpha =1}$, where $\eta $ is
determined by Eq. (\ref{eq:4.48}).

If we replaced $\gamma _{b}(T)$ in Eq. (\ref{eq:7.32}) by the bare coupling
constant $\gamma _{b}$ for $d=2$, we would obtain the previously reported
linear $T$ dependence of the non-analytic corrections\cite{coffey,
chubukov1,sarma,Catelani}. This replacement means neglecting the
renormalization of the interaction constants discussed in Sec.\ref%
{subsec:renintamplitudes}. If we set the function $\gamma _{b}(T)$
equal to the bare value $\gamma _{b}$ in $d=3$ we would obtain the
correction $\delta \chi ^{3d}(T)$ proportional to $T^{2},$ which
is regular in $T^{2}$. This
means that the first non-analytical $T^{2}\ln \left( \varepsilon _{\infty }/T\right)$ term in $3d$ is of the order $%
\gamma ^{3}$.

In the limit of small $X\left(T\right)\ll 1$ the temperature
dependence of $\gamma^2_b(T)$ takes the form
\be
\gamma^2_b(T){\sim}\gamma^2_b-2\gamma^3_bc_d\ln\frac{\eps_\infty}{T}\label{eq:gammasq},
\quad X(T)\ll 1,
\ee
where
\be
c_d=\mu_d\int
\frac{d^{d-1}\bfr_\perp}{r_0^{d-1}}\;\overline{f}^2_\perp\left(\frac{\bfr_\perp}{r_0}\right)
\ee
The factor $c_d$ depends on the precise form of the cut-off and
can be estimated only. It is roughly of the order of unity. Eq.
(\ref{eq:gammasq}) shows that the first logarithmic in temperature
corrections contain the prefactor $\gamma _{b}^{3}$ both in $2$
and $3$ dimensions. This rather high order in the coupling
constant $\gamma _{b}$ is, apparently, the reason why the
logarithmic corrections to the susceptibility have not been
noticed previously in the diagrammatic
expansions \cite{coffey, chubukov1,sarma,Catelani} (see, however, \cite%
{shekhter} for $2d$).

In the limit of large $X\left(T\right)\gg 1$ one finds the
following asymptotic temperature dependence of $\gamma^2_b(T)$
\be
\gamma^2_b(T){\propto}
\left(\ln\frac{\eps_\infty}{T}\right)^{-2},\quad X(T)\gg
1.\label{eq:add1}
\ee
More explicit formulae can only be written using a model cut-off
function and this will be done in the next section.

\subsubsection{Results for a model cut-off function}
We choose the following model cut-off function $\bar{f}_{\perp
}\left( r_{\perp }/r_{0}\right) $:
\be
\bar{f}_{\perp }\left( \frac{r_{\perp }}{r_{0}}\right)
=\frac{1}{\Omega _{d-1}}\exp \left( -\frac{r_{\perp
}}{r_{0}}\right)\label{eq:7.31},
\ee
where $\Omega _{d-1}$ is the $d-1$ dimensional solid angle.

Performing the remaining integration for this case one obtains the following
temperature dependence for the effective backward scattering constants
\begin{eqnarray}
&&\gamma _{b}^{d=2}(T)=\frac{2\gamma _{b}\ln \left[ 1+X(T)/2\right] }{X(T)}
\label{eq:7.35} \\
&&\gamma _{b}^{d=3}(T)=-\frac{2\pi \gamma _{b}\;{\mathrm{L}i}_{2}\;\left[ -{X%
}(T)/{2\pi }\right] }{{X}(T)}  \label{eq:7.36}
\end{eqnarray}%
where ${\mathrm{L}i}_{2}(x)=\sum_{k=1}^{\infty }x^{k}/k^{2}$ is the
polylogarithm function.

In the limit of small $X\left( T\right) \ll 1$ the temperature
dependence of the susceptibility computed with the model cut-off
function takes the form
\begin{eqnarray}
\delta \chi ^{2d}(T) &=&2\eta^2\gamma_b^2\frac{T}{\eps_F}\chi_0^{2d}\left(1-2\gamma_b\ln\frac{\eps_\infty}{T}\right) \label{b1} \\
\delta \chi ^{3d}\left( T\right) &=& \frac{\pi ^{2}}{3}\eta
^{2}\gamma _{b}^{2}\frac{T^{2}}{\varepsilon _{F}^{2}}\;\chi
_{0}^{3d}\left(1-\gamma_b\ln\frac{\eps_\infty}{T}\right)
\label{b2}
\end{eqnarray}
where we put $r_0^{-1}\sim p_F$ for simplicity. It should be
stressed once again that the coefficient of the logarithmic
correction cannot be determined rigorously within our model.

In the opposite limit of very low temperatures, $X\left( T\right)
\gg 1,$ asymptotic expressions for the corrections to the
susceptibility can be written using the model cut-off function of
Eq. (\ref{eq:7.31})
\be
\delta \chi ^{2d}\left(T\right) &=&\frac{1}{2}\eta^2\frac{T}{\eps_F}\chi_0^{2d}\frac{\ln^2\left(4\gamma_b\ln\frac{\eps_\infty}{T}\right)}{\ln^2\frac{\eps_\infty}{T}} \\
\delta \chi ^{3d}\left( T\right) &=& \frac{\pi ^{2}}{48}\eta
^{2}\frac{T^{2}}{\varepsilon _{F}^{2}}\;\chi
_{0}^{3d}\frac{\ln^4\left(4\pi\gamma_b\ln\frac{\eps_\infty}{T}\right)}{\ln^2\frac{\eps_\infty}{T}}.
\ee
Again we used $r_0^{-1}\sim p_F$ for simplicity. The asymptotic
behavior $1/\left(\ln(\eps_\infty/T)\right)^2$ in these equations
is not very sensitive to the form of the function $f\left(
\mathbf{k}\right) $ as can be seen from Eq.~(\ref{b0}).

\section{\label{sec:discussion}Discussion}

We have calculated non-analytical logarithmic in temperature contributions
to the spin susceptibility of a $d$-dimensional electron gas for $d=1,2,3$.
We used the bosonization method recently developed in Ref. \cite{aleiner}
and demonstrated that it can give results not only for the specific heat as
in Ref. \cite{aleiner} but also for the spin susceptibility.

The main contribution to this quantity comes from effective spin modes that
interact with each other, which leads to the non-analytic logarithmic
contributions. Although we consider isotropic systems, the low temperature
behavior is determined by spin excitations moving antiparallel to each
other. As a result, the non-analytic contributions are determined by the
backward scattering showing that there are one dimensional features also in
the dimensions $d=2,3$.

The final form of the temperature corrections to the susceptibility in two
and three dimensions is given by Eqs. (\ref{b0}-\ref{eq:7.33}). Although in $%
2d$ the correction to the susceptibility $\chi $ is very similar to the
correction to the quantity $C\left( T\right) /T$, where $C\left( T\right) $
is the specific heat\cite{aleiner}, they are quite different in $3d$. The
first logarithmic contribution to $C\left( T\right) /T$ is of the order $%
\gamma _{b}^{2},$ which is a well known result for $3d$ (Refs.
\onlinecite{eliash,doniach,brink,amit,chubukov4}). At the same
time, the expansion of the susceptibility in the logarithms starts
with the term of the order of $\gamma _{b}^{3}$, which shows that
the non-analytical in temperature corrections exist for this
quantity in three dimensions, too.

Using the bosonization scheme of Ref. \cite{aleiner} we have also reproduced
the temperature dependent correction in one dimension, Eq. (\ref{eq:6.9}),
that has been obtained long ago\cite{dl} using a renormalization group
approach for the initial electron model.

The temperature dependent correction to the susceptibility in 2$d$
was calculated recently by Shekhter and Finkelstein
\cite{shekhter} using direct diagrammatic expansions for the
initial electron model. In the approach of Ref. \cite{shekhter},
which was tailored for the calculation of the spin susceptibility
in $d=2$, the renormalization of the effective backward scattering
amplitude is attributed to all Cooper channel harmonics, while no
cut-off function was used. In the formalism of
Ref.~\onlinecite{aleiner} which we studied no decoupling in the
Cooper channel is introduced in addition to the particle-hole
channel in order to avoid overcounting in the region of phase
space close to backward scattering, which turned out to be most
important (For a more detailed discussion of the role of the
Cooper channel in the bosonization approach see Sec. VII C of Ref.
\onlinecite{aleiner}). In fact, the renormalization of the
backward scattering amplitude is obtained in this way as well,
non-zero angular harmonics are however not included.

It is important to mention that in some cases not all
non-analytical corrections are accounted for by the backward
scattering. Interesting contributions of the type $T^3\ln T$ to,
e.g. specific heat in three dimensions, are given by three-loop
diagrams in the language of the electronic Green functions (Ref.
\onlinecite{chubukov4}) and they cannot be expressed in terms of
the backward or forward scattering. However, these corrections are
proportional to higher powers of the interaction constant and are
smaller than those given by the backward scattering unless the
temperature is very low. In the latter regime the effective
backscattering amplitude is very small, Eq. (\ref{eq:add1}), and
the contribution of three and more loop diagrams can become the
most important one. Contributions that are not reduced to the
backward scattering was discussed in Ref. \onlinecite{shekhter}
for the spin susceptibility in two dimensions.

\begin{acknowledgments}
We acknowledge stimulating discussions with I.L. Aleiner. We thank A.
Shekhter and A. M. Finkel'stein for bringing to our attention their results%
\cite{shekhter} prior to publication. G.S. is thankful to the
Feinberg Graduate School, to the grant \textit{Transregio 12
\textquotedblleft Symmetries and Universality in Mesoscopic
Systems\textquotedblright\ } and GRK 384 of the DFG for financial
support. The work was completed during the visit of K.B.E. to the
Weizmann Institute who acknowledges a support of the EU
Transnational Access program and the hospitality of the Institute.
\end{acknowledgments}

\newpage \appendix

\section{Derivation of Formula Eq.~(\ref{eq:4.28}) for $\mathcal{Z}%
_s[\mathbf{h}]$ by explicit construction}

\label{app:derivation} In this appendix we explicitly construct the
supersymmetric representation of $\mathcal{Z}_s[\mathbf{h}]$ in Eq.~(\ref%
{eq:4.28}).

It is a straightforward application of the results of Ref.~%
\onlinecite{hermit} that $\mathcal{Z}_s[\mathbf{h}]$ of Eq.~(\ref{eq:4.1})
in the main text can be rewritten as
\begin{eqnarray}
\mathcal{Z}_s[\mathbf{h}]=\exp\left(2\nu\int_{\hat{X},\hat{X}^{\prime }}u%
\mathbf{h}_\mathbf{n}(x)\;L^{-1}_{X,X^{\prime }}\;\partial_{\tau^{\prime }}%
\mathbf{h}_\mathbf{n}^{\prime }(x^{\prime })\right),  \label{eq:a1}
\end{eqnarray}
where the form of $L^{-1}_{X,X^{\prime }}$ will be specified in the
following. Here $X=(\mathbf{r},\tau,\mathbf{n},u)$ and the hat in $\int_{%
\hat{X}}$ indicates that integration $\int_{\hat{n}}$ in this formula is
over the full solid angle.
\begin{eqnarray}
\left({L}^{-1}_{X,X^{\prime }}\right)_{\alpha,\beta}&=&-\frac{i}{2}\Big[%
\left<\mathbf{S}^2_{\alpha,X}\mathbf{S}^{1*}_{\beta,X^{\prime
}}\right>+\left<\mathbf{S}^1_{\alpha,X} \mathbf{S}^{1*}_{\beta,X^{\prime
}}\right>  \nonumber \\
&& -\left<\mathbf{S}^2_{\alpha,X}\mathbf{S}^{2*}_{\beta,X^{\prime
}}\right>-\left<\mathbf{S}^1_{\alpha,X}\mathbf{S}^{2*}_{\beta,X^{\prime
}}\right>\Big]  \label{eq:a2}
\end{eqnarray}
Here $\alpha, \beta$ are spin indices and averaging is defined as
\begin{eqnarray}
\left\langle\dots\right\rangle=\int \mathcal{D}({\bm \varphi},\bar{{\bm %
\varphi}})\left(\dots\right)\mbox{e}^{-\mathcal{L}[\overline{{\bm \varphi}},{%
\bm \varphi}]}  \label{eq:a3}
\end{eqnarray}
where
\begin{eqnarray}
\mathcal{L}[\bar{{\bm \varphi}},{\bm \varphi}]&=&-i\int_{\hat{X}} \overline{{%
\bm \varphi}}_X \Lambda \left[\hat{M}+i\delta\right] {\bm \varphi}_X
\label{eq:a4}
\end{eqnarray}
and supervector ${\bm \varphi}$ has been defined in Eqs.~(\ref{eq:4.14}), (\ref%
{eq:4.15}).
\begin{eqnarray}
\overline{{\bm \varphi}}={\bm \varphi}^\dagger \Lambda,\qquad
\Lambda=\sigma_3^{(H)}
\end{eqnarray}
The fermionic part of ${\bm \varphi}$, $\overline{{\bm \varphi}}$ takes care
of the normalization via identity
\begin{eqnarray}
\int \mathcal{D}({\bm \chi},\overline{{\bm \chi}})\mbox{e}^{-\mathcal{L}[%
\overline{{\bm \chi}},{\bm \chi}]}=\left[\int \mathcal{D}(\mathbf{S},%
\overline{\mathbf{S}})\mbox{e}^{-\mathcal{L}[\overline{\mathbf{S}},\mathbf{S}%
]}\right]^{-1}
\end{eqnarray}
Finally
\begin{eqnarray}
\Lambda\hat{M}_\mathbf{n}&=&\left(%
\begin{array}{cc}
\hat{L}^{\prime }_\mathbf{n} & i\hat{L}^{\prime \prime }_\mathbf{n} \\
-i\hat{L}^{\prime \prime }_\mathbf{n} & -\hat{L}^{\prime }_\mathbf{n}%
\end{array}%
\right)_H
\end{eqnarray}
We repeat that $\hat{L}^{\prime }=(\hat{L}+\hat{L}^\dagger)/2$ and $\hat{L}%
^{\prime \prime }=-i(\hat{L}-\hat{L}^\dagger)/2$ are hermitian. The explicit
form of $\Lambda\hat{M}$ is
\begin{eqnarray}
\Lambda\hat{M}_\mathbf{n}=iv_0\mathbf{n}\nabla-\Lambda_1\partial_\tau+2iu%
\hat{h}_\mathbf{n}
\end{eqnarray}
The matrix $\Lambda_1$ acts in $H$ space and is written in Eq.~(\ref{eq:4.24}%
). Restricting the angular integration to just one half sphere we can cast
formula Eq.~(\ref{eq:a4}) in a new form by introducing supervector ${\bm \phi%
}$ as
\begin{eqnarray}
{\bm \phi}(\mathbf{n})=\left(%
\begin{array}{cc}
{\bm \varphi}(\mathbf{n}) &  \\
{\bm \varphi}(-\mathbf{n}) &
\end{array}%
\right)_n,\quad\overline{{\bm \phi}(\mathbf{n})}={\bm \phi}^\dagger(\mathbf{n%
})\Lambda
\end{eqnarray}
$\mathcal{L}$ has to be modified accordingly.
\begin{eqnarray}
\mathcal{L}\rightarrow \mathcal{L}[\overline{{\bm \phi}},{\bm \phi}%
]&=&-i\int_{X} \overline{{\bm \phi}}_X \Lambda \left[\hat{\mathbbm{M}}%
+i\delta\right] {\bm \phi}_X
\end{eqnarray}
where
\begin{eqnarray}
\hat{\mathbbm{M}}_\mathbf{n}=\left(%
\begin{array}{cc}
\hat{M}_\mathbf{n} & 0 \\
0 & \hat{M}_{-\mathbf{n}}%
\end{array}%
\right)
\end{eqnarray}
The explicit form of $\Lambda \hat{\mathbbm{{M}}}$ is
\begin{eqnarray}
\Lambda\hat{\mathbbm{M}}_\mathbf{n}=iv_0\mathbf{n}\Sigma_3\nabla-\Lambda_1%
\partial_\tau+2iu\hat{\mathbbm{H}}_\mathbf{n}
\end{eqnarray}
Here we introduced $\Sigma_3=\sigma_3^{(n)}$ and
$\mathbbm{H}_\mathbf{n}(x)$ of Eq.~(\ref{eq:4.25}).

Finally the number of field components in ${\bm \phi}$, $\bar{{\bm \phi}}$
is doubled once more by introducing the electron-hole ($eh$) sector. This
can be done by introducing new vector
\begin{eqnarray}
{\bm \psi}=\frac{1}{\sqrt{2}}\left(%
\begin{array}{cc}
{\bm \phi}^* \\
{\bm \phi}
\end{array}
\right)_{eh},\qquad \overline{{\bm \psi}}={\bm
\psi}^\dagger\Lambda
\end{eqnarray}
Now
\begin{eqnarray}
\mathcal{L}\rightarrow\mathcal{L}[\overline{{\bm \psi}},{\bm \psi}%
]&=&-i\int_{X} \overline{{\bm \psi}}_X \Lambda \left[\hat{\mathcal{M}}%
+i\delta\right] {\bm \phi}_X
\end{eqnarray}
where
\begin{eqnarray}
\hat{\mathcal{M}}_\mathbf{n}&=&\left(%
\begin{array}{cc}
\hat{\mathbbm{M}}_\mathbf{n} & 0 \\
0 & \hat{\mathbbm{M}}^T_\mathbf{n}%
\end{array}%
\right)_{TR}
\end{eqnarray}
Note that the transposition for $\hat{\mathbbm{M}}^T$ includes derivatives.
Here matrix $\tau_3=\sigma_3^{(eh)}$ acts in $eh$ space. The explicit form
is
\begin{eqnarray}
\Lambda\hat{\mathcal{M}}_{\mathbf{n},u}=-iv_0\tau_3\Sigma_3\mathbf{n}%
\nabla-\Lambda_1\partial_\tau-2i\tau_3\hat{\mathbbm{H}}_{\mathbf{n}}
\end{eqnarray}
Now we can write an appropriate generalization of Eq.~(\ref{eq:a1}), Eq.~(%
\ref{eq:a3}) and make contact to formula Eq.~(\ref{eq:4.28}) in the main
text. We write
\begin{eqnarray}
&&\mathcal{L}[\overline{{\bm \psi}},{\bm \psi}]=-i2\nu\int_X \overline{{\bm %
\psi}}_X\left(\mathcal{H}+i\delta\Lambda\right){\bm \psi}_X
\end{eqnarray}
where $\mathcal{H}=\Lambda\hat{\mathcal{M}}$ (factor of $2\nu$ is introduced
for convenience) and the averaging with respect to this Lagrangian is
defined as
\begin{eqnarray}
\left\langle\dots\right\rangle=\int \mathcal{D}({\bm \psi},\overline{{\bm %
\psi}})\left(\dots\right)\mbox{e}^{-\mathcal{L}[\overline{{\bm \psi}},{\bm %
\psi}]}
\end{eqnarray}
Using
\begin{eqnarray}
\mathcal{F}_\mathbf{h}(X)=\partial_X\mathbbm{H}_\mathbf{n}(x)\mathcal{F}%
_0,\quad \overline{\mathcal{F}_\mathbf{h}(X)}=\left(C\mathcal{F}_\mathbf{h}%
(X)\right)^T
\end{eqnarray}
where $\mathcal{F}_0$ is defined in Eq.~(\ref{eq:4.25}) and
\begin{eqnarray}
&&\partial_X(\alpha)=\left(%
\begin{array}{cc}
1 & 0 \\
0 & u[\alpha\partial_\tau+(1-\alpha)iv_0\mathbf{n}\nabla\Sigma_3]%
\end{array}%
\right)
\end{eqnarray}
one verifies that
\begin{eqnarray}
&&\overline{{\bm \psi}}\mathcal{F}_\mathbf{h}=\overline{\mathcal{F}_\mathbf{h%
}}{\bm \psi}=\frac{1}{2}\times \\
&&\left(\hat{\mathcal{O}}_\mathbf{n}(\alpha)\mathbf{h}_\mathbf{n}\left(%
\mathbf{S}_\mathbf{n}^{1*}-\mathbf{S}_\mathbf{n}^{2*}\right)+\mathbf{h}_%
\mathbf{n}\left(\mathbf{S}_\mathbf{n}^1+\mathbf{S}_\mathbf{n}^2\right)+\left(%
\mathbf{n}\leftrightarrow -\mathbf{n}\right)\right)  \nonumber
\end{eqnarray}
where
\begin{eqnarray}
\hat{\mathcal{O}}_\mathbf{n}(\alpha)=u(\alpha\partial_\tau+(1-\alpha)iv_0%
\mathbf{n}\nabla)
\end{eqnarray}
Using $\left\langle S^i_{\alpha,X} S^{j*}_{\beta,X^{\prime
}}\right\rangle\propto \delta_{\mathbf{n},\mathbf{n}^{\prime
}}\delta_{u,u^{\prime }}$ one obtains
\begin{eqnarray}
\mathcal{I}[\mathbf{h}]=\int_{XX^{\prime }}\overline{\mathcal{F}_\mathbf{h}%
(X)}\left\langle{\bm \psi}_X\overline{{\bm \psi}_{X^{\prime }}}\right\rangle%
\mathcal{F}_\mathbf{h}(X^{\prime })=\frac{1}{2}\int_{XX^{\prime }}h_{\mathbf{%
n}}^\gamma(x)  \nonumber \\
\left\langle\left(S_{\gamma,X}^1+S_{\gamma,X}^2\right)\left(S_{%
\beta,X}^{1*}-S_{\beta,X}^{2*}\right)\right\rangle\hat{\mathcal{O}}_\mathbf{n%
}(\alpha)h_\mathbf{n}^\beta(x^{\prime })  \nonumber \\
\end{eqnarray}
Summation over spin indices $\gamma,\beta$ is implied. Using further Eq.~(%
\ref{eq:a2}) one finds
\begin{eqnarray}
\mathcal{I}[\mathbf{h}]&=&\frac{i}{2\nu}\int du dx (d\mathbf{n})\; \Big[%
\mathbf{h}_\mathbf{n}(x)\left(u\hat{L}^{-1}_{\mathbf{n},u}\left(\partial_\tau%
\mathbf{h}_\mathbf{n}\right)(x)\right)  \nonumber \\
&+&\mathbf{h}_\mathbf{n}(x)\left(u(1-\alpha)\hat{L}^{-1}_{\mathbf{n},u}(iv_0%
\mathbf{n}\nabla-\partial_\tau)\mathbf{h}_\mathbf{n}(x)\right)\Big]
\nonumber \\
&&
\end{eqnarray}
The last line can be simplified by noticing
\begin{eqnarray}
\hat{L}^{-1}_{\mathbf{n},u}(iv_0\mathbf{n}\nabla-\partial_\tau)\mathbf{h}_%
\mathbf{n}(x)=\mathbf{h}_\mathbf{n}(x)
\end{eqnarray}
This equality holds, since $\hat{h}\mathbf{h}=\mathbf{h}\times\mathbf{h}=0$.
The result is
\begin{eqnarray}
\mathcal{Z}_s[\mathbf{h}]&=&\exp\left(-4i\nu^2 \int_{XX^{\prime }}\;%
\overline{F_\mathbf{h}(X)}\left\langle{\bm \psi}_{X}\overline{{\bm \psi}%
_{X^{\prime }}}\right\rangle F_\mathbf{h}(X^{\prime })\right)  \nonumber \\
&&\times\exp\left(-\nu(1-\alpha)\int_{\hat{\mathbf{n}}, x}\mathbf{h}_\mathbf{%
n}^2(x)\right)
\end{eqnarray}
This formula is used in the main text, Eq.~(\ref{eq:4.28}).

\section{Diagrams of Fig.~\ref{highdsus}}

\label{app:diagrams} In this appendix we give analytic expressions for the
diagrams displayed in Fig.~\ref{highdsus}. We introduce
\begin{eqnarray}
&&{\bm \Delta}_i^{\sigma_1\sigma_2}(z_1,z_2,\mathbf{k}) \\
&&\qquad=\int d^d\mathbf{r}\mbox{e}^{-i\mathbf{k}\mathbf{r}}{\bm \Delta}%
_i^{\sigma_1\sigma_2}(\widehat{\mathbf{n}_1\mathbf{n}_2},u_1,u_2,\mathbf{r}%
_\perp)\overline{f}(\mathbf{r})  \nonumber
\end{eqnarray}
for ${\bm \Delta}$ and use similar notation for amplitudes $\mathcal{B}$ and
$\Gamma$. To simplify expressions let us write
\begin{eqnarray}
{\bm \Delta}_\mathcal{F}(z_1,z_2,\mathbf{k})=\sum_{i=1}^4\sum_{\sigma_k=\pm}{%
\bm \Delta}_i^{\sigma_1\sigma_2}(z_1,z_2,\mathbf{k})\left(\mathcal{F}%
_{\sigma_1}\overline{\mathcal{F}_{\sigma_2}}\right)^i  \nonumber \\
\end{eqnarray}
and suppress the trivial dependence on $(z_1,z_2)$. Then the relevant
expressions read
\begin{eqnarray}
&&\delta\chi_1(\mathbf{q},\varepsilon)=-4\eta^2 T\sum_\omega \int d\mathbf{p}
dz_1dz_2\;u_1^2u_2^2\; \\
&&\times\mbox{str}\Big(\Delta_\mathcal{F}(\mathbf{p})\mathbbm{T}^{(1)}_{%
\mathbf{n}_1}(p,q)\Delta_\mathcal{F}(\mathbf{p}-\mathbf{q})\overline{%
\mathbbm{T}}^{(1)}_{\mathbf{n}_2}(p,q)\Big)  \nonumber
\end{eqnarray}
\begin{eqnarray}
&&\delta\chi_2(\mathbf{q},\varepsilon)=-8\eta^2T\sum_\omega\int d\mathbf{p}
dz_1dz_2\;u_1^3u_2 \\
&&\qquad\times\mbox{str}\Big({\bm \Delta}_{\mathcal{F}}(\mathbf{p})%
\mathbbm{T}_{\mathbf{n}_1}^{(2a)}(p,q){\bm \Delta}_\mathcal{F}(\mathbf{p})%
\mathbbm{T}_{\mathbf{n}_2}^{(2b)}(p)\Big)  \nonumber
\end{eqnarray}
\begin{eqnarray}
&&\delta{\chi_3}(\mathbf{q},\varepsilon)=8\eta^2\sum_{i=1}^{4}T\sum_\omega%
\int d\mathbf{p} dz_1dz_2\;u_1^2u_2^2 \\
&&\qquad\times \Gamma_i(\mathbf{p}-\mathbf{q})\;\overline{\mathcal{F}_0}%
\mathbbm{T}^{(3)}_{\mathbf{n}_1}(p,q){\bm \Delta}^i_\mathcal{F}(\mathbf{p})%
\overline{\mathbbm{T}}_{\mathbf{n}_2}^{(3)}(p,q)\mathcal{F}_0  \nonumber
\end{eqnarray}

\begin{eqnarray}
&&\delta{\chi_4}(\mathbf{q},\varepsilon)=-8\eta^2\sum_{i=1}^4\;T\sum_\omega%
\int d\mathbf{p} dz_1dz_2\;u_1^2u_2^2\;  \nonumber \\
&&\qquad\times\mathcal{B}_i^{\sigma_1}(\mathbf{p}-\mathbf{q})\mathcal{B}%
_i^{\sigma_2}(\mathbf{p}) \\
&&\qquad\times\overline{\mathcal{F}_0}\mathbbm{T}^{(4a)}_{\mathbf{n}%
_1}(p,q)\left(\mathcal{F}_{\sigma_1}\overline{\mathcal{F}_{\sigma_2}}%
\right)^i\mathbbm{T}^{(4b)}_{\mathbf{n}_2}(p,q)\mathcal{F}_0  \nonumber
\end{eqnarray}

\begin{eqnarray}
&&\chi_5(\mathbf{q},\varepsilon)=-16\eta^2\sum_{i=1}^{4}\sum_{\sigma_1}T%
\sum_\omega\int d\mathbf{p} dz_1dz_2\;u_1^2u_2^2  \nonumber \\
&&\quad\times\mathcal{B}_i^{\sigma_1}(\mathbf{p}-\mathbf{q})\;\overline{%
\mathcal{F}_{\sigma_1}}\mathbbm{T}_{\mathbf{n}_1}^{(5a)}(p,q){\bm \Delta}_%
\mathcal{F}^i(\mathbf{p})\mathbbm{T}_{\mathbf{n}_2}^{(5b)}(p,q)\mathcal{F}_0
\nonumber \\
\end{eqnarray}

\begin{eqnarray}
\chi_6(\mathbf{q},\varepsilon)=-16\eta^2\sum_{\sigma_1}\sum_{i=1}^4T\sum_%
\omega\int d\mathbf{p} dz_1dz_2\;u_1u_2^3  \nonumber \\
\mathcal{B}_i^{\sigma_1}(\mathbf{p}-\mathbf{q}) \;\overline{\mathcal{F}%
_{\sigma_1}}\mathbbm{T}^{(6a)}_{\mathbf{n}_1}(p){\bm \Delta}_\mathcal{F}^i(%
\mathbf{p})\mathbbm{T}_{\mathbf{n}_2}^{(6b)}(p,q)\mathcal{F}_0
\end{eqnarray}

\begin{eqnarray}
&&\chi_7(\mathbf{q},\varepsilon)=8\eta^2\sum_{\sigma_1,\sigma_2=\pm}%
\sum_{i=1}^4\;T\sum_\omega\int d\mathbf{p} dz_1dz_2\;u_1^3u_2\;  \nonumber \\
&&\;\times\mathcal{B}^{\sigma_1}_i(\mathbf{p})\mathcal{B}^{\sigma_2}_i(%
\mathbf{p}) \overline{\mathcal{F}}_{0}\mathbbm{T}_{\mathbf{n}%
_1}^{(7a)}(p,q)\left(\mathcal{F}_{0}\overline{\mathcal{F}_{\sigma_1}}%
\right)^i\mathbbm{T}_{\mathbf{n}_2}^{(7b)}(p)\mathcal{F}_{\sigma_2},
\nonumber \\
\end{eqnarray}
where
\begin{eqnarray}
\mathbbm{T}^{(1)}_{\mathbf{n}_1}(p,q)&=&\left[(i\omega+\mathbf{v}_1\mathbf{p}%
\Sigma_3)-\tau_+(i\varepsilon+\mathbf{v}_1\mathbf{q}\Sigma_3)\right]
\nonumber \\
&&\times\mathcal{G}_{\mathbf{n}_1}(p)\mathcal{G}_{\mathbf{n}_1}(p-q) \\
\mathbbm{T}_{\mathbf{n}_1}^{(2a)}(p,q)&=&\tau_3(i\omega+\mathbf{v}_1\mathbf{p%
}\Sigma_3)\mathcal{G}^2_{\mathbf{n}_1}(p)\mathcal{G}_{\mathbf{n}_1}(p+q) \\
\mathbbm{T}_{\mathbf{n}_2}^{(2b)}(p)&=&\tau_3(i\omega+\mathbf{v}_2\mathbf{p}%
\Sigma_3)\mathcal{G}_{\mathbf{n}_2}(p) \\
\mathbbm{T}^{(3)}_{\mathbf{n}_1}(p,q)&=&\Big[\tau_+(i\varepsilon+\mathbf{v}_1%
\mathbf{q}\Sigma_3)+\tau_-(i\omega+\mathbf{v}_1\mathbf{p}\Sigma_3)\Big]
\nonumber \\
&&\times\mathcal{G}_{\mathbf{n}_1}(p)\mathcal{G}_{\mathbf{n}_1}(q) \\
\mathbbm{T}^{(4a)}_{\mathbf{n}_1}(p,q)&=&\Big[(i\varepsilon+\mathbf{v}_1%
\mathbf{q}\Sigma_3)-\tau_-(i\omega+\mathbf{v}_1\mathbf{p}\Sigma_3)\Big]
\nonumber \\
&&\qquad\times\mathcal{G}_{\mathbf{n}_1}(q)\mathcal{G}_{\mathbf{n}_1}(q-p) \\
\mathbbm{T}^{(4b)}_{\mathbf{n}_2}(p,q)&=&\Big[\tau_+(i\omega+\mathbf{v}_2%
\mathbf{p}\Sigma_3)+\tau_-(i\varepsilon+\mathbf{v}_2\mathbf{q}\Sigma_3)\Big]
\nonumber \\
&&\qquad\times\mathcal{G}_{\mathbf{n}_2}(p)\mathcal{G}_{\mathbf{n}_2}(q) \\
\mathbbm{T}_{\mathbf{n}_1}^{(5a)}(p,q)&=&\Big[(i\omega+\mathbf{v}_1\mathbf{p}%
\Sigma_3)-\tau_+(i\varepsilon+\mathbf{v}_1\mathbf{q}\Sigma_3)\Big]
\nonumber
\\
&&\qquad\times \mathcal{G}_{\mathbf{n}_1}(p-q)\mathcal{G}_{\mathbf{n}_2}(p)
\\
\mathbbm{T}_{\mathbf{n}_2}^{(5b)}(p,q)&=&\Big[\tau_+(i\omega+\mathbf{v}_2%
\mathbf{p}\Sigma_3)+\tau_-(i\varepsilon+\mathbf{v}_2\mathbf{q}\sigma_3)\Big]
\nonumber \\
&&\qquad\times \mathcal{G}_{\mathbf{n}_2}(p)\mathcal{G}_{\mathbf{n}_2}(q) \\
\mathbbm{T}^{(6a)}_{\mathbf{n}_1}(p)&=&\tau_3(i\omega+\mathbf{v}_1\mathbf{n}%
_1\mathbf{p}\Sigma_3)\mathcal{G}_{\mathbf{n}_1}(p) \\
\mathbbm{T}_{\mathbf{n}_2}^{(6b)}(p,q)&=&\Big[\tau_+(i\omega+\mathbf{v}_2%
\mathbf{p}\Sigma_3)+\tau_-(i\varepsilon+\mathbf{v}_2\mathbf{q}\Sigma_3)\Big]
\nonumber \\
&&\qquad\times\tau_3\mathcal{G}_{\mathbf{n}_2}(p)\mathcal{G}_{\mathbf{n}%
_2}(p+q)\mathcal{G}_{\mathbf{n}_2}(q) \\
\mathbbm{T}_{\mathbf{n}_1}^{(7a)}(p,q)&=&\tau_3(i\varepsilon+\mathbf{v}_1%
\mathbf{p}\Sigma_3)\mathcal{G}^2_{\mathbf{n}_1}(q)\mathcal{G}_{\mathbf{n}%
_1}(p+q) \\
\mathbbm{T}_{\mathbf{n}_2}^{(7b)}(p)&=&\tau_3(i\omega+\mathbf{v}_2\mathbf{p}%
\Sigma_3)\mathcal{G}_{\mathbf{n}_2}(p)
\end{eqnarray}
Four dimensional notation was used $p=(\omega,\mathbf{p})$, $q=(\varepsilon,%
\mathbf{q})$.

\end{document}